\documentclass[12pt,a4paper]{article}
\usepackage[utf8]{inputenc}
\usepackage{amsmath,graphicx,dcolumn,subcaption,mathtools,amssymb}
\usepackage{braket}
\usepackage{siunitx}
\usepackage{appendix}
\usepackage[mathscr]{eucal}
\usepackage{bm}
\usepackage{bbm}
\usepackage{cancel}
\usepackage{color}
\usepackage{rotating}
\usepackage{xcolor}
\usepackage[most]{tcolorbox}
\usepackage{comment}

\usepackage[numbers,sort&compress]{natbib}
\usepackage[]{float}
\usepackage[font=small, labelfont=bf]{caption}

\usepackage{slantsc}
\usepackage{xspace}

\usepackage{setspace}
\allowdisplaybreaks

\usepackage{hyperref}
\usepackage{footnotebackref}

\providecommand{\href}[2]{#2}

\newcommand\as{\alpha_{\mathrm{S}}}

\def\to{\rightarrow}

\newcommand\Matrix{{\sc Matrix}\xspace}

\newcommand\OpenLoops{{\sc OpenLoops}\xspace}
\newcommand\Recola{{\sc Recola}\xspace}

\newcommand{\VVamp}{{\sc VVamp}\xspace}

\def\refeq#1{\mbox{Eq.\,\eqref{#1}}}

\def\refeqtwo#1#2{\mbox{Eqs.\ \eqref{#1} and \eqref{#2}}}
\def\refeqtoeq#1#2{\mbox{Eqs.\ \eqref{#1} -- \eqref{#2}}}
\def\reffi#1{\mbox{Figure~\ref{#1}}}
\def\reffitwo#1#2{\mbox{Figures~\ref{#1} and \ref{#2}}}
\def\reffis#1#2{\mbox{Figures~\ref{#1}--\ref{#2}}}
\def\refta#1{\mbox{Table~\ref{#1}}}
\def\reftatwo#1#2{\mbox{Tables~\ref{#1} and \ref{#2}}}

\def\refse#1{\mbox{Section~\ref{#1}}}

\def\refapp#1{\mbox{App.~\ref{#1}}}

\newcommand\GeV{\ensuremath{{\rm GeV}}\xspace}
\newcommand\TeV{\ensuremath{{\rm TeV}}\xspace}

\newcommand\rcut{\ensuremath{r_{\mathrm{cut}}}\xspace}
\newcommand\qT{\ensuremath{q_T}\xspace}
\newcommand\muF{\ensuremath{\mu_F}\xspace}
\newcommand\muR{\ensuremath{\mu_R}\xspace}
\newcommand\muIR{\ensuremath{\mu_{\rm IR}}\xspace}
\newcommand\tmuR{\ensuremath{\tilde{\mu}_R}\xspace}
\newcommand\tmuIR{\ensuremath{\tilde{\mu}_{\rm IR}}\xspace}
\newcommand\tQ{\ensuremath{\tilde{Q}}\xspace}

\newcommand\Vgamma{\ensuremath{V\gamma}\xspace}
\newcommand\Zgamma{\ensuremath{Z\gamma}\xspace}
\newcommand\Wgamma{\ensuremath{W\gamma}\xspace}

\newcommand\WW{\ensuremath{WW}\xspace}

\newcommand\Wgammagamma{\ensuremath{W\gamma\gamma}\xspace}

\newcommand\Zgammagamma{\ensuremath{Z\gamma\gamma}\xspace}
\newcommand\WWgamma{\ensuremath{WW\gamma}\xspace}

\newcommand\WZgamma{\ensuremath{WZ\gamma}\xspace}

\newcommand\VVV{\ensuremath{VVV}\xspace}
\newcommand\VVZ{\ensuremath{VVZ}\xspace}
\newcommand\WWW{\ensuremath{WWW}\xspace}
\newcommand\WWZ{\ensuremath{WWZ}\xspace}
\newcommand\WZZ{\ensuremath{WZZ}\xspace}
\newcommand\ZZZ{\ensuremath{ZZZ}\xspace}

\newcommand\ttW{\ensuremath{t \bar t W}\xspace}

\newcommand\ttH{\ensuremath{t \bar t H}\xspace}

\newcommand{\pplmlpa}{\ensuremath{pp\to \ell^-\ell^+\gamma}\xspace}
\newcommand{\ppnlnlxa}{\ensuremath{pp\to \nu_{\ell}\bar\nu_{\ell}\gamma}\xspace}
\newcommand{\pplmnlxa}{\ensuremath{pp\to \ell^-\bar\nu_{\ell}\gamma}\xspace}
\newcommand{\pplpnla}{\ensuremath{pp\to \ell^+\nu_{\ell}\gamma}\xspace}

\newcommand{\qqxlmlpa}{\ensuremath{q \bar{q} \to \ell^{-}\ell^{+}\gamma}\xspace}

\newcommand{\pplmlppnlpnlxa}{\ensuremath{pp\to \ell^-\ell^{\prime +}\nu_{\ell{'}}\bar{\nu}_{\ell}\gamma}\xspace}
\newcommand{\ppemxnmnexa}{\ensuremath{pp\to e^{-}\mu^{+}\nu_{\mu}\bar{\nu}_e\gamma}\xspace}
\newcommand{\ppmexnenmxa}{\ensuremath{pp\to \mu^{-}e^{+}\nu_{e}\bar{\nu}_{\mu}\gamma}\xspace}
\newcommand{\qqxlmlppnlpnlxa}{\ensuremath{q \bar{q} \to \ell^-\ell^{\prime +}\nu_{\ell{'}}\bar{\nu}_{\ell}\gamma}\xspace}

\newcommand{\ppttxa}{\ensuremath{pp\to t\bar t\gamma}\xspace}
\newcommand{\pplmlppnlpnlxbbxa}{\ensuremath{pp\to \ell^-\ell^{\prime +}\nu_{\ell{'}}\bar{\nu}_{\ell}b\bar{b}\gamma}\xspace}

\newcommand\Hone{\ensuremath{H^{(1)}}\xspace}
\newcommand\HoneSA{\ensuremath{H^{(1)}_{\rm SA}}\xspace}
\newcommand\Htwo{\ensuremath{H^{(2)}}\xspace}
\newcommand\HtwoSA{\ensuremath{H^{(2)}_{\rm SA}}\xspace}

\newcommand\Hn{\ensuremath{H^{(n)}}\xspace}

\newcommand\noRW{\ensuremath{{\rm bare}}\xspace}

\newcommand\HtwoSAnoRW{\ensuremath{H^{(2)}_{\rm SA,\,\noRW}}\xspace}

\newcommand\HnSAnoRW{\ensuremath{H^{(n)}_{\rm SA,\,\noRW}}\xspace}
\newcommand\MzeroRW{\ensuremath{|\mathcal{M}_0|^2\textnormal{-RW}}\xspace}

\newcommand\HnSAMzeroRW{\ensuremath{H^{(n)}_{\rm SA,\MzeroRW}}\xspace}
\newcommand\MoneMzeroRW{\ensuremath{\mathcal{M}_1\mathcal{M}_0\textnormal{-RW}}\xspace}
\newcommand\HtwoSAMoneMzeroRW{\ensuremath{H^{(2)}_{\rm SA,\MoneMzeroRW}}\xspace}
\newcommand\MoneRW{\ensuremath{|\mathcal{M}_1|^2\textnormal{-RW}}\xspace}
\newcommand\HtwoSAMoneRW{\ensuremath{H^{(2)}_{\rm SA,\MoneRW}}\xspace}
\newcommand\calH{\ensuremath{\mathcal{H}}\xspace}
\newcommand\deltacalH{\ensuremath{\delta\mathcal{H}}\xspace}

\newcommand\dsHone{\ensuremath{d\sigma_{\hspace*{-0.2em}H^{(1)}}}\xspace}
\newcommand\dsHoneSA{\ensuremath{d\sigma_{\hspace*{-0.2em}H^{(1)}_{\rm SA}}}\xspace}
\newcommand\dsHtwo{\ensuremath{d\sigma_{\hspace*{-0.2em}H^{(2)}}}\xspace}
\newcommand\dsHtwoSA{\ensuremath{d\sigma_{\hspace*{-0.2em}H^{(2)}_{\rm SA}}}\xspace}

\newcommand\dsHn{\ensuremath{d\sigma_{\hspace*{-0.2em}H^{(n)}}}\xspace}

\newcommand\dsHtwoSAMzeroRW{\ensuremath{d\sigma_{\hspace*{-0.2em}H^{(2)}_{\rm SA,\,\MzeroRW}}}\xspace}
\newcommand\dsHtwoSAMoneRW{\ensuremath{d\sigma_{\hspace*{-0.2em}H^{(2)}_{\rm SA,\,\MoneRW}}}\xspace}

\newcommand\dsHtwoSAMSQovertwo{\ensuremath{d\sigma_{\hspace*{-0.2em}H^{(2)}_{\rm SA}(Q/2)}}\xspace}
\newcommand\dsHtwoSAMSQtwo{\ensuremath{d\sigma_{\hspace*{-0.2em}H^{(2)}_{\rm SA}(2Q)}}\xspace}
\newcommand\dsHtwoSAnoRW{\ensuremath{d\sigma_{\hspace*{-0.2em}H^{(2)}_{\rm SA,\,\noRW}}}\xspace}

\newcommand{\pTgamma}{\ensuremath{p_{T,\gamma}}\xspace}

\newcommand{\absetagamma}{\ensuremath{|\eta_{\gamma}|}\xspace}

\newcommand{\pTe}{\ensuremath{p_{T,e}}\xspace}
\newcommand{\etae}{\ensuremath{\eta_{e}}\xspace}
\newcommand{\absetae}{\ensuremath{|\etae|}\xspace}
\newcommand{\pTem}{\ensuremath{p_{T,e^-}}\xspace}

\newcommand{\pTmu}{\ensuremath{p_{T,\mu}}\xspace}
\newcommand{\etamu}{\ensuremath{\eta_{\mu}}\xspace}
\newcommand{\absetamu}{\ensuremath{|\etamu|}\xspace}

\newcommand{\pTlep}{\ensuremath{p_{T,\ell}}\xspace}

\newcommand{\absetalep}{\ensuremath{|\eta_{\ell}|}\xspace}

\newcommand{\pTlepone}{\ensuremath{p_{T,\ell_1}}\xspace}

\newcommand{\pTleplep}{\ensuremath{p_{T,\ell\ell}}\xspace}

\newcommand{\pTmiss}{\ensuremath{p_{T,{\rm miss}}}\xspace}

\newcommand{\dsigma}{\ensuremath{d\sigma}\xspace}
\newcommand{\sigmaLO}{\ensuremath{\sigma_{\rm LO}}\xspace}
\newcommand{\dsigmaLO}{\ensuremath{d\sigma_{\rm LO}}\xspace}

\newcommand{\DeltasigmaNLO}{\ensuremath{\Delta\sigma_{\rm NLO}}\xspace}

\newcommand{\DeltasigmaNNLO}{\ensuremath{\Delta\sigma_{\rm NNLO}}\xspace}

\newcommand{\dsigmaR}{\ensuremath{d\sigma_{\rm R}}\xspace}

\newcommand{\dsigmaCT}{\ensuremath{d\sigma_{\rm CT}}\xspace}

\newcommand\DeltaSA{\ensuremath{\delta_{\rm SA}}\xspace}
\newcommand\DeltaSAHone{\ensuremath{\delta^{\Hone}_{\rm SA}}\xspace}
\newcommand\DeltaSAmuIR{\ensuremath{\delta^{\rm \mu_{\rm IR}}_{\rm SA}}\xspace}
\newcommand\DeltaSARW{\ensuremath{\delta^{\rm RW}_{\rm SA}}\xspace}

\usepackage{etoolbox}
\makeatletter
\patchcmd{\@sect}{#8}{\boldmath #8}{}{}
\let\ori@chapter\@chapter
\def\@chapter[#1]#2{\ori@chapter[\boldmath#1]{\boldmath#2}}
\makeatother

\newlength{\plotheightstd}
\newlength{\plotheightapp}
\begin{document}
\hypersetup{pageanchor=false}
\begin{titlepage}
\begin{flushright}
ZU-TH 28/26\\
\end{flushright}

\renewcommand{\thefootnote}{\fnsymbol{footnote}}
\vspace*{0.5cm}

\begin{center}
{\Large \bf NNLO QCD corrections to\\[0.3cm] $\boldsymbol{WW\gamma}$ production at the LHC}
\end{center}

\par \vspace{2mm}
\begin{center}
{\bf Paolo Garbarino}$^{(a)}$, {\bf Massimiliano Grazzini}$^{(a)}$ and   {\bf Stefan Kallweit}$^{(a)}$

\vspace{5mm}

$^{(a)}$ Physik Institut, Universit\"at Z\"urich, Winterthurerstrasse 190, 8057 Z\"urich, Switzerland\\[0.2cm]

\vspace{5mm}

\end{center}

\par \vspace{2mm}
\begin{center} {\large \bf Abstract} 

\end{center}
\begin{quote}
\pretolerance 10000
Triboson production processes are crucial to study quartic gauge-boson couplings.
We present the computation of the radiative corrections to \WWgamma production at the
next-to-next-to-leading order~(NNLO) in QCD. The leptonic decays of the $W$ bosons and off-shell effects are fully included.
The calculation is exact, apart from the finite part of the two-loop amplitudes, which
is evaluated in a soft-photon approximation.
We validate our approach by using \Zgamma production as a reference process, where we take advantage of the
availability of the exact two-loop amplitudes to derive a conservative error estimate for our approximation,
and then apply it to \WWgamma production.
For typical selection cuts, at the centre-of-mass energy \mbox{$\sqrt{s}=13$\,TeV},
the NNLO corrections increase the next-to-leading order (NLO) result by about $16\%$,
and the perturbative uncertainties are reduced to the $\pm4\%$ level.
The uncertainty from the soft approximation turns out to be at the few per mille level, largely subdominant compared to the
residual perturbative uncertainties, both for the fiducial cross section and the most relevant differential distributions.
\end{quote}

\vspace*{\fill}
\begin{flushleft}
August 2026
\end{flushleft}
\end{titlepage}

\clearpage
\pagenumbering{arabic}
\setcounter{page}{1}
\hypersetup{pageanchor=true}


\renewcommand{\thefootnote}{\fnsymbol{footnote}}

\section{Introduction}
\label{sec:introduction}
The precise understanding of the electroweak (EW) symmetry breaking mechanism is a
paramount task of the physics programme of the LHC at CERN.
While diboson production has been and is further studied in great detail to
measure the trilinear couplings of the EW gauge bosons, triboson production
processes are --- together with diboson production in weak-boson scattering --- the key
to investigate their quartic couplings.
Any deviations from the Standard Model (SM) predictions would point towards new-physics effects.
Given the purely EW structure of these processes,
with three powers of the EW coupling already entering at the production stage,
triboson cross sections are significantly smaller than those for diboson production.
After previous searches for triboson signatures
by the ATLAS~\cite{ATLAS:2016jeu,ATLAS:2017bon,ATLAS:2019dny} and
CMS~\cite{CMS:2014cdf} Collaborations, first measurements of triboson production
processes could be achieved: Triphoton production has been measured by ATLAS
at a centre-of-mass energy of 8\,\TeV~\cite{ATLAS:2017lpx}.
\Wgammagamma final states have been studied by ATLAS at 8\,\TeV~\cite{ATLAS:2015ify}
and 13\,\TeV~\cite{ATLAS:2023avk}, as well as 
\Zgammagamma final states at 8\,\TeV~\cite{ATLAS:2016qjc}
and 13\,\TeV~\cite{ATLAS:2022wmu}. CMS has presented measurements of both
\Zgammagamma and \Wgammagamma production at
8\,\TeV~\cite{CMS:2017tzy} and 13\,\TeV~\cite{CMS:2021jji}.
First observations of final states with a massive vector-boson pair in
association with a photon were recently reported in 13\,\TeV analyses, for both \WZgamma and \WWgamma production,
by ATLAS~\cite{ATLAS:2023zkw, ATLAS:2025yxf} and CMS~\cite{CMS:2025oey, CMS:2023rcv}.
Moreover, final states with three massive bosons have been observed in 13\,\TeV analyses: CMS reported the observation
of \VVV final states in a combined analysis of \WWW, \WZZ,
\WWZ and \ZZZ production, based mostly on the fully leptonic decay channels
(also including same-sign dilepton+two-jet signatures)~\cite{CMS:2020hjs},
and more recently a \WWZ measurement using data at 13 and 13.6\,\TeV~\cite{CMS:2025hlu}. 
ATLAS presented observations of the \WWW channel~\cite{ATLAS:2022xnu}, and of \VVZ final states
in a combined analysis~\cite{ATLAS:2024nab}.

All these measurements have been carried out almost exclusively in the leptonic decay channels
due to the significantly lower backgrounds. Also from a theoretical point of view
those decay modes can be predicted with highest precision because the
complexity of QCD corrections is technically almost unchanged if leptonic decays are included.
Next-to-leading-order~(NLO) QCD
corrections~\cite{Lazopoulos:2007ix,Hankele:2007sb,Binoth:2008kt,Campanario:2008yg,Bozzi:2009ig,Bozzi:2010sj,Bozzi:2011wwa,Baur:2010zf,Bozzi:2011en,Campbell:2012ft}
are known to be essential to get at least close
to reasonable predictions for LHC triboson measurements, given that at leading order~(LO)
these processes proceed only through quark--antiquark annihilation. Thus, the impact of the
gluon content of the protons is accounted for only starting at NLO QCD, often resulting in
very large positive corrections. Recently, NLO QCD corrections to triboson production have also been computed
in the SM effective field theory~\cite{Celada:2024cxw}.

The calculation of NLO EW corrections is complicated by the fact that the leptonic
decays do not factorize. Predictions for on-shell triboson production, in part including
decays in a factorized form like a narrow-width approximation, were
presented in Refs.~\cite{Nhung:2013jta,Shen:2015cwj,Shen:2016ape,Wang:2016fvj,Dittmaier:2017bnh,Wang:2017wsy,Zhu:2020ous}.
Although the calculation of EW corrections has been widely automated by several
groups~\cite{Frederix:2018nkq,Proceedings:2018jsb}, performing a full off-shell
calculation of NLO EW corrections remains a formidable task, in particular for the
six-particle final states arising from the decays of three massive bosons. Here,
only the experimentally most relevant cases have been completed so far:
for fully leptonic decays of all three massive gauge bosons,
off-shell NLO EW corrections have been achieved only for
\WWW production~\cite{Schonherr:2018jva,Dittmaier:2019twg}.
Results for the final states with one of the three massive bosons decaying hadronically,
together with two same-sign~\cite{Denner:2024ufg}
and three charged leptons~\cite{Denner:2024ndl}, are also available.
Complete NLO EW corrections for triboson processes involving
two or more final-state photons were studied in Ref.~\cite{Greiner:2017mft},
and for \WZgamma production in Ref.~\cite{Cheng:2021gbx}, including leptonic gauge-boson decays.

For diboson processes it has been shown that the next-to-next-to-leading order~(NNLO)
QCD corrections are significant and positive, in particular for processes involving
direct photons~\cite{Catani:2011qz,Grazzini:2013bna,Grazzini:2015nwa,Campbell:2016lzl,Grazzini:2017mhc,Campbell:2017aul,Catani:2018krb,Campbell:2021mlr}.
Given the similarity of the production mechanisms between diboson and triboson processes,
a similar pattern may be expected for the latter as well. 
Indeed, for triphoton production~\cite{Chawdhry:2019bji,Kallweit:2020gcp}%
\footnote{These results are based on a leading-colour approximation of the involved two-loop amplitudes.
Complete amplitudes including subleading-colour contributions have been made available in the
meanwhile~\cite{Abreu:2023bdp}.}, huge NNLO corrections of $\mathcal{O}(60\%)$
wrt.\ NLO predictions at the level of fiducial cross sections have been found,
and it was shown that good agreement with the data measured by
ATLAS~\cite{ATLAS:2017lpx} could only be achieved by including those corrections.
Recently, first NNLO QCD predictions for \Wgammagamma~\cite{Garbarino:2025bfg} production were presented,
relying on the leading-colour two-loop amplitudes computed in Ref.~\cite{Badger:2024sqv}.
In that work, it was shown that NNLO corrections are again sizable, increasing the
NLO fiducial cross section by $\mathcal{O}(20\%)$.
The extension of these calculations to other triboson processes is limited by the
knowledge of corresponding two-loop amplitudes.
For the triboson class with
two off-shell legs, which is the relevant case for \WWgamma production, only
the planar two-loop master integrals are known by now~\cite{Abreu:2024yit}.
Given the remarkable progress of recent years, full two-loop amplitudes
for the complete class of triboson production processes
appear to be within reach in the not-too-distant future.

In this paper, we follow a different path, which is based on the observation that
for most NNLO QCD calculations performed so far the impact of the (properly regularised)
two-loop amplitudes has turned out to be numerically minor. Consequently, even quite rude
approximations of these amplitudes promise to provide sufficiently accurate 
predictions,
as long as the rest of the calculation is performed exactly.
We adopt this approach and perform an NNLO QCD calculation
for \WWgamma production --- more precisely, the complete calculation for the corresponding
different-flavour leptonic final state --- within the \Matrix framework~\cite{Grazzini:2017mhc}.
Only the yet unknown two-loop amplitudes do not enter our calculation exactly,
but in an appropriate soft-photon approximation.
To judge the quality of this approximation and to provide an error estimate for it
on a fully differential level, we study a sample diboson process, namely \Zgamma production in its leptonic decay mode.
Here, the full two-loop amplitudes are available~\cite{Gehrmann:2011ab},
and thus an exact calculation
at NNLO QCD accuracy can be performed~\cite{Grazzini:2013bna,Grazzini:2015nwa,Campbell:2017aul}.
This way we are not only able to test the performance of the soft approximation,
but also to confront our procedure to estimate its error with the true NNLO result.
The outcome of this study gives
us confidence that the approach is able to provide reliable predictions within
this error estimate also in cases where exact amplitudes are not yet available, and that these results
can be considered NNLO accurate since the systematic approximation error is much smaller than the
residual perturbative uncertainties.
  
The paper is organized as follows: In \refse{sec:calculation} we provide information about how our
calculation is performed, with a particular focus on the soft-photon approximation that we apply on
the yet unknown two-loop amplitudes, describing how we estimate the related error on our final results.
In \refse{sec:validationZA}, this procedure is validated in a calculation
for \Zgamma hadroproduction, and our approximation is confronted with the known exact NNLO QCD results.
Based on this validation, in \refse{sec:resultsWWA} we apply the same approach to \WWgamma hadroproduction
and thereby provide the first predictions for a triboson process involving two off-shell massive weak bosons
at NNLO QCD accuracy. We conclude our discussion in \refse{sec:summary}.
To further confirm the validity of our soft-photon approximation and in particular its error estimate,
we provide a selection of additional differential distributions, together with details on the error estimates, for our
validation process, \Zgamma production, in \refapp{app:validationZA} and for the actual process of interest,
\WWgamma production, in \refapp{app:resultsWWA}.

\section{Calculational details}
\label{sec:calculation}
In this section, we discuss how the calculation at NNLO QCD accuracy is performed.
Our validation process, \mbox{\pplmlpa}, as well as the actual process of interest, \mbox{\pplmlppnlpnlxa},
belong to the class of hadroproduction processes of colourless final-state systems and can thus
be addressed by applying the standard \qT-subtraction formalism~\cite{Catani:2007vq},
as implemented in the \Matrix framework~\cite{Grazzini:2017mhc} (see \refse{sec:qTsubtraction}).
Note that we will usually refer to these processes as \Zgamma and \WWgamma production,
respectively, for brevity. However, we always consider the complete hadronic production of the leptonic final
states, in amplitudes and phase space, i.e.\ we take into account all off-shell effects, resonant and
non-resonant contributions, interferences, spin correlations, etc., without making use
of any kind of resonance approximation.
The whole calculation is carried out exactly for our validation process, \Zgamma production, up to NNLO QCD
accuracy, whereas for \WWgamma production we apply a soft-photon approximation
only on the finite remainders of the two-loop amplitudes (see \refse{sec:softphotonapproximation}),
while keeping the rest of the computation exact.

\subsection[NNLO QCD accuracy through the \qT-subtraction method]{NNLO QCD accuracy through the $\boldsymbol{q_T}$-subtraction method}
\label{sec:qTsubtraction}
In general, if only QCD corrections are considered, cross sections can be written as
\begin{equation}
\sigma=\sigmaLO+\DeltasigmaNLO+\DeltasigmaNNLO+...\,,
\end{equation}
where \sigmaLO is the LO cross section, \DeltasigmaNLO the NLO QCD
correction, \DeltasigmaNNLO the NNLO QCD correction, and so forth.

Besides the challenge of obtaining the relevant scattering
amplitudes, in particular those at two-loop level, the implementation of a complete
NNLO calculation needs to deal with the presence of infrared~(IR) divergences at
intermediate stages.
In this work, NNLO IR singularities are handled and cancelled by using the \qT-subtraction
formalism~\cite{Catani:2007vq}, which is suitable for the hadroproduction of any
colourless final-state system $F$.
According to the \qT-subtraction formalism, the differential cross section \dsigma
can be evaluated as
\begin{equation}
\label{eq:mainqTsubtraction}
d\sigma={\cal H}\otimes d\sigmaLO+\left[\dsigmaR-\dsigmaCT\right]\, .
\end{equation}
The first term on the right-hand side of \refeq{eq:mainqTsubtraction} corresponds
to the contribution with vanishing transverse momentum \qT of $F$, i.e.\ \mbox{$\qT=0$}.
It is obtained through a convolution, with
respect to the longitudinal-momentum fractions $z_1$ and $z_2$ of the colliding partons,
of the perturbatively computable function ${\cal H}$ with the LO cross section \dsigmaLO.
The real contribution \dsigmaR corresponds to the cross section to
produce the colourless system $F$ accompanied by additional QCD radiation that provides a recoil
with finite transverse momentum \qT.
When \dsigma is evaluated at NNLO, \dsigmaR
is obtained through an NLO calculation for $F+$jet production
by using the dipole subtraction
formalism~\cite{Catani:1996jh,Catani:1996vz,Catani:2002hc}.
The role of the counterterm \dsigmaCT is to cancel the singular behaviour of
\dsigmaR in the limit \mbox{$\qT\to 0$},
rendering the square bracket term in \refeq{eq:mainqTsubtraction} finite.

Our computation is implemented within a suitable extension of the \Matrix framework~\cite{Grazzini:2017mhc}.
The required tree-level and one-loop amplitudes are obtained with
\OpenLoops~\cite{Cascioli:2011va, Buccioni:2017yxi,Buccioni:2019sur}
and \Recola~\cite{Actis:2016mpe,Denner:2017wsf,Denner:2016kdg}.
In order to numerically evaluate the contribution in the square bracket of
\refeq{eq:mainqTsubtraction}, a technical cut-off \rcut is introduced on the dimensionless
variable \mbox{$r=\qT/Q$}, where $Q$ is the invariant mass of the colourless system.
The final result is extracted by computing
the cross section at fixed values of \rcut and numerically performing the extrapolation \mbox{$\rcut\to 0$}.
More details on the procedure and its uncertainties can be found in
Ref.~\cite{Grazzini:2017mhc}. We note that this extrapolation procedure
is performed on a bin-wise level for distributions, thereby providing also differential results
that are free of power corrections in \rcut, but contain an extrapolation error to account for
the uncertainties related to this procedure.

\begin{figure}[t]
\centering
\includegraphics[height=0.22\textheight]{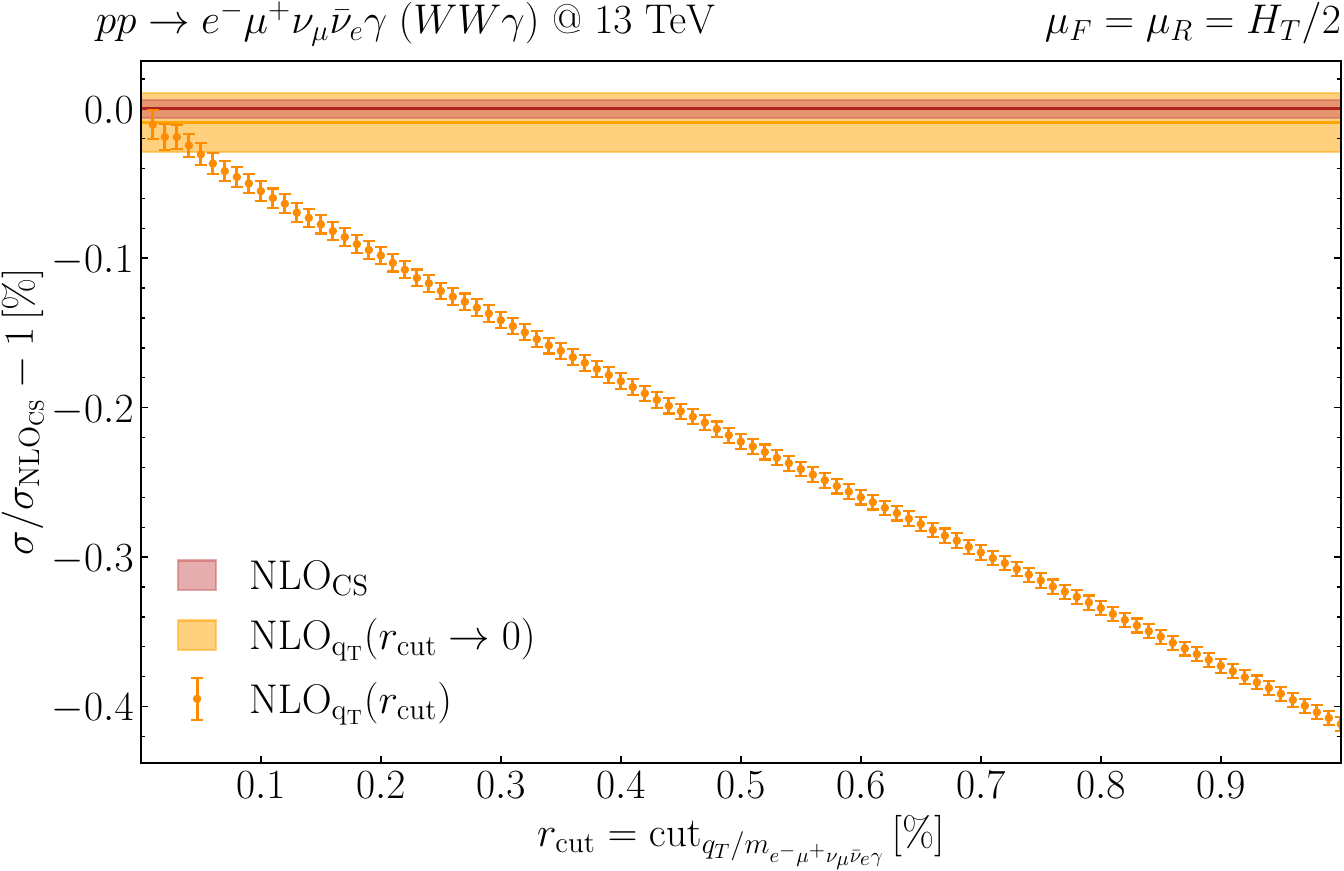}
\hfill
\includegraphics[height=0.22\textheight]{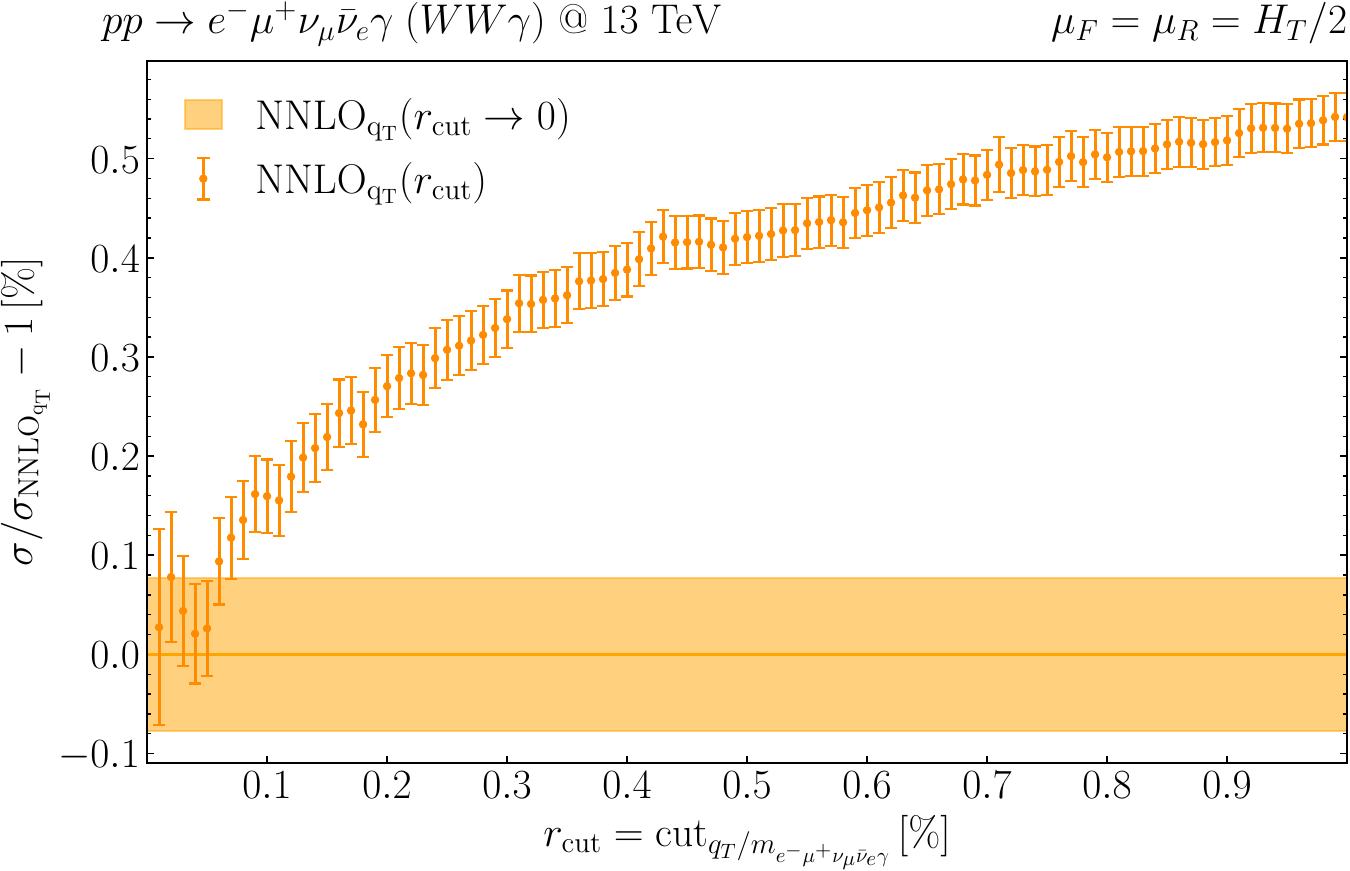}
\vspace*{1ex}
\caption{Illustration of the \rcut dependence at NLO (left) and NNLO (right) for \WWgamma production for the setup used by ATLAS in Ref.~\cite{ATLAS:2025yxf}.} 
\label{fig:qTdependenceWWA} 
\end{figure}

We anticipate that the \rcut dependence is well
under control for both processes under consideration, with numerical errors at the
per mille level for the fiducial cross sections, as illustrated in \reffi{fig:qTdependenceWWA}
for \WWgamma production in the setup used by ATLAS in Ref.~\cite{ATLAS:2025yxf} (see \refse{sec:fidresu}).
For reference, we also show the NLO cross section calculated by using \qT subtraction (${\rm NLO}_{q_T}$), compared to the
\rcut-independent result obtained with dipole subtraction (${\rm NLO}_{\rm CS}$). 
The significant \rcut dependence of the cross section is a well-known feature of processes
involving isolated photons~\cite{Grazzini:2017mhc,Ebert:2019zkb};
attempts to reduce this dependence have been discussed
in Ref.~\cite{Campbell:2024hjq}. We refrain from further elaborating on this situation here,
since we find the numerical control we achieve over those power corrections sufficient for
our purposes.

\subsection{Soft-photon approximation for two-loop amplitudes}
\label{sec:softphotonapproximation}
The finite part of the $n$-loop virtual amplitude enters the N$^{n}$LO cross section ($n=1,2$) through its interference
with the Born level amplitude. The hard function $H$, related to ${\cal H}$
in \refeq{eq:mainqTsubtraction} through
\begin{equation}\label{eq:HHdeltaH}
{{\cal H}=H\delta(1-z_1)\delta(1-z_2)+\delta{\cal H}}\,,
\end{equation}
is defined through its perturbative coefficients in an expansion in powers of the QCD coupling $\as(\muR)$ as
\begin{equation}\label{eq:Hn}
\Hn(\muIR) = \left.\frac{2{\rm Re}\left({\cal M}^{(n)}_{\rm fin}(\muR, \muIR) {\cal M}^{(0)*} \right)}
{\left|{\cal M}^{(0)}\right|^2}\right|_{\muR=Q}\,.
\end{equation}
Here, \muR is the renormalisation scale, and ${\cal M}_{\rm fin}^{(n)}$
are the perturbative coefficients of the finite parts of the renormalised virtual $n$-loop amplitudes
for the partonic processes under consideration (\mbox{\qqxlmlpa} and
\mbox{\qqxlmlppnlpnlxa} for \Zgamma and \WWgamma, respectively),
after the subtraction of IR singularities at the scale \muIR.
Following previous works on \ttH~\cite{Catani:2022mfv,Devoto:2024nhl} and \ttW~\cite{Buonocore:2023ljm} production,
the finite part is defined according to the
conventions of Ref.~\cite{Becher:2009cu}.
In order to obtain an approximation of the NNLO coefficients \Htwo, we use a soft-photon approximation~(SA),
as outlined in the following.

\subsubsection{Soft-photon factorisation}
\label{sec:softfact}
It has long been known that when a photon with an energy $k_0$ much smaller than the characteristic energies of the
other external charged particles is radiated in a hard
scattering process, the scattering amplitude factorises as
\begin{equation}
\label{eq:softM}
\mathcal{M}(p_1,p_2,p_3,\ldots, p_n,k)\approx \sqrt{4\pi\alpha}\,\sum_{i=1}^n Q_i\,\frac{p_i\cdot \varepsilon^*(k)}{p_i\cdot k}\,
\mathcal{M}(p_1,p_2,p_3,\ldots, p_n)\,,
\end{equation}
where the symbol '$\approx$' means that on the right-hand side we have neglected contributions that are less
singular than $1/k_0$ in the soft limit.
In \refeq{eq:softM}, $Q_i$ is the charge of the radiating particle (in units of $e>0$) in case the particle is outgoing
and its opposite in case it is incoming, and $\varepsilon^\ast(k)$ is the polarization vector of the emitted photon.
Squaring the amplitude, summing over the photon polarisations and using gauge invariance, we obtain
\begin{align}
\label{eq:softM2}
\left|\mathcal{M}(p_1,p_2,p_3,\ldots, p_n,k)\right|^2 \approx&\; -4\pi\alpha
\sum_{i,j=1}^{n}Q_iQ_j\,\frac{p_i\cdot p_j}{\left(p_j\cdot k\right)\left(p_i\cdot k\right)}
\left|\mathcal{M}(p_1,p_2,p_3,\ldots, p_n)\right|^2\, .
\end{align}
In \refeqtwo{eq:softM}{eq:softM2}, $\mathcal{M}(p_1,p_2,p_3,\ldots p_n,k)$ denotes the scattering amplitude
of the full partonic process, and ${\mathcal{M}}(p_1,p_2,p_3,\ldots, p_n)$ that of the corresponding
reduced process in which the photon has been removed.

The factorisation formula in \refeq{eq:softM} holds in the strict soft-photon limit, \mbox{$k\to0$}.
Away from that regime, a procedure to absorb the photon momentum is required in order to evaluate the
reduced amplitude, where the remaining momenta are kept on shell.
Such procedure requires a projection that for a given set of momenta $\{p_i,k\}$
defines a projected set of momenta $\{\tilde{p}_i\}$.
In this paper, we use the so-called \qT-recoil prescription~\cite{Catani:2015vma}, where 
the photon momentum is absorbed symmetrically by the initial-state partons,
thereby modifying also the partonic centre-of-mass energy from \mbox{$Q^2=(p_1+p_2)^2$}
to \mbox{$\tilde{Q^2}=(\tilde{p}_1+\tilde{p}_2)^2$} for the reduced process,
while all other final-state momenta remain unchanged.
This way, invariant masses of possible resonances that do not involve the photon are preserved.%
\footnote{The impact of varying the exact procedure within this preferred class of projections is minor.
We thus do not introduce an explicit projection uncertainty in our error estimate.}

Since virtual QCD corrections do not affect the soft factorisation formula up to the two-loop order%
\footnote{Equivalently, the soft QED eikonal current does not receive corrections up to two-loop order.
From three-loop order, massless fermion-loop contributions~\cite{Ma:2023gir} appear that modify the
factorisation picture.}, \refeq{eq:softM} holds beyond the tree level and can thus be used
to obtain an approximation of the finite remainders which enter
the \Hn coefficients (see Eq.~\eqref{eq:Hn}) up to \mbox{$n=2$},
\begin{align}
\label{eq:softapproxHO}
&\mathcal{M}^{(n)}_{\rm fin}(p_1,p_2,p_3,\ldots, p_n,k;\muR, \muIR)\approx\nonumber\\
&\qquad\quad\sqrt{4\pi\alpha} \,\sum_{i=1}^n Q_i\,\frac{p_i\cdot \varepsilon^*(k)}{p_i\cdot k}\,{\mathcal{M}}^{(n)}_{\rm fin}({\tilde p}_1,{\tilde p}_2,p_3,\ldots,p_n;\tmuR, \tmuIR)\,,\qquad n=1,2\,,
\end{align}
where $\{\tilde{p}_1,\tilde{p}_2,p_3,\ldots,p_n\}$ are the on-shell projected momenta
obtained from $\{p_1,p_2,p_3,\ldots,p_n\}$ through the \qT-recoil prescription.%
\footnote{Note that we drop the tilde on $\{p_3,\ldots,p_n\}$ since they are kept unchanged by
our particular recoil prescription.} We use \refeq{eq:softapproxHO} with \mbox{$\muR=\muIR=Q$} and
\mbox{$\tmuR=\tmuIR=\tilde{Q}$} to define our approximate result.%
\footnote{We always evaluate the strong coupling as \mbox{$\as(\muR=Q)$} also in the reduced amplitudes.}

We note that due to the freedom in defining the finite part of the virtual amplitudes and
in choosing the subtraction scale, this approximation procedure is not unique. 
While in an exact computation the dependence on the subtraction scale \muIR always cancels out exactly in the final result,
an approximate treatment of the virtual contribution
generally induces a residual dependence on the subtraction scale and scheme
since different finite terms are evaluated either exactly or approximately.
We account for these ambiguities of the approach in our error estimate discussed below,
by including an uncertainty obtained through a variation of the subtraction scale \muIR.

\subsubsection{Approximation of two-loop amplitudes}
The SA works very well when approaching the soft-photon limit.
This is confirmed numerically for our validation process
at tree level, at one-loop order, and even at two-loop order, where we
take advantage of the availability of the exact \Vgamma two-loop amplitudes~\cite{Gehrmann:2011ab}
inside the \Matrix framework~\cite{Grazzini:2013bna,Grazzini:2015nwa,Grazzini:2017mhc}.
Using \refeq{eq:softapproxHO}, the soft approximation can be obtained in terms of the
Drell--Yan amplitude~\cite{Matsuura:1988sm}, and
when the photon is soft compared to the other particles in the process,
we find excellent agreement with the exact result.
Since the full two-loop amplitudes for \WWgamma production are not known yet,
we can perform this validation only up to one-loop order here. At two-loop order, 
we rely on the $WW$ amplitudes of Refs.~\cite{Gehrmann:2014bfa,Caola:2014iua},
implemented in \VVamp~\cite{hepforge:VVamp} and interfaced to \Matrix for massive diboson
processes~\cite{Cascioli:2014yka,Gehrmann:2014fva,Grazzini:2015wpa,Grazzini:2015hta,Grazzini:2016ctr,Grazzini:2016swo,Grazzini:2017mhc,Grazzini:2017ckn,Kallweit:2018nyv,Grazzini:2018owa,Kallweit:2019zez,Grazzini:2020stb,Grazzini:2021iae},
to compute the SA predictions.

Given the universality of soft factorisation, we expect the quality of the SA not to depend on either the
process under consideration or the perturbative order, but rather on the size of the power-suppressed
contributions neglected in \refeqtwo{eq:softM2}{eq:softapproxHO}.
In processes in which a direct photon is produced in association with heavy vector bosons,
a minimum transverse momentum is always imposed on the photon, and the bulk of the events is produced
at transverse momenta close to this threshold. In this phase-space region,
power corrections are not expected to be large.
There is, however, no obvious reason why the SA should provide good proxies of
the exact amplitudes if applied for phase-space points
far away from the soft-photon limit, i.e.\ beyond its region of validity.
We thus follow the approach adopted for other processes~\cite{Catani:2022mfv,Buonocore:2023ljm,Devoto:2024nhl}
and {\it reweight} our approximated amplitudes with
the ratios of exact to approximated amplitudes calculated at lower orders.
Such a procedure corresponds to applying our approximation not directly to the relevant amplitude,
but to a {\it ratio} of amplitudes.
While there is no formal proof for this procedure to work, we can test its
performance at the NLO level for each process under consideration and
use the resulting insights to make predictions for the NNLO level.
Moreover, for validation processes like \Zgamma production, where we
have the exact NNLO results available, we can directly investigate
the performance at this highest order.
Since \Zgamma production has the same partonic structure as \WWgamma production, and even
the same pattern of soft-photon radiation --- a quark--antiquark pair in the initial state and
two charged leptons in the final state --- we expect the SA to perform comparably in the two processes.
We note that improvements through similar reweighting procedures are also observed
in other contexts without a rigorous formal justification,
like in the large-$m_t$ approximation in Higgs boson production via gluon fusion,
both at the inclusive~\cite{Kramer:1996iq} and the
differential~\cite{Jones:2018hbb} level.

We now turn to discussing possible ways to reweight the approximate amplitudes
at the respective orders, keeping in mind that we eventually aim for an
approximation only at the two-loop level,
whereas the one-loop level is only studied to get insights
into the performance of the SA and to eventually derive an error estimate for the approach.

For reference, and to illustrate improvements obtained through the reweighting procedure,
we will also include the SA without any reweighting ({\it bare} SA)
in the following numerical studies.
At one-loop and two-loop level, the corresponding \Hn coefficients (see \refeq{eq:Hn}) read
\begin{equation}
\label{eq:HnnoRW}
\HnSAnoRW(\muIR) = \left.\frac{2{\rm Re}\left(\mathcal{M}_{\rm SA, fin}^{({n})}(\tmuIR,\tmuR)\mathcal{M}_{\rm SA}^{(0)\ast}\right)}{\left|\mathcal{M}^{(0)}\right|^2}\right|_{\tmuR=\tQ}\,,\qquad n=1,2\,.
\end{equation}
We see that in \refeq{eq:HnnoRW} only the numerator entering the \Hn coefficient is approximated.
To the purpose of limiting the impact of the approximation, the natural approach
is to approximate both the numerator and the denominator.
We refer to this procedure as \textit{\MzeroRW},
for which the corresponding coefficients \Hn read
\begin{equation}
\label{eq:M0M0-RW}
\HnSAMzeroRW(\muIR) = \left.\frac{2{\rm Re}\left(\mathcal{M}_{\rm SA, fin}^{({n})}(\tmuIR,\tmuR)\mathcal{M}_{\rm SA}^{(0)\ast}\right)}{\left|\mathcal{M}_{\rm SA}^{(0)}\right|^2}\right|_{\tmuR=\tQ}\,,\qquad n=1,2\,.
\end{equation}
We remind the reader that the \Hn coefficients enter the cross section calculation as multiplicative corrections
to the exact squared Born matrix element, $|\mathcal{M}^{(0)}|^2$ (see \refeq{eq:mainqTsubtraction}),
so in this approach the interferences between the approximate loop- and tree-level
amplitudes are effectively reweighted by a factor $|\mathcal{M}^{(0)}|^2/|\mathcal{M}_{\rm SA}^{(0)}|^2$.
Trivially, this reweighting applied at the tree level would make the corresponding SA result exact.
Following the idea to use a reweighting that would make the next-lower order exact, we can define
another procedure at the two-loop level, labelled
\textit{\MoneMzeroRW}, so that the corresponding coefficient \Htwo reads
\begin{equation}
\label{eq:MoneMzeroRW}
\HtwoSAMoneMzeroRW(\muIR) = \left.\HtwoSAnoRW(\muIR)\times\frac{2{\rm Re}\left(\mathcal{M}_{\rm fin}^{(1)}(Q,\muR)\mathcal{M}^{(0)\ast}\right)}{2{\rm Re}\left(\mathcal{M}_{\rm SA, fin}^{(1)}(\tQ,\tmuR)\mathcal{M}_{\rm SA}^{(0)\ast}\right)}\right|_{\muR=Q,\,\tmuR=\tQ}\,.
\end{equation}
Note that we define the reweighting factor through amplitudes evaluated at \mbox{$\muIR=Q$} and \mbox{$\tmuIR=\tQ$},
respectively, also  when $\HtwoSAMoneMzeroRW(\muIR)$ is calculated at a scale \mbox{$\muIR\neq Q$}.
Eventually, we introduce a third reweighting procedure, referred to as \textit{\MoneRW}, which is motivated
by the idea that it would make the one-loop squared contribution to \calH exact, if treated in SA in the first place.%
\footnote{This contribution enters the \deltacalH term in \refeq{eq:HHdeltaH} exactly in our calculation, since it can be directly obtained from the general one-loop amplitude providers \OpenLoops~\cite{Cascioli:2011va, Buccioni:2017yxi,Buccioni:2019sur}
or \Recola~\cite{Actis:2016mpe,Denner:2017wsf,Denner:2016kdg}.} The corresponding coefficient \Htwo reads
\begin{equation}
\label{eq:MoneRW}
\HtwoSAMoneRW(\muIR) = \left.\HtwoSAnoRW(\muIR)\times\frac{\left|\mathcal{M}_{\rm fin}^{(1)}(Q,\muR)\right|^2}{\left|\mathcal{M}_{\rm SA,fin}^{(1)}(\tQ,\tmuR)\right|^2}\right|_{\muR=Q,\,\tmuR=\tQ}\,.
\end{equation}
Again, the reweighting factor is defined at fixed scales \mbox{$\muIR=Q$} and \mbox{$\tmuIR=\tQ$}.

We recall that the structure of the soft factorisation formula in \refeq{eq:softM2} remains unchanged up to two-loop order. Consequently, the soft factors cancel between numerator and denominator in all the
reweighting procedures in \refeqtoeq{eq:M0M0-RW}{eq:MoneRW}.
Since it is not obvious in the first place which of the reweighting procedures performs best,
we rely on the numerical results for our validation process,
discussed in \refse{sec:validationZA}, to choose a nominal prediction.
We anticipate that this will be the \textit{\MoneMzeroRW} approach.
When not explicitly discussing the reweighting prescriptions,
we will usually refer to this preferred SA prediction for the \Htwo
coefficient as \HtwoSA and to the corresponding contribution to the (differential)
cross section as \dsHtwoSA for simplicity.
The other reweighting prescriptions will be used to define the error estimate in the following section.

\subsubsection{Error estimate}
\label{sec:errors}

Our strategy in the construction of an overall error estimate of our approximation of the
two-loop virtual contribution is to make use of as much
information as possible that can be collected without knowing the exact 
two-loop amplitudes. From those ingredients we construct an error estimate that is meant to be sufficiently
conservative to cover the exact result.
If this error is much smaller than the customary scale-variation uncertainty, which, as is well known,
provides just a lower limit of the residual perturbative uncertainty,
we can consider our final NNLO result fully predictive.

We take into account several ways to estimate the approximation error
and combine them by taking their envelope, on a bin-by-bin basis for
kinematic distributions, to predict our final error to be assigned to the SA results.

For all relevant cases, we are able to
assess the performance of the SA prediction at NLO, including Born reweighting as in \refeq{eq:M0M0-RW}.
We thus construct a first error estimate based on the assumption that
the relative accuracy of the contribution of the coefficient \HtwoSA
at NNLO, \dsHtwoSA, cannot be expected to be better than
that from \HoneSA at NLO, \dsHoneSA. 
To be conservative, we impose a tolerance factor of two on the relative accuracy
of \dsHoneSA and assign this error estimate symmetrically to our prediction for \dsHtwoSA,
referred to as \DeltaSAHone:
\begin{align}
\DeltaSAHone &= 2 \times \left| \frac{\dsHoneSA}{\dsHone} -1 \right| \times \Bigl| \dsHtwoSA \Bigr| \,.
\label{eq:H1-based_error}
\end{align}
We note that neither of the
\dsHn contributions, exact or approximate, are positive definite,
so both accidental under- and overestimates of
this ingredient of the error estimate may appear. We assume the former to be compensated
by the other ingredients discussed in the following;
to avoid the latter, in general we introduce an upper bound in vicinity
of zero transitions of the exact contribution $d\sigma_{\Hone}$.%
\footnote{For the processes and distributions considered in this paper, no such zero transitions appeared.}

All three reweighting approaches discussed in the previous section
provide equally acceptable predictions. As previously mentioned, motivated by the studies on
our reference process (see \refse{sec:validationZA}) we choose the \MoneMzeroRW SA as our reference prediction and
take the others into account in the construction of the error estimate.
More precisely, we assign the maximum of their absolute deviations from our
nominal prediction symmetrically,
labelling it \DeltaSARW:
\begin{align}
\DeltaSARW &= \max\left( \Bigl| \dsHtwoSAMzeroRW - \dsHtwoSA \Bigr|, \Bigl| \dsHtwoSAMoneRW - \dsHtwoSA \Bigr| \right).
\label{eq:RW-based_error}
\end{align}

Last, we take into account the ambiguity of the chosen approximation scale \muIR
at which the SA is applied.%
\footnote{As pointed out in \refse{sec:softfact}, if the exact result were available, the corresponding
scale-dependent terms would exactly compensate for any choice of \muIR.
However, with an approximation, this is no longer the case.}
Since a natural choice is the
invariant mass of the produced colourless system, i.e.\ \mbox{$\muIR=Q$}, we estimate
this ambiguity by varying \muIR by a factor of two around \mbox{$\muIR=Q$},
and assign the maximum of the absolute deviations from our nominal result as an
uncertainty, dubbed \DeltaSAmuIR:
\begin{align}
\DeltaSAmuIR &= \max\left( \Bigl| \dsHtwoSAMSQovertwo + (Q/2\!\to\!Q) - \dsHtwoSA \Bigr|, \Bigl| \dsHtwoSAMSQtwo + (2Q\!\to\!Q) - \dsHtwoSA \Bigr| \right) .
\label{eq:muIR-variation_error}
\end{align}
Since the amplitude that enters \Htwo needs to be evaluated for \mbox{$\muIR=Q$} by
construction, the exact running from the approximation scale \mbox{$\muIR=Q/2\,(2Q)$}
to \mbox{$\muIR=Q$} is added, indicated by $(Q/2\,(2Q) \to Q)$.
This term is implicitly understood in the following. Correspondingly, \DeltaSAmuIR
is effectively determined from the difference of the \muIR evolution
of \Htwo and the \tmuIR evolution of \HtwoSA,
namely \mbox{$\muIR=Q/2\,(2Q)\to Q$} and \mbox{$\tmuIR=\tQ/2\,(2\tQ)\to \tQ$}, respectively,
which can both be calculated exactly without knowing the involved two-loop amplitudes.
We note that, besides the subtraction scale, also the choice of the scheme to define the
finite part of the virtual amplitude introduces additional arbitrariness in the approximation procedure.
We explicitly checked that defining the finite remainder of the two-loop amplitude in the
\qT scheme~\cite{Catani:2013tia}, rather than in the scheme of Ref.~\cite{Becher:2009cu},
leads to differences that are much smaller than the impact of \muIR variations,
both inclusively and differentially. Therefore, we will neglect the effect of the scheme choice in the following.

We expect the resulting error estimate, \DeltaSA, which is taken in each bin
as the envelope of \DeltaSAHone, \DeltaSARW, and \DeltaSAmuIR, to cover all the relevant sources of
uncertainties. To support this expectation, in the next section we will investigate
its performance for our validation process.

\section[Validation of soft-photon approximation for \Zgamma]{Validation of soft-photon approximation for $\boldsymbol{Z\gamma}$}
\label{sec:validationZA}
We consider \Zgamma production in the default setup of \Matrix~\cite{Grazzini:2017mhc}
for the validation of the SA and, in particular, its
error estimate derived in the previous section.
We choose the process \mbox{\pplmlpa} for the discussion here,
but have undertaken similar studies also for the other processes of the \Vgamma class,
i.e.\ \mbox{\ppnlnlxa} as well as \mbox{\pplmnlxa} and \mbox{\pplpnla}.
The applied fiducial cuts are
motivated by the ATLAS analysis of Ref.~\cite{ATLAS:2013way} performed at \mbox{$\sqrt{s}=7\,\TeV$}
and thus define a typical phase-space region for measurements of this process. 

Explicitly, we consider proton--proton collisions at a centre-of-mass energy of \mbox{$\sqrt{s}=13\,\TeV$}.
The isolated photon has to fulfill \mbox{$\pTgamma>15\,\GeV$} and \mbox{$\absetagamma<2.37$},
and to be isolated against hadronic energy by a smooth-cone isolation~\cite{Frixione:1998jh}
with parameters \mbox{$n=1$}, \mbox{$\epsilon=0.5$}, and \mbox{$\delta_0=0.4$}.
We require \mbox{$\pTlep>25\,\GeV$} and \mbox{$|\eta_\ell|<2.47$} for
the charged leptons. Resolved jets are understood to have
\mbox{$p_{T,{\rm jet}}>30\,\GeV$} and \mbox{$|\eta_{\rm jet}|<4.4$}, after applying the anti-$k_T$
algorithm~\cite{Cacciari:2008gp} with \mbox{$\Delta R=0.4$}, and we require a separation of
isolated photons and leptons against such resolved jets of \mbox{$\Delta R_{\ell/\gamma,{\rm jet}}>0.3$},
as well as a separation between the photon and the charged leptons of \mbox{$\Delta R_{\gamma,\ell}>0.7$}.
Finally, we impose an invariant-mass cut of \mbox{$m_{\ell^-\ell^+}>40\,\GeV$}.
We stick to the scale choice applied in Ref.~\cite{Grazzini:2017mhc}, i.e.\
we set renormalisation and factorisation scales for the central predictions to
\mbox{$\muR=\muF=\sqrt{m^2_Z+\pTgamma^2}$} and consider the customary seven-point variations,
varying $\muF$ and $\muR$ by a factor of two around their common central value with the
constraint \mbox{$0.5 \leq \muF/\muR\leq 2$}.

We employ the $G_\mu$ scheme, and 
the input parameters are set to the PDG values~\cite{Patrignani:2016xqp}:
\begin{align}
G_F &= 1.16639\times 10^{-5}\,\GeV^{-2}\,,\nonumber\\
m_W&=80.385\,\GeV\,, \qquad\Gamma_W=2.0854\,\GeV\,,\nonumber\\
m_Z &= 91.1876\,\GeV\,,\qquad \Gamma_Z=2.4952\,\GeV\,.\nonumber
\end{align}
We use a diagonal Cabibbo--Kobayashi--Maskawa~(CKM) matrix.
The on-shell mass of the top quark is \mbox{$m_t=173.2\,\GeV$},
and its width is set to \mbox{$\Gamma_t=1.44262\,\GeV$}.
We treat the bottom quark as massless
and adopt the parton distribution functions~(PDFs) of
\texttt{NNPDF40\_as\_01180}~\cite{NNPDF:2021njg} with the corresponding five-flavour running
at the respective order, all provided by the LHAPDF interface~\cite{Buckley:2014ana}.
We use the complex-mass scheme~\cite{Denner:2005fg} throughout,
and thus compute the EW mixing angle as
\mbox{$\cos\theta_W^2=\mu_W^2/\mu_Z^2$}
with \mbox{$\mu_V^2=m_V^2-i\Gamma_V\,m_V$}, \mbox{$V=W,Z$}. We calculate the EW coupling through
\mbox{$\alpha=\sqrt{2}\,G_F \left|\mu_W^2\left(1-\mu_W^2/\mu_Z^2\right)\right|/\pi$}, and replace,
following the reasoning of Ref.~\cite{Buccioni:2019sur},
one factor of $\alpha$ by \mbox{$\alpha_0=1/137.036$} for each identified final-state photon.
This holds in particular for the explicit factor of $\alpha$ in the factorisation
formulae in \refeqtoeq{eq:softM}{eq:softapproxHO}.

To discuss the performance of the SA, and in particular
its error estimate, we mainly concentrate on two representative differential distributions,
namely the photon transverse momentum, \pTgamma, and its absolute pseudo-rapidity, \absetagamma.
We note that the \pTgamma distribution is the distribution where the SA may be expected to
perform worst, since in its tail we explicitly look into the region where the photon
is farthest from being soft. The purpose of choosing it is to illustrate that the
SA-related error is sufficiently well under control even in this case, although
the approach is far from providing an accurate estimate of the exact
two-loop amplitude. The pseudo-rapidity distribution of the photon, on the other hand,
is dominated by low-energy photons, and the SA may thus be expected to perform better in general.
We note, however, that configurations where the photon goes really soft
must be excluded in any analysis where the photon is to be observed for
both experimental and theoretical reasons. Consequently, the region where the exact
amplitude is reproduced at the per mille level and better by the SA ---
which has been explicitly checked in dedicated calculations ---
is never directly approached.
In order to provide a more complete picture of the performance of the approximation throughout the
fiducial phase space, we present a selection of additional kinematic distributions in \refapp{app:validationZA}.

\begin{figure}[t]
\centering
\includegraphics[height=\plotheightstd]{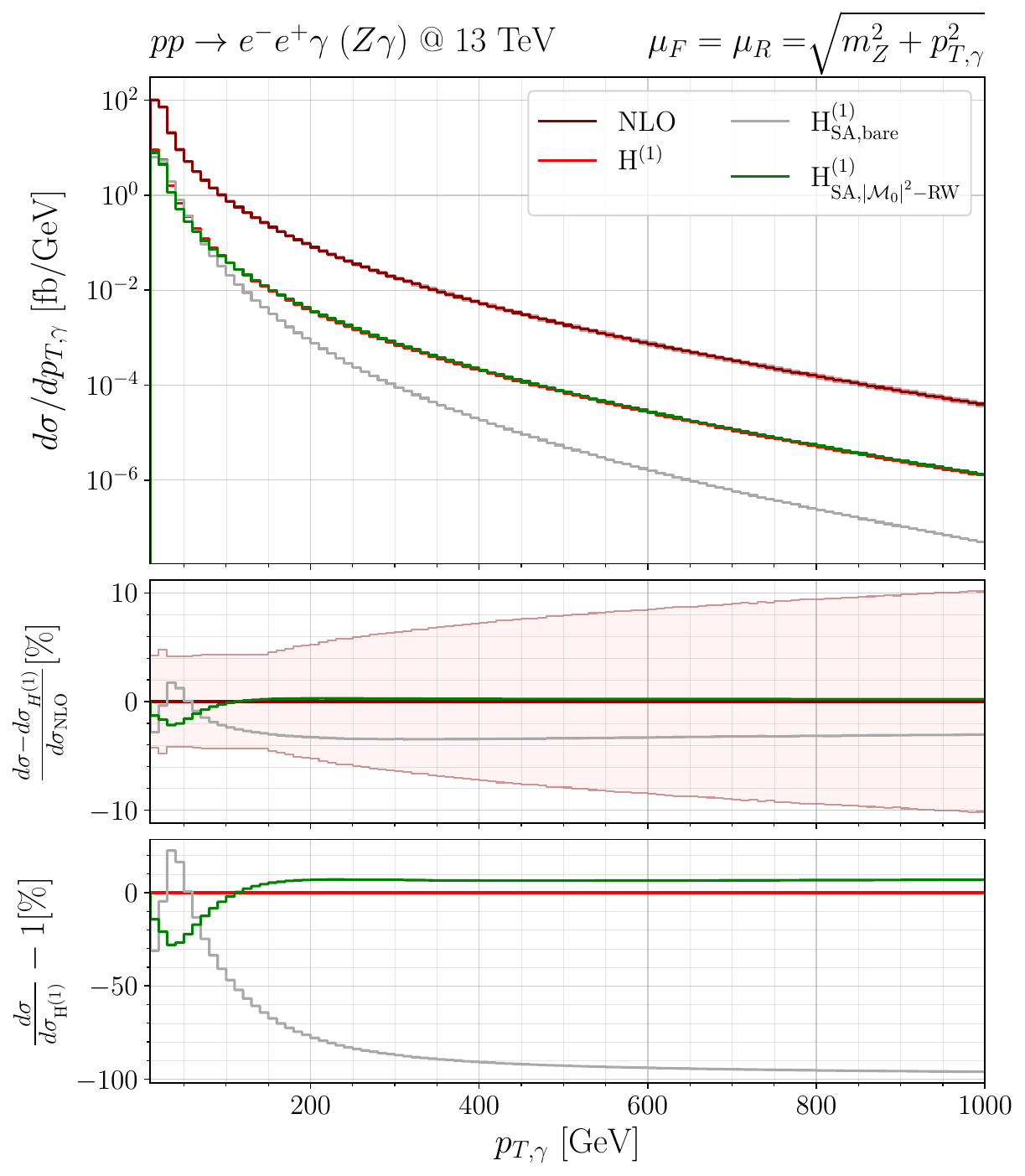}
\hfill
\includegraphics[height=\plotheightstd]{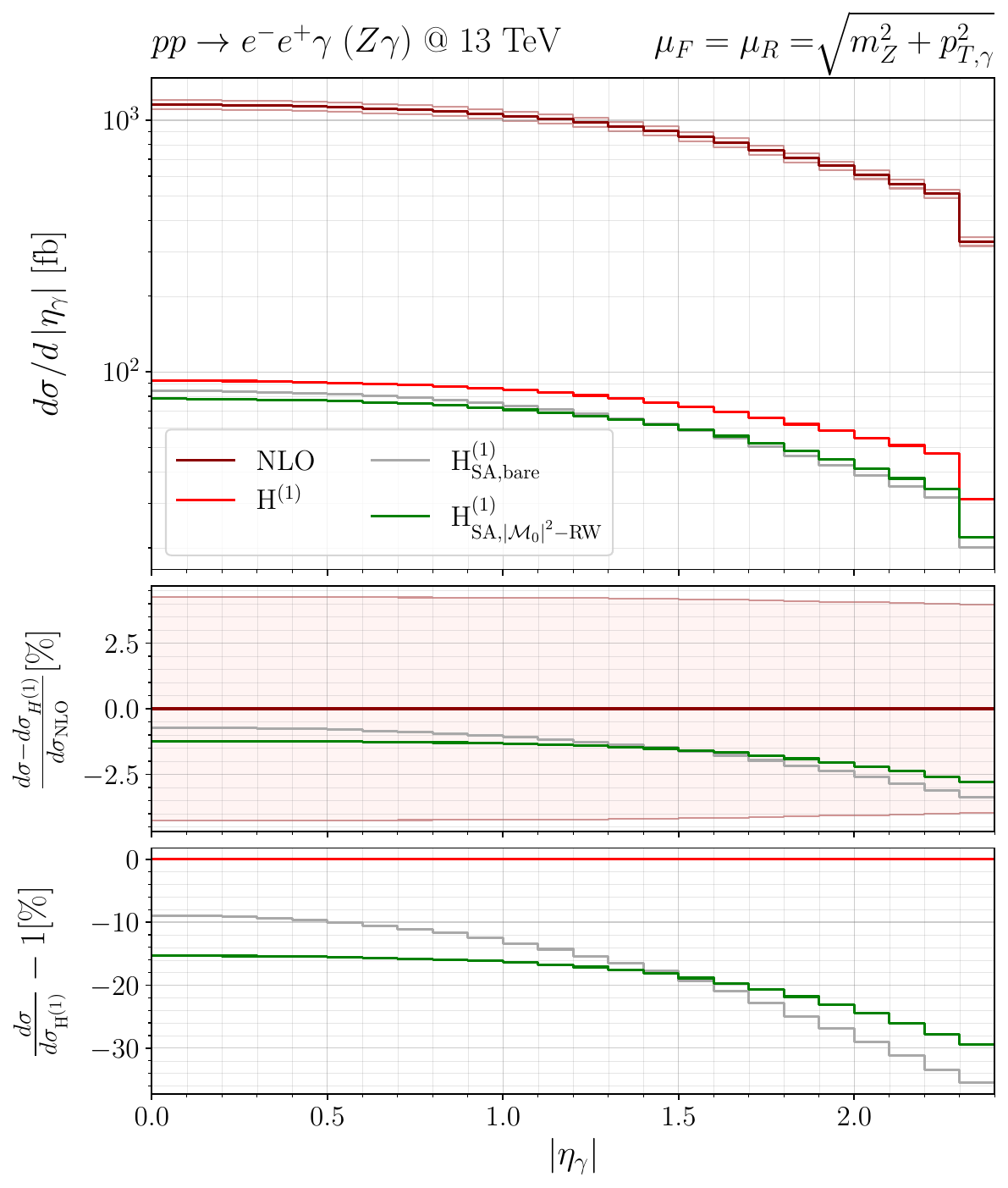}
\vspace*{1ex}
\caption{\label{fig:errorZAH1based} Distributions to illustrate the performance
  of the SA at NLO, which is used to derive the \Hone-based error estimate, \DeltaSAHone,
  in the photon transverse momentum, \pTgamma (left), and its absolute pseudo-rapidity, \absetagamma (right),
  for \Zgamma production.
  The upper panels show the NLO result with its conventional seven-point scale-variation band as well 
  as the integrated \Hone coefficient in three variants:
  exact, \noRW SA, and \MzeroRW SA. In the central panels, the deviations of the two \dsHoneSA variants from the exact \dsHone are displayed, relative to the exact NLO prediction.
We also show a projection of the NLO scale-variation band about unity as a reference for the size of the SA effect.
  In the lower panels, the two \dsHoneSA variants are shown, relative to
  the exact \dsHone.
}
\end{figure}
We now turn to discussing the performance of the SA at NLO.
In \reffi{fig:errorZAH1based} we show, for the two sample distributions, in the upper panels the NLO predictions
as well as the contribution of the \Hone coefficient, \dsHone, exactly, in the \noRW SA and in the \MzeroRW SA.
In the central panels, we depict the deviations of the two
\dsHoneSA variants from the exact \dsHone, relative to the exact NLO result, and add a projection of the
NLO scale-variation band about unity to give a reference for the higher-order uncertainties.
In the lower panels, we show the same deviations, but relative to the exact \dsHone.

For the \pTgamma distribution (left plot), we find that both SA predictions only reproduce the exact \dsHone at the
level of $30\%$ in the low-\pTgamma region, where the bulk of the cross section is, with the
\MzeroRW SA underestimating it by almost as much as the \noRW overestimates it.
Nevertheless, the impact on the full NLO result remains at the level of less than $3\%$,
which is still inside the NLO scale-variation band. In the tail of the distribution, the
\MzeroRW SA reproduces the exact \dsHone remarkably well,
to better than $10\%$ throughout, whereas the \noRW SA substantially underestimates it, by almost $100\%$.
Since the impact of \dsHone on the NLO result quickly decreases as \pTgamma increases,
reaching a few per cent in the tail, both approximations reproduce the exact NLO result very well,
with the \MzeroRW SA being essentially indistinguishable from the exact result in the tail.

For the absolute pseudo-rapidity distribution of the photon, \absetagamma (right plot), we find that, in the bulk region, 
the \noRW and the \MzeroRW SA perform better than for \pTgamma, underestimating the exact \dsHone
by only $10\%$ and $15\%$, respectively.
This slighlty better performance of the \noRW SA, however, is mostly due to the sign change of
its relative deviation from the exact result in the low-\pTgamma region, and should thus be considered accidental.
Towards larger \absetagamma values, the underestimate gets more pronounced for both SA variants,
but the \MzeroRW SA turns out to reproduce the shape of the distribution significantly better.
Relative to the full NLO results, for both SA variants the deviations remain at the
level of only a few per cent over the full \absetagamma range, even covered by the NLO scale-uncertainty band.

Overall, we observe that the phase-space dependence of the \noRW SA is more pronounced than with a reweighting applied.
Since this pattern is confirmed by all the additional distributions in
\reffis{ZAplots_pT_gamma}{ZAplots_dR_em_gamma_dphi_em_gamma_dy_em_gamma} (left columns) of \refapp{app:validationZA},
we build our
\Hone-based error estimate only upon the \MzeroRW SA prediction, according to \refeq{eq:H1-based_error}.

\begin{figure}[t]
\centering
\includegraphics[height=\plotheightstd]{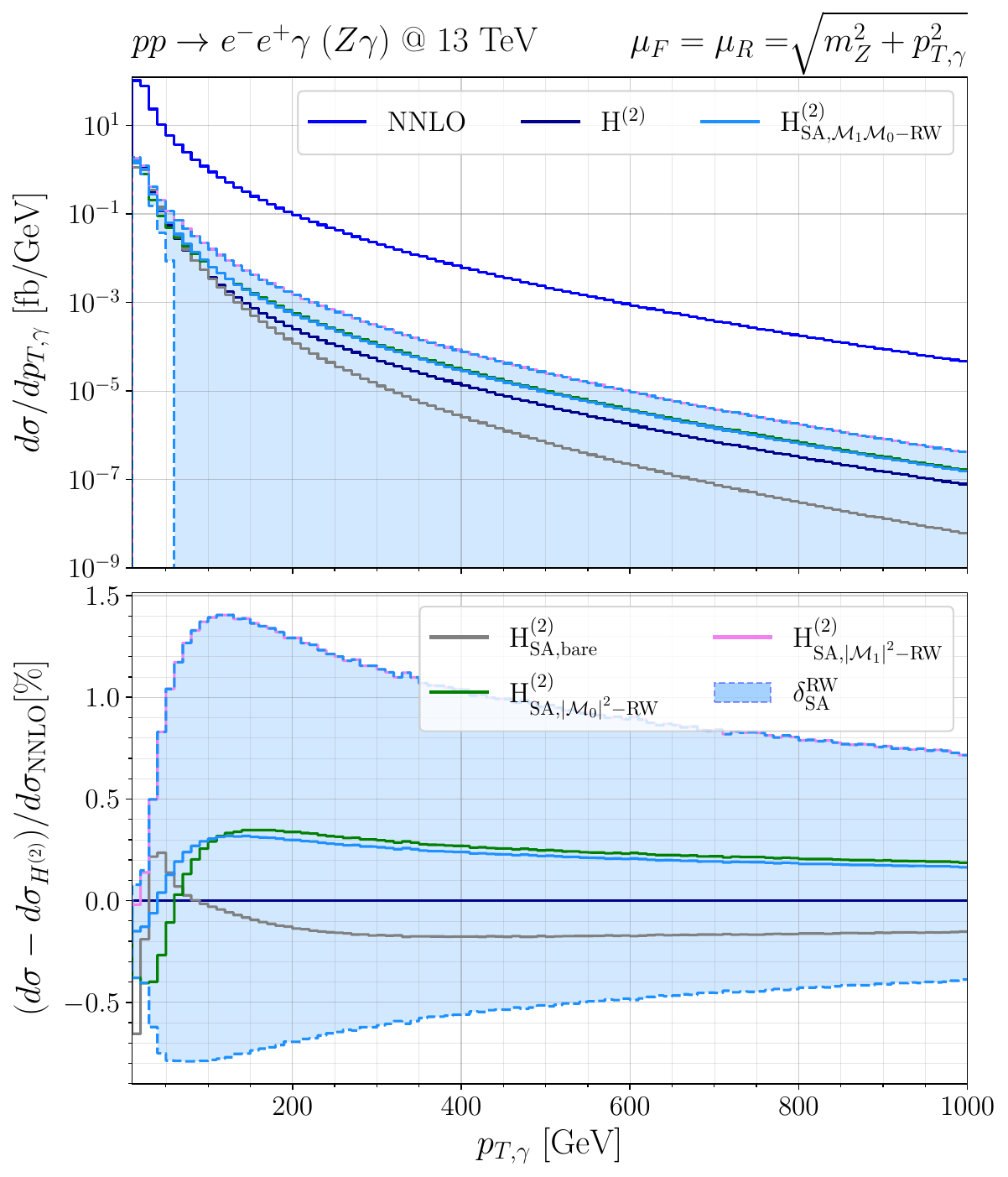}
\hfill
\includegraphics[height=\plotheightstd]{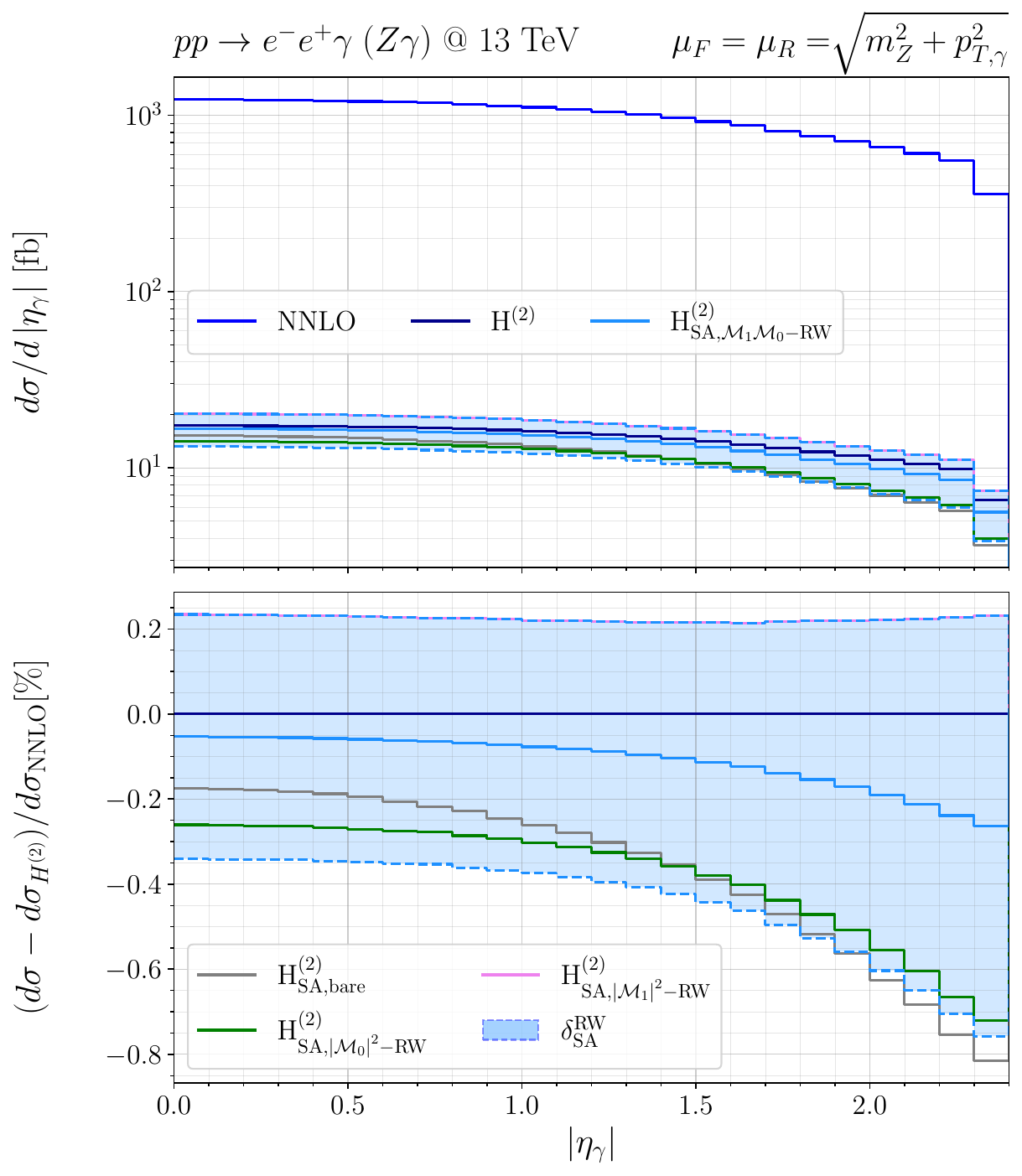}
\vspace*{1ex}
\caption{\label{fig:errorZARWbased} Distributions to illustrate the derivation of the reweighting-based
  error estimate, \DeltaSARW, in the photon transverse momentum,
  \pTgamma (left), and its absolute pseudo-rapidity, \absetagamma (right), for \Zgamma production.
  The upper panels show the exact NNLO result as well 
  as five variants of the integrated \Htwo coefficient:
  exact, \noRW SA, and the three reweighted variants of the SA.
  The lower panels show the deviations of these four variants from the exact \dsHtwo, 
  relative to the exact NNLO prediction. Also depicted in both panels is \DeltaSARW,
  as defined in the main text, indicated as a band
  around the nominal \MoneMzeroRW SA prediction.
}
\end{figure}

We continue the discussion with our motivation to choose the reweighting based on the
interference of tree-level and (IR-subtracted) one-loop virtual amplitudes,
\textit{\MoneMzeroRW}, as our nominal prediction, in
\reffi{fig:errorZARWbased}.
The upper panels show the NNLO result together with the contributions of the integrated \Htwo
coefficient, calculated exactly and using the \noRW as well as the three reweighted SA variants,
respectively, in order to illustrate the overall relevance of \dsHtwo for the full NNLO prediction.
The lower panels show the deviations of the four approximate variants of \dsHtwoSA from the exact \dsHtwo,
relative to the exact NNLO prediction. The final
reweighting-based error estimate, \DeltaSARW, as defined in \refeq{eq:RW-based_error}, is indicated as a band
around the best SA prediction, \MoneMzeroRW.

We find for the \pTgamma distribution that \dsHtwo contributes roughly $1\%$ to the NNLO
cross section in the low-\pTgamma region, where the bulk of the cross section is, but its importance drops
quite rapidly to about $10^{-3}$ of the differential NNLO cross section in the tail.
This effect is analogous to that observed at NLO and
originates from the fact that, in this phase-space region, the dominant contribution comes from
real radiation recoiling against the high-$p_T$ photon.
Considering the bulk region, we find that all variants of \dsHtwoSA give rise to
deviations of less than $\pm0.7\%$ from the exact NNLO result, while the \textit{\MoneMzeroRW} SA deviates by less than
$\pm0.3\%$. In the tail of the distribution,  all reweighted
predictions substantially overestimate the exact \dsHtwo, even by an order of magnitude
in the case of the \textit{\MoneRW}. However, the aforementioned smallness of \Htwo in the tail of the \pTgamma
distribution turns this --- on its own quite substantial --- deviation into a few per mille effect
on the NNLO cross section.
The \noRW SA undershoots the exact \dsHtwo in the tail of the \pTgamma
distribution, by more than an order of magnitude in the TeV region. Apparently, all reweightings
significantly overcompensate this underestimate.
Naively, the \noRW SA thus seems to reproduce the NNLO result better, mostly
because neglecting \Htwo completely in this region would already be a good approximation.
As a result, we find that all approximations of \Htwo are quite poor in the tail
of the \pTgamma distribution; nevertheless, our best \textit{\MoneMzeroRW} SA prediction
is able to reproduce the exact NNLO result better than $\pm0.3\%$ over the full range. 
The error estimate based on the symmetrised envelope of the other two reweightings
barely exceeds $\pm1\%$ of the NNLO cross section and covers the
exact result everywhere.

For the pseudo-rapidity distribution of the photon, the size of the contribution of the exact \Htwo coefficient is
$\mathcal{O}(1\%)$ of the NNLO result over the full range, which is expected since it is
dominated by the low-\pTgamma region throughout. We confirm our NLO observation that the reweighted predictions
in general reproduce the shape of the exact distribution much better than the \noRW.
Moreover, the \textit{\MoneMzeroRW} SA performs best concerning the normalisation of the distribution,
leading to a maximal discrepancy of about $0.3\%$ relative to the NNLO result. The size of the error estimate based
on the symmetrised envelope of the other two reweighting prescriptions is smaller
than $\pm0.5\%$ in the full range, and again covers the exact result everywhere.

Even the \noRW SA turns out to perform reasonably well for these
two sample distributions. We nevertheless exclude it from the reweighting-based error estimate because it
fails to provide a reliable approximation in other phase-space regions,
in particular around resonances involving the softly approximated photon.
\begin{figure}[t]
\centering
\includegraphics[height=\plotheightstd]{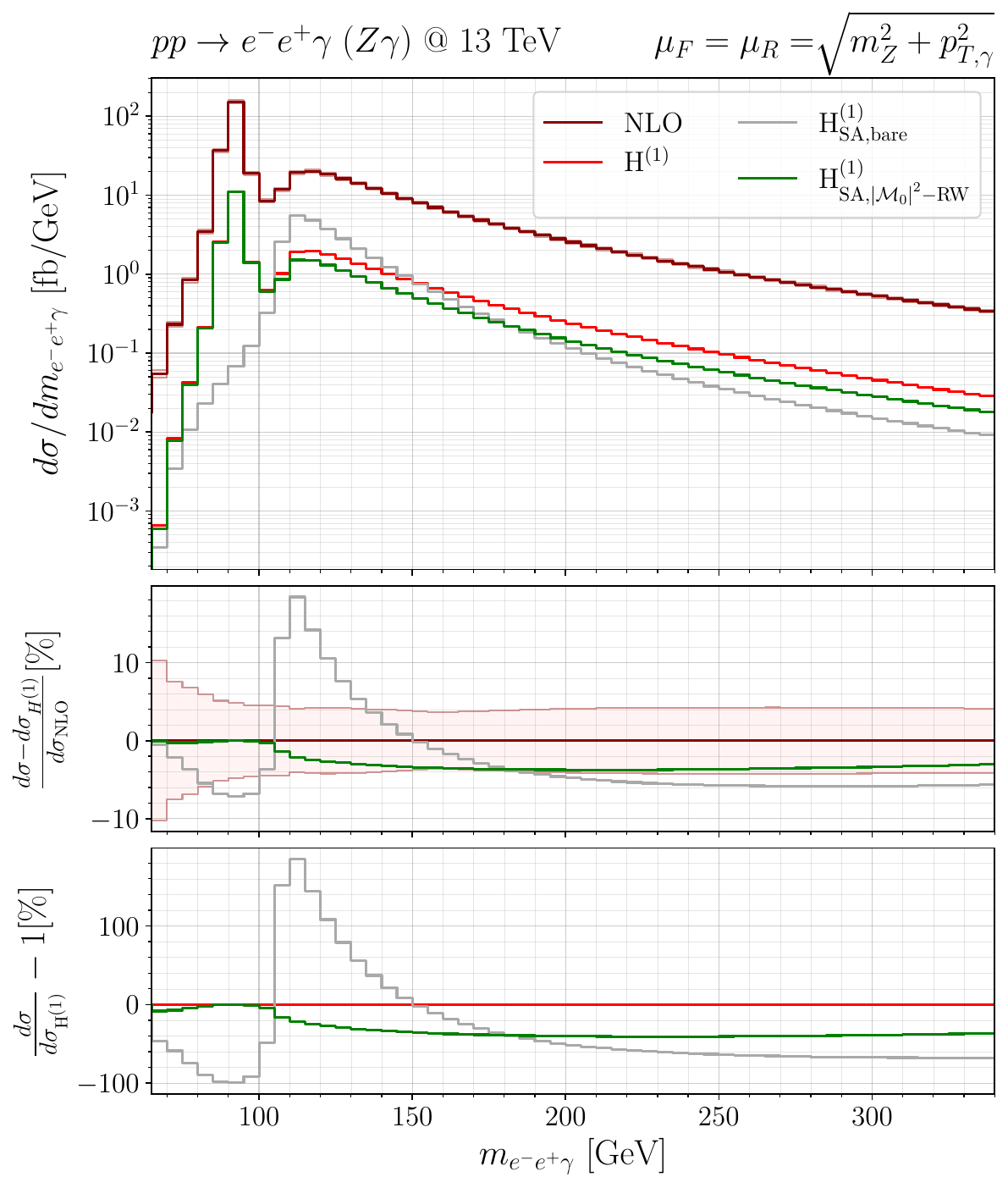}
\hfill 
\includegraphics[height=\plotheightstd]{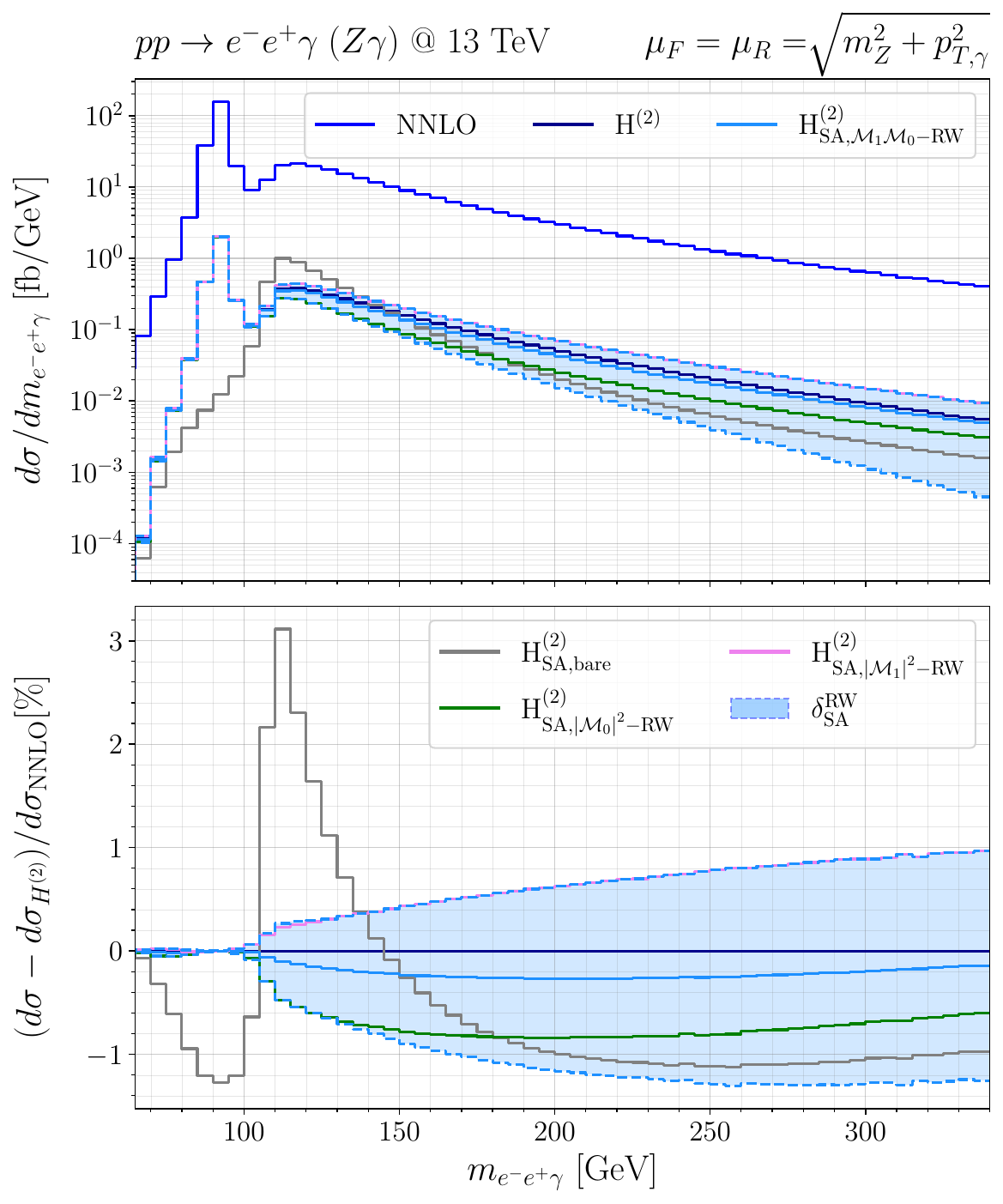}
\vspace*{1ex}
\caption{\label{fig:errorZAm_eegamma} Distribution
  in the invariant mass of the dilepton--photon system, $m_{e^-e^+\gamma}$, at NLO (left) and NNLO (right),
  for \Zgamma production, to illustrate the impact of the reweighting procedure around the resonances.
  Details as in \reffitwo{fig:errorZAH1based}{fig:errorZARWbased}, respectively.
}
\end{figure}
In \reffi{fig:errorZAm_eegamma} we illustrate this feature for the invariant mass of the
dilepton--photon system, $m_{e^-e^+\gamma}$.
Around both the main peak at $m_Z$ and the second, broader peak, the \noRW SA fails to
reproduce the resonance structure, exceeding the NLO scale-variation band (left plot),
and deviating by several per cent from the exact NNLO result (right plot), respectively.
In contrast, the \MzeroRW SA reproduces the exact NLO result within a few percent over the
entire phase space, and the \textit{\MoneMzeroRW} SA does so at NNLO within a few per mille.

In summary, we find that the \textit{\MoneMzeroRW} SA performs
best among the reweighted predictions.
We thus choose it as our reference prediction, as anticipated, and construct the reweighting-based error estimate,
\DeltaSARW, from the other reweighted SA variants, according to \refeq{eq:RW-based_error}.

\begin{figure}[t]
\centering
\includegraphics[height=\plotheightstd]{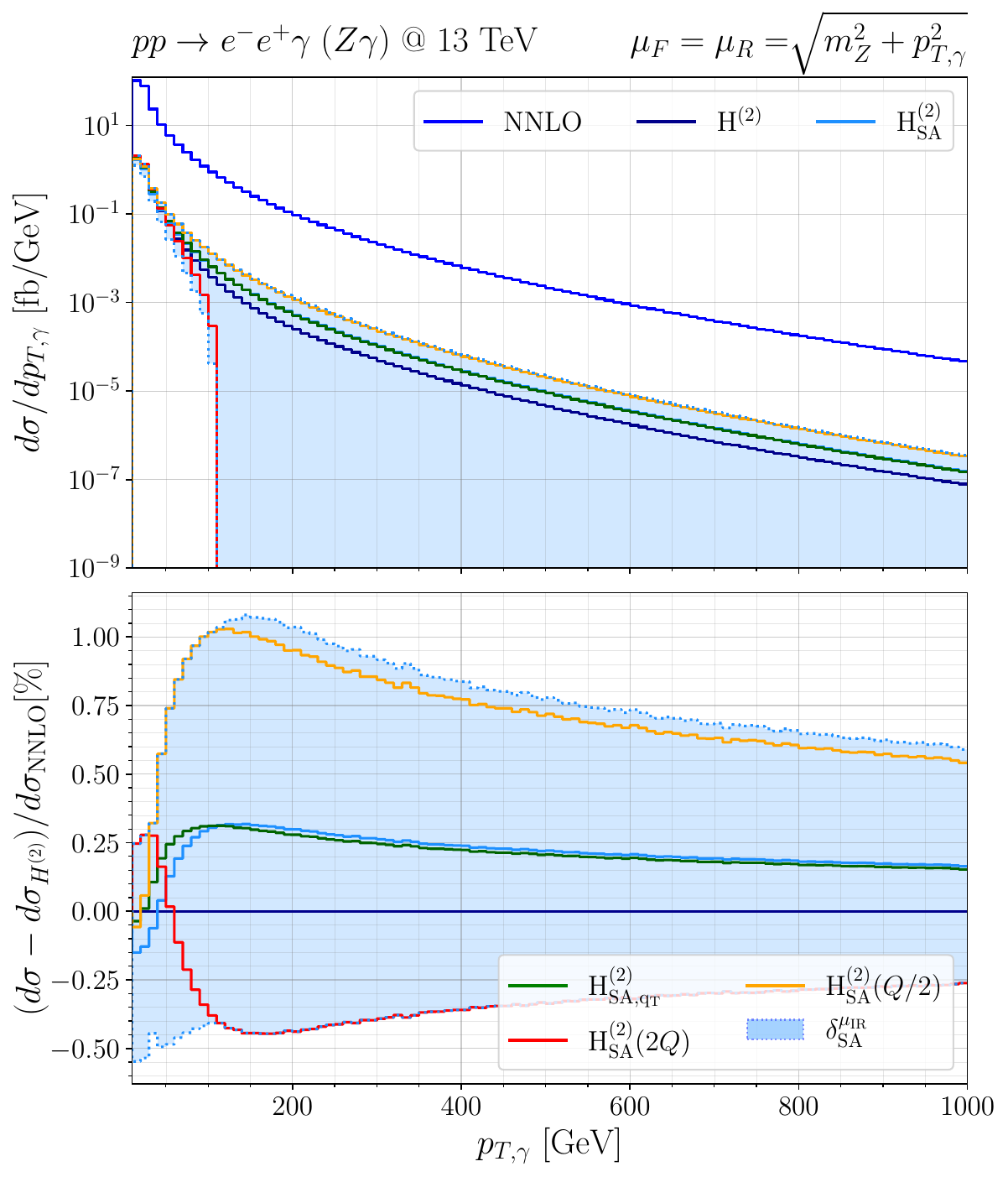}
\hfill
\includegraphics[height=\plotheightstd]{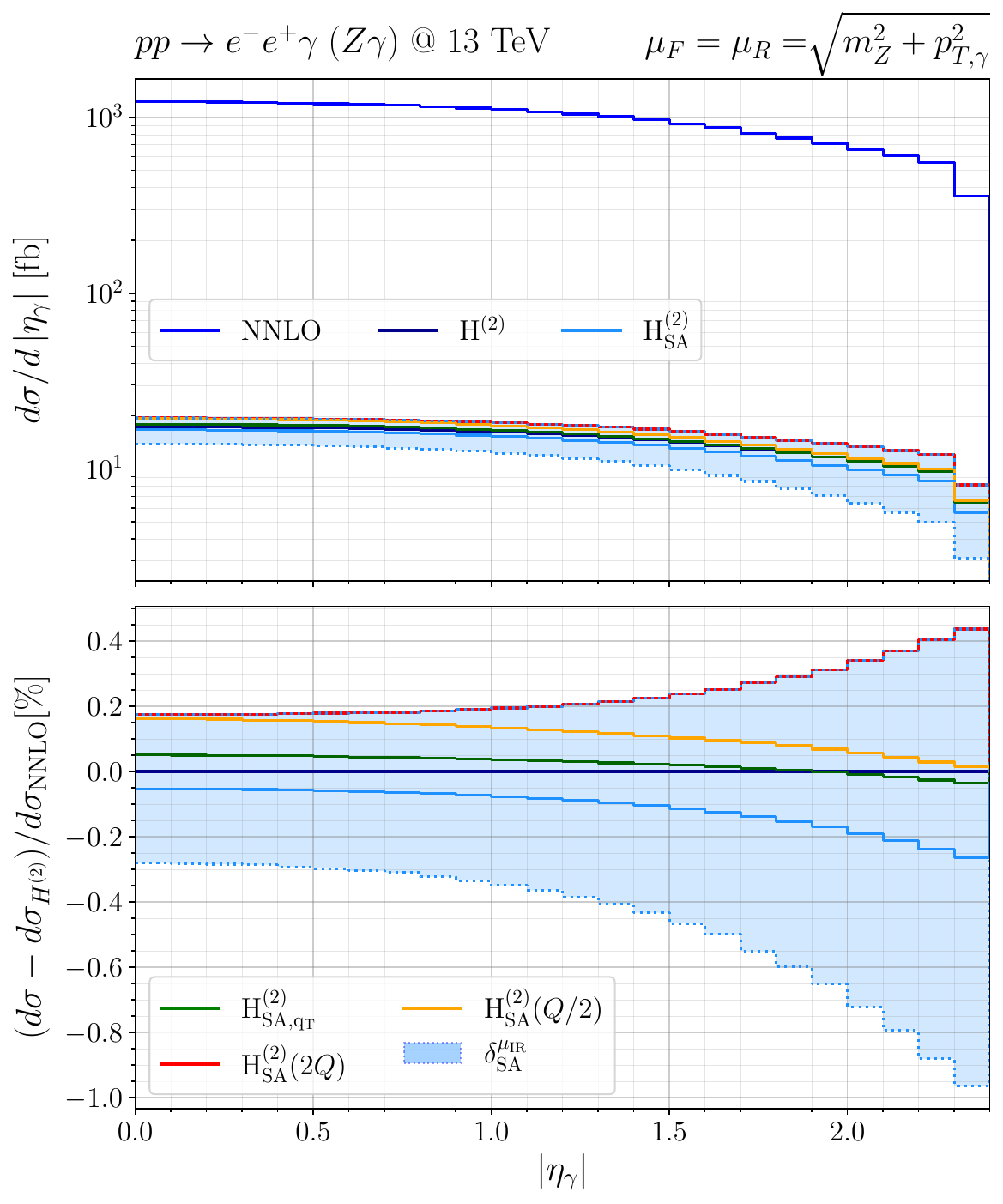}
\vspace*{1ex}
\caption{\label{fig:errorZAmuapproxbased} Distributions to illustrate the derivation of the
  error estimate based on the variation of the approximation scale, \DeltaSAmuIR, in the photon transverse momentum,
  \pTgamma (left), and its absolute pseudo-rapidity, \absetagamma (right), for \Zgamma production.
  The upper panels show the exact NNLO result as well 
  as five variants of the integrated \Htwo coefficient:
  exact and four variants of the SA, namely for the approximation
  scales \mbox{$\muIR=Q\,\textnormal{(default)},Q/2,2Q$}
  in our standard subtraction scheme~\cite{Becher:2009cu},
  and for \mbox{$\muIR=Q$} in the \qT scheme.
  The lower panels show the deviations of the same four \dsHtwoSA variants from the exact \dsHtwo,
  relative to the exact NNLO prediction. Also depicted in both panels
  is \DeltaSAmuIR, as defined in the main text, indicated as a band
  around the nominal SA prediction.
}
\end{figure}
Finally, we consider the uncertainties related to the subtraction scale \muIR
at which the exact amplitude is approximated, shown in \reffi{fig:errorZAmuapproxbased}.
While the invariant mass of the produced system is the natural choice,
\mbox{$\muIR=Q$}, as it is also the scale at which \Hn enters the calculation
in the \qT-subtraction framework, this choice is in principle arbitrary. Exploiting
the running of both the exact and the approximate amplitudes with \muIR,
which involves only amplitudes at lower loop orders, we can determine the impact of different subtraction scales
without the need to calculate the two-loop amplitudes explicitly.
We thus estimate the uncertainty of our approximation due to the arbitrariness of the choice of \muIR
by varying it by a factor of two up and down around
our nominal prediction at \mbox{$\muIR=Q$}. As defined in \refeq{eq:RW-based_error},
we take the symmetrised envelope of these variations about the nominal prediction as
\DeltaSAmuIR.
The scheme in which the finite part of
the virtual amplitude is defined is another source of arbitrariness in the procedure.
By showing the corresponding curve in the \qT scheme alongside
our default curve in \reffi{fig:errorZAmuapproxbased},
we illustrate that the effect is typically small compared to that of the \muIR
variation. As anticipated in \refse{sec:softphotonapproximation},
we will thus not consider it as an additional source of uncertainty.

\begin{figure}[t]
\centering
\includegraphics[height=\plotheightstd]{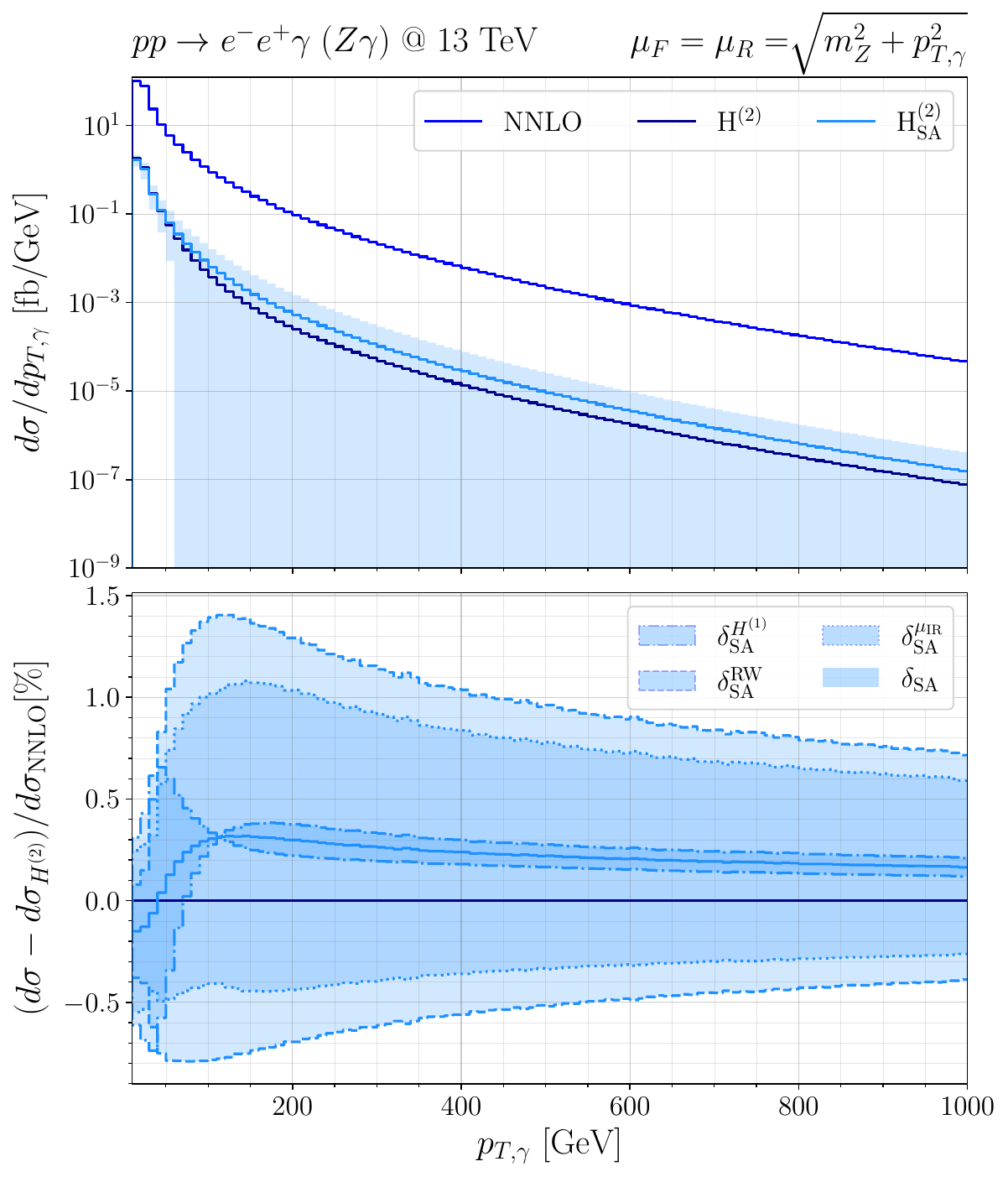}
\hfill
\includegraphics[height=\plotheightstd]{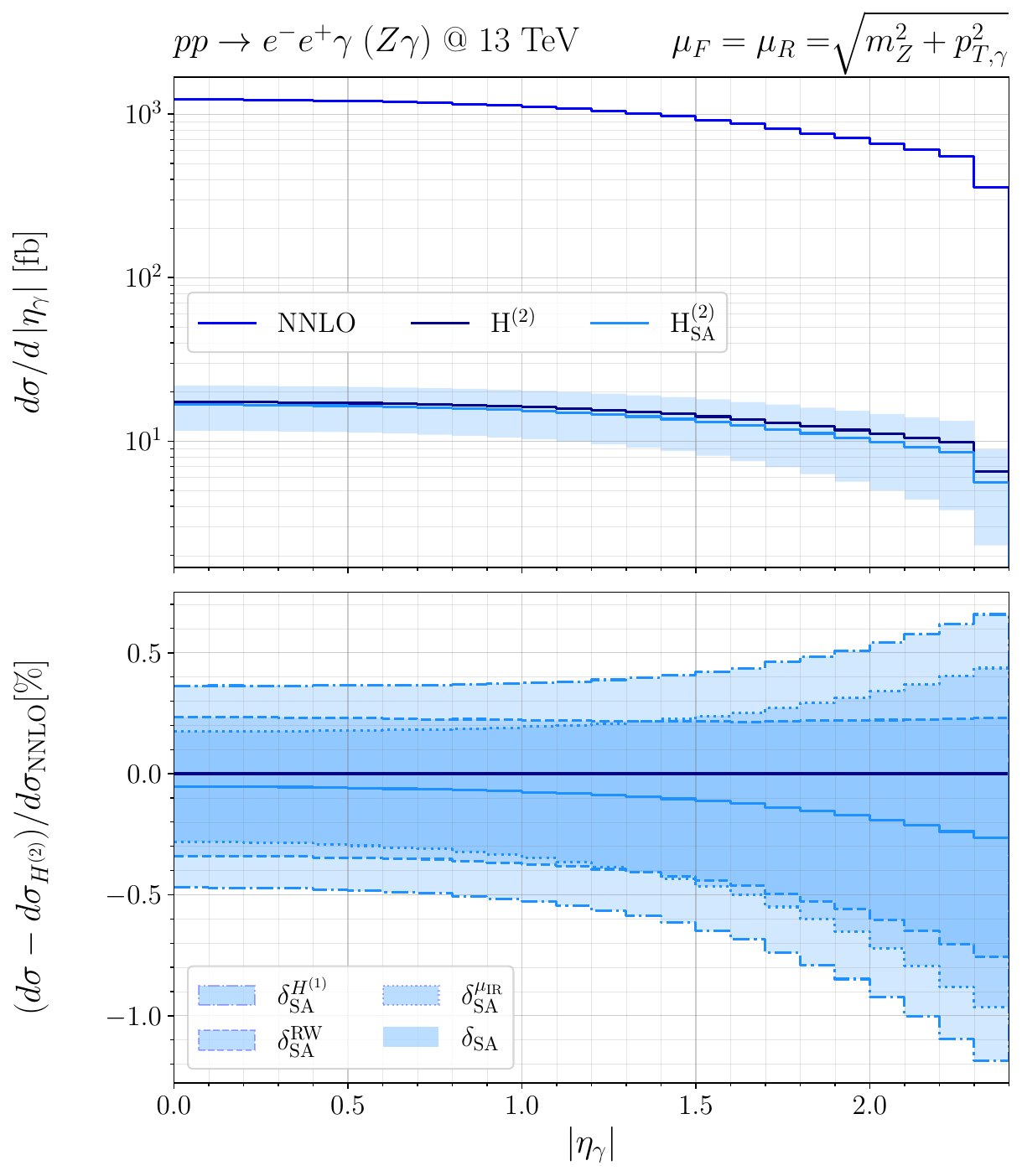}
\vspace*{1ex}
\caption{\label{fig:errorZAcombined} Distributions to illustrate the construction of the
  combined error estimate, \DeltaSA, in the photon transverse momentum,
  \pTgamma (left), and its absolute pseudo-rapidity, \absetagamma (right),
  for \Zgamma production.
  The upper panels show the exact NNLO result as well 
  as the exact integrated \Htwo
  coefficient and the nominal \HtwoSA variant.
  The lower panels show the deviation of \dsHtwoSA from \dsHtwo, relative to the exact NNLO prediction,
  with the three individual error estimates
  \DeltaSAHone, \DeltaSARW, and \DeltaSAmuIR as bands
  around \dsHtwoSA.
  Also indicated in both panels is \DeltaSA, which is defined as the envelope
  of the three individual error estimates and thus corresponds to the entire shaded area.
}
\end{figure}
In \reffi{fig:errorZAcombined} we show the combined error estimate
\DeltaSA, defined as the envelope of the three individual error bands
\DeltaSAHone, \DeltaSARW, and \DeltaSAmuIR, in absolute terms in the upper panels,
and relative to the full NNLO cross section in the lower.
The exact NNLO predictions are also shown for reference.
Even when taking into account the additional distributions in 
\reffis{ZAplots_pT_gamma}{ZAplots_dR_em_gamma_dphi_em_gamma_dy_em_gamma} (central columns) of
\refapp{app:validationZA}, we find that the exact result is covered by our combined
error estimate throughout the phase space. We further note that in almost all regions
each individual estimate would already cover the exact result --- with the notable exception of \DeltaSAHone
in the \pTgamma distribution --- and that taking into account the envelope of all individual error bands
often overestimates the true error.
We nevertheless keep all three of them to be sufficiently conservative.

\begin{figure}[t]
\centering
\includegraphics[height=\plotheightstd]{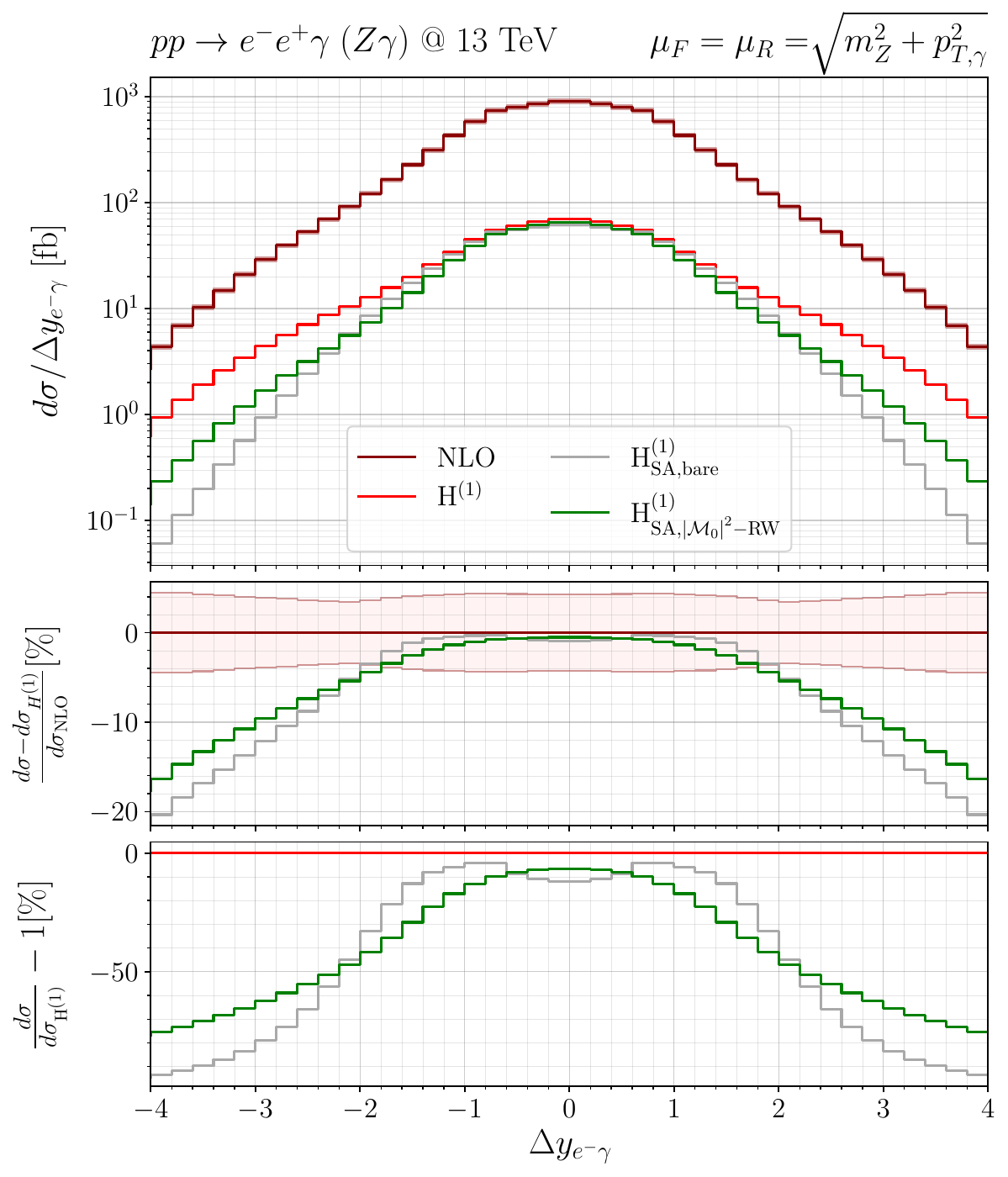}
\hfill
\includegraphics[height=\plotheightstd]{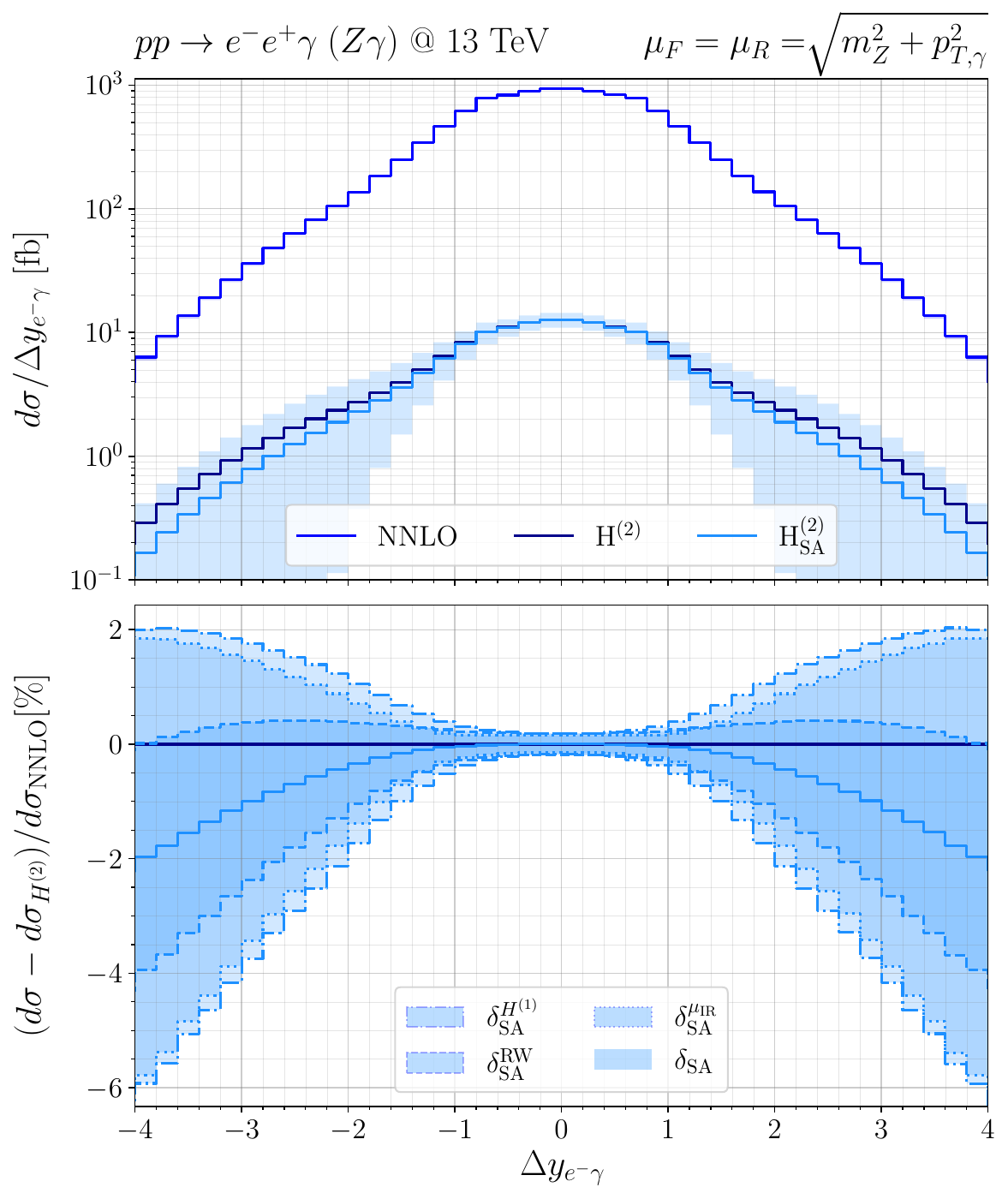}
\vspace*{1ex}
\caption{\label{fig:ZAdy_e_gamma} Distribution
  in the rapidity separation between the electron and the photon, at NLO (left) and NNLO (right),
  for \Zgamma production, to illustrate the SA and its error estimate in regions of bad performance.
  Details as in \reffitwo{fig:errorZAH1based}{fig:errorZAcombined}, respectively.
}
\end{figure}

To show that the final error estimate constructed in this way is indeed conservative,
we present in \reffi{fig:ZAdy_e_gamma} the distribution in the rapidity separation between the electron and the photon,
both at NLO and at NNLO.
Within our survey of distributions, the SA performs worst for
this type of observable, particularly at large rapidity separations.
We find, however, that this pattern is already present at NLO, where \dsHoneSA deviates from \dsHone by
up to about $15\%$ of the NLO prediction for large rapidity separations, far beyond the NLO scale-variation band.
Consequently, the corresponding error estimate \DeltaSAHone increases,
so that \dsHtwoSA still reproduces the exact \dsHtwo within the assigned error.
This observation gives us confidence that our prescription is sufficiently conservative and
would be able to identify any loss of precision that might arise in certain regions of phase space.
Note that, while \DeltaSAHone dominates the overall error \DeltaSA in our envelope prescription,
even \DeltaSARW and \DeltaSAmuIR would individually cover the exact result as well.

\begin{figure}[t]
\centering
\includegraphics[height=\plotheightstd]{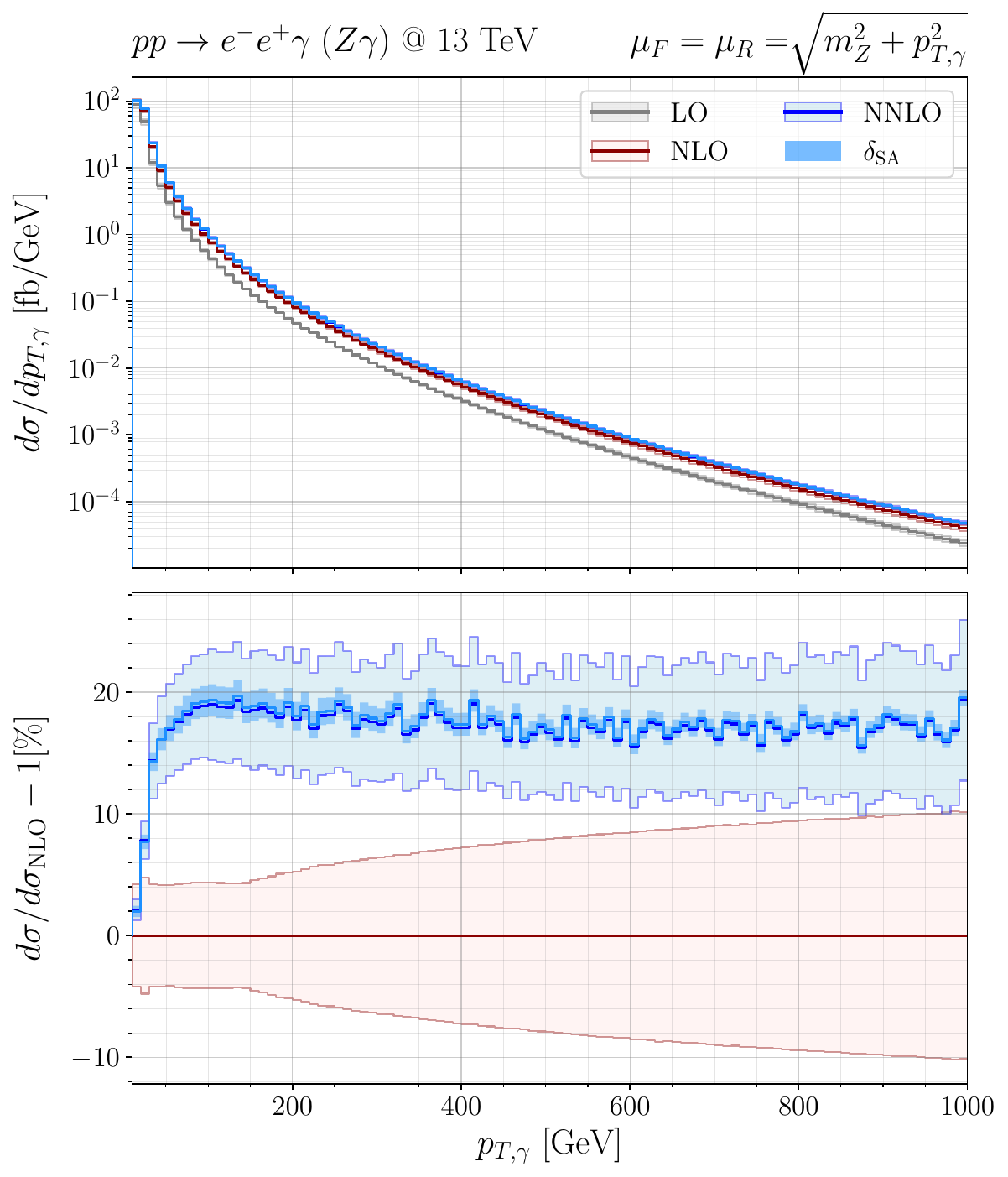}
\hfill 
\includegraphics[height=\plotheightstd]{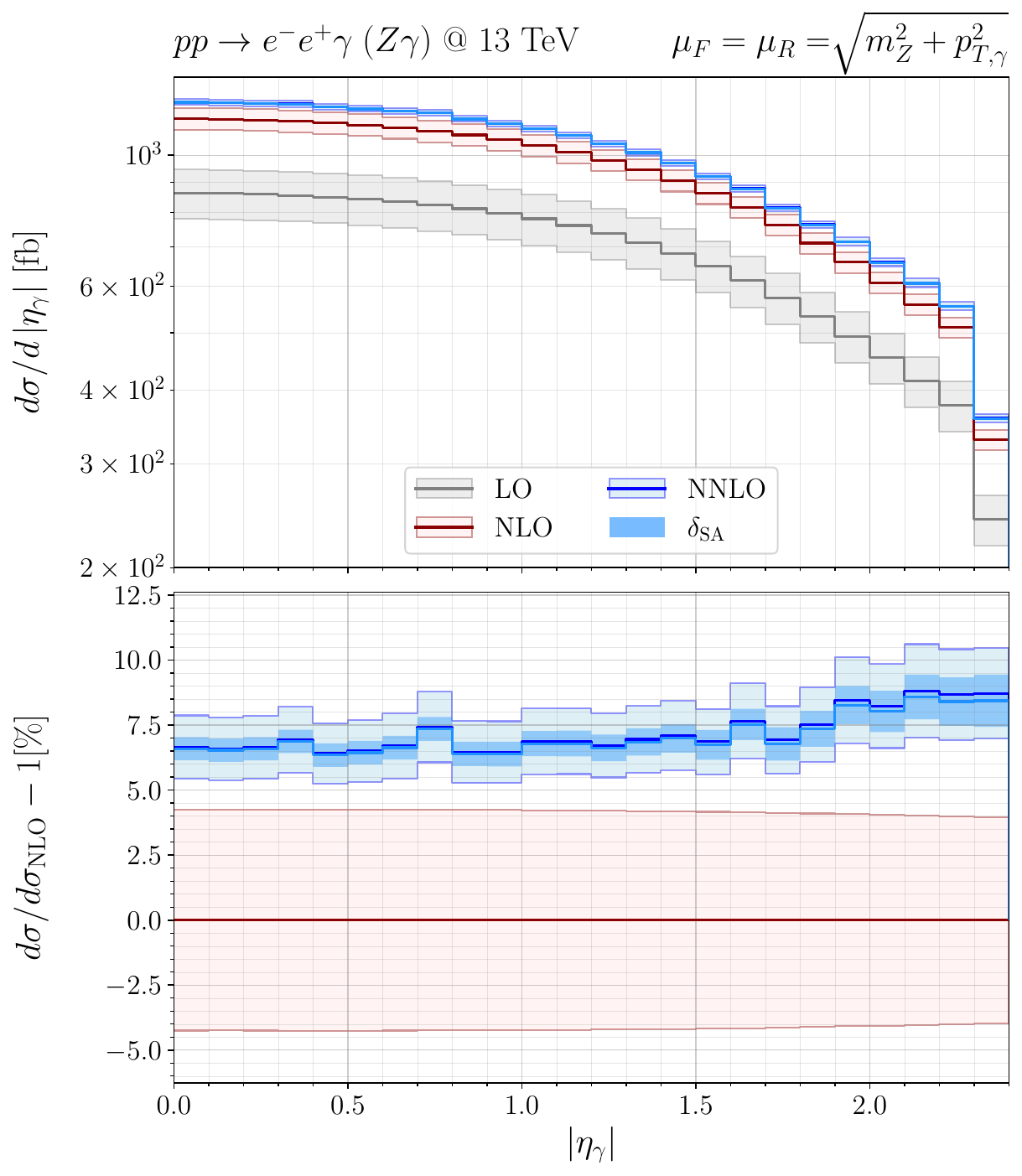}
\vspace*{1ex}
\caption{\label{fig:errorZAKfactors} Distributions in the
  photon transverse momentum, \pTgamma (left),
  and its absolute pseudo-rapidity,
  \absetagamma (right), for \Zgamma production.
  The upper panels show the absolute predictions at LO, NLO, and NNLO accuracy,
  together with their conventional seven-point scale-variation bands.
  The lower panels depict the NLO and NNLO results relative to the former,
  both again with their scale-uncertainty bands.
  Additionally, we indicate the NNLO prediction computed using the nominal SA variant for the two-loop coefficient,
  together with the final SA error estimate, \DeltaSA.
}
\end{figure}

\reffi{fig:errorZAKfactors} puts this approximation-based error
estimate into the context of the perturbative uncertainties due to missing higher orders,
estimated through the customary seven-point scale variation.
In the upper panels we provide the absolute predictions at LO, NLO, and NNLO together with
their respective scale-variation bands, which clearly underestimate the true
perturbative uncertainties at least at LO and NLO.
In the lower panels we show the NNLO results relative to NLO, together with the corresponding scale-variation bands.
Furthermore, we provide the combined SA error estimate
as an additional blue band. The corresponding uncertainty is clearly subdominant.
When including the whole survey of distributions shown
in \reffis{ZAplots_pT_gamma}{ZAplots_dR_em_gamma_dphi_em_gamma_dy_em_gamma} (right columns) of \refapp{app:validationZA},
we find that the same conclusion holds almost everywhere in phase space.
Even in the regions where we have demonstrated that the SA performs worst,
namely those characterised by large rapidity separations between a lepton and the photon,
as discussed in connection with \reffi{fig:ZAdy_e_gamma},
the combined SA error estimate is covered by the NNLO scale-uncertainty band.
Its size amounts to about half of the NNLO scale-uncertainty band, and occasionally even slightly more in those regions.
However, it is worth noting that the SA error estimate is typically more than twice as large as
the actual deviation of the nominal SA prediction from the exact NNLO result in all these cases.
Further examples where the SA error estimate becomes almost as large as the scale-uncertainty band
are found in a few phase-space regions where the scale uncertainty falls below the per cent level.
Such configurations call for a more sophisticated assessment of perturbative uncertainties
rather than indicating a breakdown in the applicability of the SA.%
\footnote{We note that in \Zgamma production accidental cancellations occur at NNLO
when considering the standard seven-point scale variation, such that the resulting bands are significantly
smaller than in otherwise similiar processes such as \Wgamma or \WWgamma production. To
overcome this issue, occasionally nine-point scale variations (i.e.\ including the otherwise excluded
antipodal variations) have been used to estimate perturbative uncertainties in \Zgamma
production (see e.g. Refs.~\cite{Grazzini:2013bna,Grazzini:2015nwa}).
Larger perturbative uncertainties than those quoted here make the SA performance even more striking.
Indeed, the nine-point scale variation would increase these uncertainties  by a factor of two to three,
particularly in those regions where they are accidentally small for the default seven-point variation.}

\renewcommand{\arraystretch}{1.5}
\begin{table}[t]
\begin{center}
\begin{tabular}{lll}
&$~~~\sigma[{\rm fb}]$ & $\sigma/\sigma_{NLO}$ \\
\hline
LO & $1641.6(1)\,^{+9.1\%}_{-9.8\%}$ & $0.75$ \\
\hline
NLO & $2190.9(1)\,^{+2.7\%}_{-4.2\%}$ & $1.00$ \\
\hline
$\rm{NLO_{SA}}$ & $2158.2(1)\,^{+2.6\%}_{-4.1\%}$ & $0.99$ \\
\hline
NNLO & $2340(2)\phantom{.0}\,^{+1.1\%}_{-1.2\%}$ & $1.07$ \\
\hline
$\rm{NNLO_{SA}}$ & $2337(2)\phantom{.0}\,^{+1.1\%}_{-1.2\%}{\;\pm\;0.5\%\,(\rm SA)}$ & $1.07$
\end{tabular}
\end{center}
\caption{\label{tab:inclusiveZA} Fiducial cross sections for
  \mbox{\pplmlpa} in the setup of this section, $\ell=e\textnormal{ or }\mu$.
  Predictions at LO, NLO, and NNLO are stated with their seven-point scale-variation uncertainties.
  Numerical uncertainties are shown in brackets. At NNLO they also include the uncertainty from the
  \mbox{$\rcut\to 0$} extrapolation. Additionally, we state the
  error estimate for the approximation of the two-loop amplitudes, \DeltaSA.
  For comparison, we also report the results one would achieve by replacing the exact \dsHone and \dsHtwo contributions
  with the corresponding SA results, dubbed $\rm{NLO_{SA}}$ and $\rm{NNLO_{SA}}$, respectively.}
\end{table}
To complete the discussion of \Zgamma production, we finally present the fiducial results in \refta{tab:inclusiveZA}.
The NLO corrections increase the LO result by about $33\%$, and the NLO cross section obtained using the SA,
i.e.\ by replacing \dsHone with \dsHoneSA, reproduces the exact result within $1.5\%$.
The correction from the $gq$ channel, opening up at NLO, is negative and
as small as $-1.5\%$ of the NLO prediction, so the $q\bar{q}$ channel
remains by far the dominant contribution.
The impact of \dsHone on the NLO cross section is about $8\%$.
Moving to the next perturbative order, we find NNLO corrections of about $+7\%$, with the new channels,
namely $gg$ and $q\bar{q}^\prime$, contributing about $+0.7\%$ and $+1.1\%$, respectively,
of the total NNLO cross section. In particular, the loop-induced $gg$ channel~\cite{Adamson:2002rm}
accounts for about $10\%$ of the NNLO corrections.
The SA reproduces the full result at the per mille level at this order. Its relative error, $\DeltaSA=\pm 0.5\%$,
remains smaller than the corresponding scale uncertainty of $\pm 1.2\%$.
The impact of \dsHtwo on the NNLO cross section is about $1.5\%$.

With these results at hand, we can now apply the very same procedure for constructing the SA and
its error estimate to \WWgamma production in the following section.

\section[NNLO QCD predictions for \WWgamma at the LHC]{NNLO QCD predictions for $\boldsymbol{WW\gamma}$ at the LHC}
\label{sec:resultsWWA}
We now turn to the actual goal of this paper, namely to provide first predictions at
NNLO QCD accuracy  for a triboson process involving two massive electroweak bosons.
We focus on the production of a \WWgamma final state, more precisely the full processes with
different-flavour leptonic $W$-boson decays,
\mbox{\ppemxnmnexa} and \mbox{\ppmexnenmxa}.

As discussed in some detail for (off-shell) \WW production at NNLO QCD in
Refs.~\cite{Gehrmann:2014fva,Grazzini:2016ctr},
in the standard five-flavour scheme with massless bottom quarks
also the predictions for \WWgamma production would be contaminated by
resonant top-quark topologies at higher orders in QCD.
More precisely, at NLO $gb$-induced partonic channels would have single-top configurations mixed in,
leading to even more sizable NLO effects than already expected for a triboson process.
Moreover, at NNLO both $gg$-induced and $q\bar{q}$-induced channels would
involve resonant top-quark pair production contributions,
which would give rise to NNLO effects of several hundred percent. Both of these huge
corrections would actually be only due to the mixed treatment of processes and
should thus be avoided to achieve a clean definition of \WWgamma production.
The diagonal CKM matrix that we use, which in particular does not allow the third
generation to mix with the two light ones, allows us to address this issue
by performing the calculation in a four-flavour scheme:
While no initial-state bottom quarks appear,
contributions with a bottom-quark pair in the final state
do not need to be generated radiatively since the involved
\mbox{$g\to b\bar{b}$} splittings are regularised by the finite bottom-quark mass.
The off-shell top-quark contributions are treated as part of an
off-shell \mbox{\ppttxa} calculation, i.e.\ \mbox{\pplmlppnlpnlxbbxa}, and do not need to be
included in \WWgamma production since the IR structures of the processes are disentangled.
We thus use a four-flavour scheme and the PDFs of
\texttt{NNPDF30\_lo\_as\_0118\_nf\_4}~\cite{Ball:2014uwa} and
\texttt{NNPDF40\_nlo/nnlo\_as\_0118\_nf\_4}~\cite{NNPDF:2021njg}
with the corresponding four-flavour running at the respective order. We set the numerical value
of the bottom-quark mass to \mbox{$m_b=4.75\,\GeV$} and the central renormalization and factorization scales to
\mbox{$\muR=\muF=H_T/2=(E_{T,\ell^{-}\bar{\nu}_{\ell}} + E_{T,\ell^{+}\nu_{\ell}} + \pTgamma)/2$}\,,
with \mbox{$E_{T,\ell\nu_\ell}=\sqrt{m_{\ell\nu_\ell}^2+p_{T\ell\nu_\ell}^2}$}.
Apart from these adaptations, we use the same schemes and parameters as in the previous section.
In the following, we present predictions for the fiducial cross sections in two setups inspired by the
ATLAS~\cite{ATLAS:2025yxf} and CMS~\cite{CMS:2023rcv} measurements, respectively,
as well as predictions for a selection of differential distributions in the latter.

\subsection{Fiducial results}
\label{sec:fidresu}
In this section, we provide fiducial results for the two experimental setups adopted in the
CMS~\cite{CMS:2023rcv} and ATLAS~\cite{ATLAS:2025yxf} \WWgamma measurements.
We start by considering the CMS analysis~\cite{CMS:2023rcv}, with proton--proton collisions at a
centre-of-mass energy of \mbox{$\sqrt{s}=13\,\TeV$}.
We require the isolated photon to fulfill \mbox{$\pTgamma>20\,\GeV$} and \mbox{$\absetagamma<2.5$}.
Here, the smooth-cone isolation~\cite{Frixione:1998jh} is defined as for \Zgamma production,
with \mbox{$n=1$}, \mbox{$\epsilon=0.5$}, and \mbox{$\delta_0=0.4$}.
The charged leptons must satisfy \mbox{$\pTe>25\,\GeV$}, \mbox{$\absetae<2.5$} and \mbox{$\pTmu>20\,\GeV$},
\mbox{$\absetamu<2.4$}, while the missing transverse momentum of the event is \mbox{$\pTmiss>20\,\GeV$}.
The separation between charged leptons and the isolated photon is required to satisfy \mbox{$\Delta R_{\gamma,\ell}>0.5$}.
The invariant mass and transverse momentum of the dilepton system have to
fulfill \mbox{$m_{\ell\ell}>10\,\GeV$} and \mbox{$\pTleplep>15\,\GeV$}, respectively.
Eventually, we require a minimum transverse mass of the $WW$ system,
\begin{equation}
m_{T,WW}=\sqrt{2\pTleplep\,\pTmiss[1-\cos\phi(\overset{\to}{p}_{T,\ell\ell},\overset{\to}{p}_{T,{\rm miss}})]}>10\,\GeV\nonumber\, .
\end{equation}
In \refta{tab:inclusiveCMS} we present the corresponding results for the fiducial cross sections
at LO, NLO, and NNLO, together with their scale uncertainties.
\renewcommand{\arraystretch}{1.5}
\begin{table}[t]
\begin{center}
\begin{tabular}{lll}
&$~~~\sigma[{\rm fb}]$ & $\sigma/\sigma_{NLO}$ \\
\hline
LO & $1.9000(1)\,^{\phantom{0}+3.1\%}_{\phantom{0}-3.8\%}$ & $0.48$ \\
NLO & $3.9637(1)\,^{\phantom{0}+5.3\%}_{\phantom{0}-4.2\%}$ & $1.00$ \\
\hline
$gg$LO & $0.1508(1)\,^{+26.0\%}_{-19.5\%}$ & $0.04$\\
\hline
NNLO & $4.596(4)\phantom{0}\,^{\phantom{0}+3.6\%}_{\phantom{0}-3.2\%}{\;\pm\;0.3\%\,(\rm SA)}$  & $1.16$\\
\hline
CMS & $5.9\pm 0.8\, (\rm stat.) \pm 0.8\, (\rm sys.) \pm 0.7\, (\rm mod.) $ \\
\end{tabular}  
\end{center}
\caption{\label{tab:inclusiveCMS} Fiducial cross sections for the combination of
  \mbox{\ppemxnmnexa} and \mbox{\ppmexnenmxa} in the CMS setup.
  Predictions at LO, NLO, and NNLO are stated with their seven-point scale-variation uncertainties.
  Numerical uncertainties are shown in brackets.
  At NNLO they also include the uncertainty from the $\rcut\to 0$ extrapolation.
  Additionally, we state the error estimate for the approximation of the two-loop amplitudes, \DeltaSA.
  We also report the loop-induced gluon-fusion contribution, denoted as $gg$LO.
  For comparison, the fiducial cross section measured by CMS~\cite{CMS:2023rcv} is provided as well. 
}
\end{table}
We report results for the combination of the two leptonic different-flavour decay channels, which individually
have slightly different cross sections due to the different cuts on electrons and muons
and the charge asymmetry present in $pp$ collisions.
At NNLO we also state the combined SA error estimate, \DeltaSA, computed as defined in \refse{sec:errors}.
The contribution of the loop-induced gluon-fusion channel is also reported.
In the last row, we provide the fiducial cross section measured by CMS,
which is in agreement with our NNLO predictions within the uncertainties.

We recognize the general pattern of QCD corrections known from several multi-boson production processes involving photons:
The NLO corrections are huge, more than $+100\%$  for our setup, and far beyond the scale-variation uncertainties at LO,
which are driven solely by factorisation scale variations.
About $73\%$ of the NLO cross section is due to the $q\bar{q}$ channel,
the $qg$ channel accounting for the remaining $27\%$, i.e.\ both channels contribute almost equally to the NLO corrections.
Using \HoneSA instead of the exact \Hone would change the NLO prediction by only about $0.5\%$.
The impact of \dsHone on the NLO cross section is found to be about $8.5\%$.

NNLO corrections increase the cross section by further $16\%$, exceeding the NLO scale-variation
uncertainties by almost a factor of four. The large impact of NNLO corrections is, however, consistent with
the results previously found for multi-boson production processes involving
photons~\cite{Catani:2011qz,Grazzini:2013bna,Grazzini:2015nwa,Campbell:2016lzl,Grazzini:2017mhc,Campbell:2017aul,Catani:2018krb,Campbell:2021mlr,Chawdhry:2019bji,Kallweit:2020gcp,Garbarino:2025bfg}.
The loop-induced gluon-fusion channel contributes about $24\%$ of the NNLO correction.
This is in line with the results known for neutral diboson production processes, in particular $WW$ production,
where the contribution of the gluon-fusion channel is of similar size~\cite{Gehrmann:2014fva,Grazzini:2016ctr}.
The two-loop contribution computed in the SA, \dsHtwoSA, accounts for about $1.7\%$ of the NNLO cross section.

To achieve a more conservative estimate of the residual perturbative uncertainties,
we always consider the symmetrised version of scale variations.
More precisely, we take the larger of the upward and downward variations,
assign it symmetrically about the nominal prediction, and leave the latter unchanged.
Nevertheless, uncertainties estimated in this way are clearly not reliable at least up to NLO. 
At NNLO the ensuing uncertainty is $\pm 3.6\%$.
Since, in contrast to NLO, all the partonic channels are open at NNLO, we do not expect particularly
large additional corrections at N$^3$LO and beyond. The NLO corrections to the gluon-fusion channel are expected to 
further increase the NNLO result, but even a $100\%$ correction on the $gg$LO
contribution would still be covered by the quoted scale-variation band.
The error from approximating the two-loop amplitudes is estimated to be $0.3\%$, smaller than
what has been found in \Zgamma production, and, in particular,
significantly smaller than the residual perturbative uncertainties.

Following the ATLAS analysis of Ref.~\cite{ATLAS:2025yxf}, we again consider
proton--proton collisions at a centre-of-mass energy of \mbox{$\sqrt{s}=13\,\TeV$}.
To approximately mimic the experimental photon isolation criteria, we apply a smooth-cone
isolation~\cite{Frixione:1997np} with \mbox{$n=1$}, \mbox{$\epsilon=0.07$}, and \mbox{$\delta_0=0.2$},
and require the isolated photon to fulfill \mbox{$\pTgamma>20\,\GeV$} and \mbox{$\absetagamma<2.37$}.
We impose the same selection cuts on electrons and muons, namely
\mbox{$\pTlep>20\,\GeV$} and \mbox{$\absetalep<2.5$}, and additionally require the leading lepton
to satisfy \mbox{$\pTlepone>27\,\GeV$}.
Moreover, we ask for \mbox{$\Delta R_{\gamma,\ell}>0.4$} for the separation between charged leptons and
the isolated photon.
The corresponding results are reported in \refta{tab:inclusiveATLAS}, summed over both
leptonic different-flavour channels, which contribute equally here due to the flavour-independent cuts.
In the last row, the corresponding fiducial cross section measured by ATLAS
is reported, which is in agreement with both NLO and NNLO predictions.

\renewcommand{\arraystretch}{1.5}
\begin{table}[t]
\begin{center}
\begin{tabular}{lll}
&$~~~\sigma[{\rm fb}]$ & $\sigma/\sigma_{NLO}$ \\
\hline
LO & $2.9806(1)\,^{\phantom{0}+3.7\%}_{\phantom{0}-4.5\%}$ & $0.49$ \\
NLO & $6.0997(4)\,^{\phantom{0}+4.9\%}_{\phantom{0}-3.9\%}$ & $1.00$ \\
\hline
$gg$LO & $0.2202(7)\,^{+25.6\%}_{-19.3\%}$ & $0.04$\\
\hline
NNLO & $7.050(5)\phantom{0}\,^{\phantom{0}+3.5\%}_{\phantom{0}-3.1\%}{\;\pm\;0.3\%\,(\rm SA)}$  & $1.16$\\
\hline
ATLAS & $6.2\pm 0.8\, (\rm stat.) \pm 0.6\, (\rm sys.)$ \\
\end{tabular}  
\end{center}
\caption{\label{tab:inclusiveATLAS}
As in Table~\ref{tab:inclusiveCMS}, but for the ATLAS setup~\cite{ATLAS:2025yxf}.}
\end{table}

Comparing \reftatwo{tab:inclusiveCMS}{tab:inclusiveATLAS}, we find basically the same pattern of the radiative corrections,
although the absolute predictions vary significantly between the applied cuts.

\subsection{Differential distributions}

We now turn to presenting differential distributions for \WWgamma production. Since there is no data available
to be compared to, we restrict our discussion to one sample case, namely
the \mbox{\ppemxnmnexa} channel in the CMS setup of Ref.~\cite{CMS:2023rcv}.

\begin{figure}[t]
\centering
\includegraphics[height=\plotheightstd]{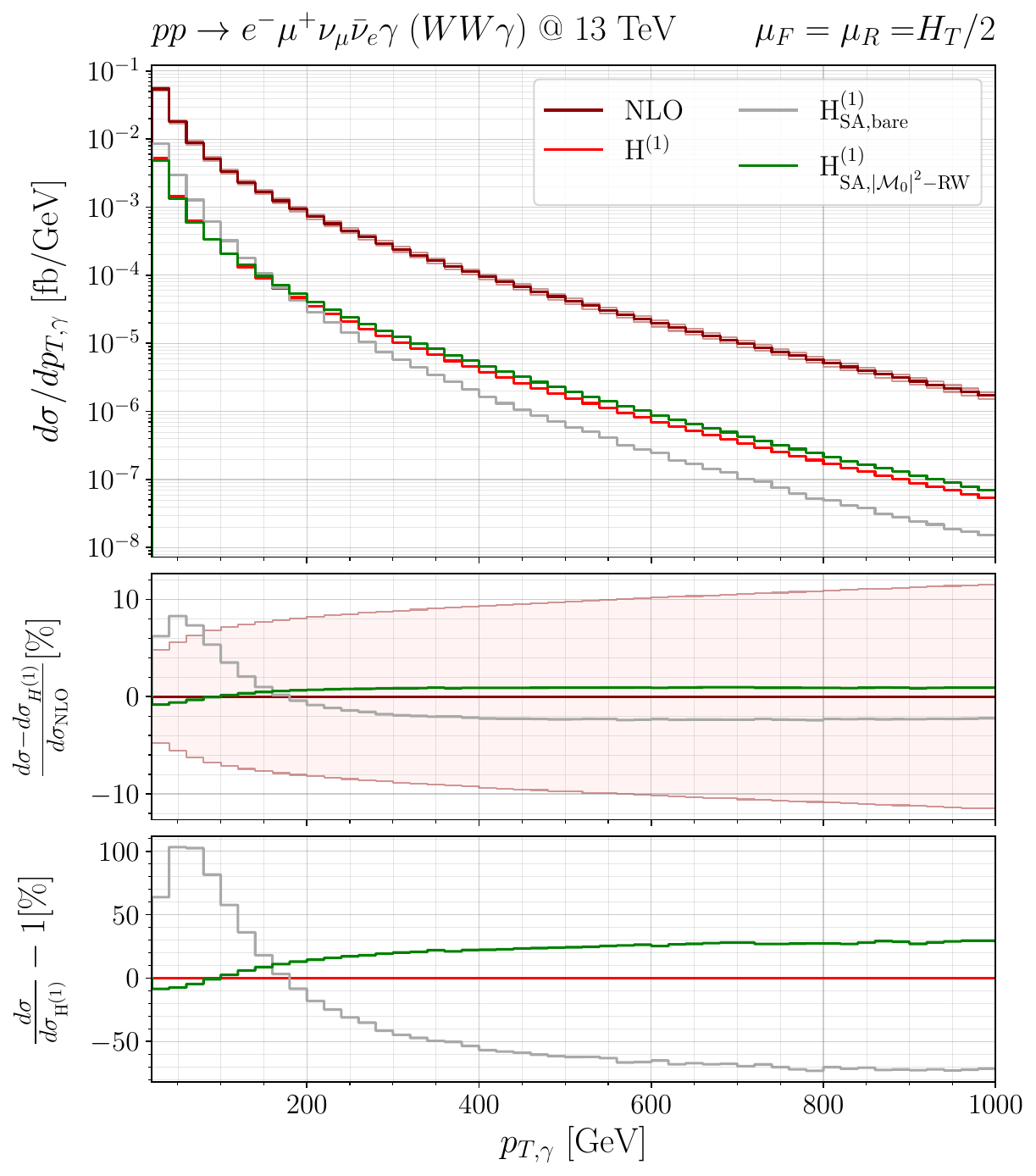}
\hfill
\includegraphics[height=\plotheightstd]{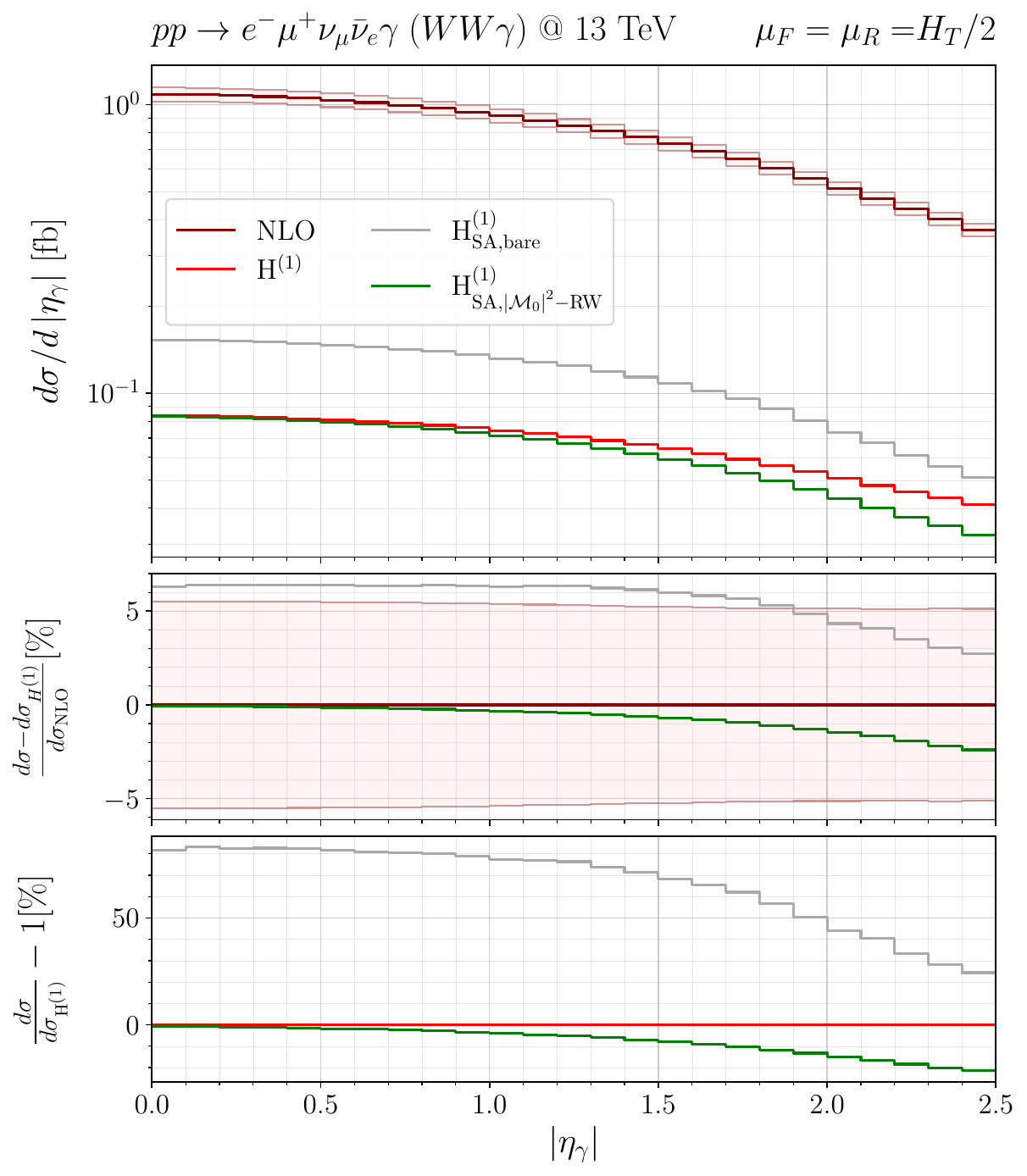}
\vspace*{1ex}
\caption{\label{fig:errorWWAH1based} Distributions to illustrate the performance
  of the SA at NLO, which is used to derive the \Hone-based error estimate, \DeltaSAHone, in the photon transverse
  momentum, \pTgamma (left), and its absolute pseudo-rapidity, \absetagamma (right), for \WWgamma production.
  Details as in \reffi{fig:errorZAH1based}.
}
\end{figure}

We start by recalling some aspects of the SA error estimate.
As in \refse{sec:validationZA} for \Zgamma production, we consider the photon transverse momentum, \pTgamma,
and its absolute pseudo-rapidity, \absetagamma, as sample distributions
to illustrate the behaviour of the SA at NLO in \reffi{fig:errorWWAH1based}.
For the \noRW SA we find a qualitatively similar behaviour to \Zgamma production (see Fig.~\ref{fig:errorZAH1based}):
a substantial overestimate of the exact result at low \pTgamma, where the bulk of the cross
section lies, and an underestimate in the high-\pTgamma region. The
\absetagamma distribution mirrors the behaviour of the low-\pTgamma region in particular
for central pseudo-rapidities. This overestimate is reduced towards larger pseudo-rapidities,
reflecting the same shape effect as observed for \Zgamma production.
It is, however, significantly more pronounced than for \Zgamma production, leading to
a larger deviation from the exact NLO result and even exceeding the scale-variation band, in
particular in the phase-space regions that contribute most to the fiducial cross section,
namely those dominated by the low-\pTgamma region.
However, with the \MzeroRW SA, which has provided a reasonable approximation of the exact NLO result
already for \Zgamma production, we find a significant improvement.
For low \pTgamma values, the \dsHone contribution is reproduced within $10\%$,
while at larger \pTgamma it is overestimated by about $30\%$.
Due to the small size of \dsHone and its decreasing impact as \pTgamma increases,
the NLO cross section is reproduced at the percent level across the full \pTgamma range.
For the \absetagamma distribution, the NLO result is reproduced at the per mille level in the central region,
while it is understimated by up to about $2\%$ at larger pseudo-rapidities.

\begin{figure}[t]
\centering
\includegraphics[height=\plotheightstd]{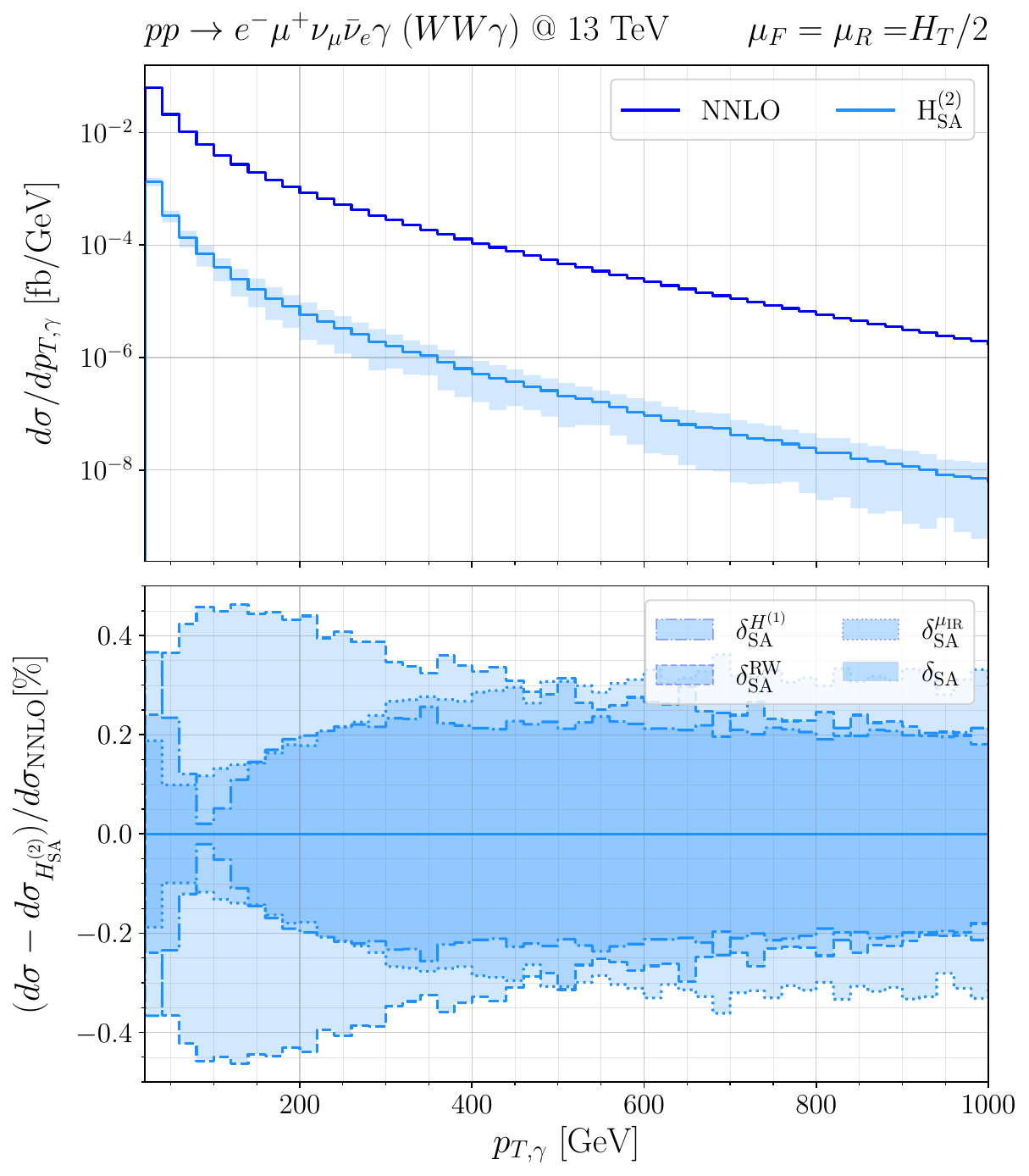}
\hfill
\includegraphics[height=\plotheightstd]{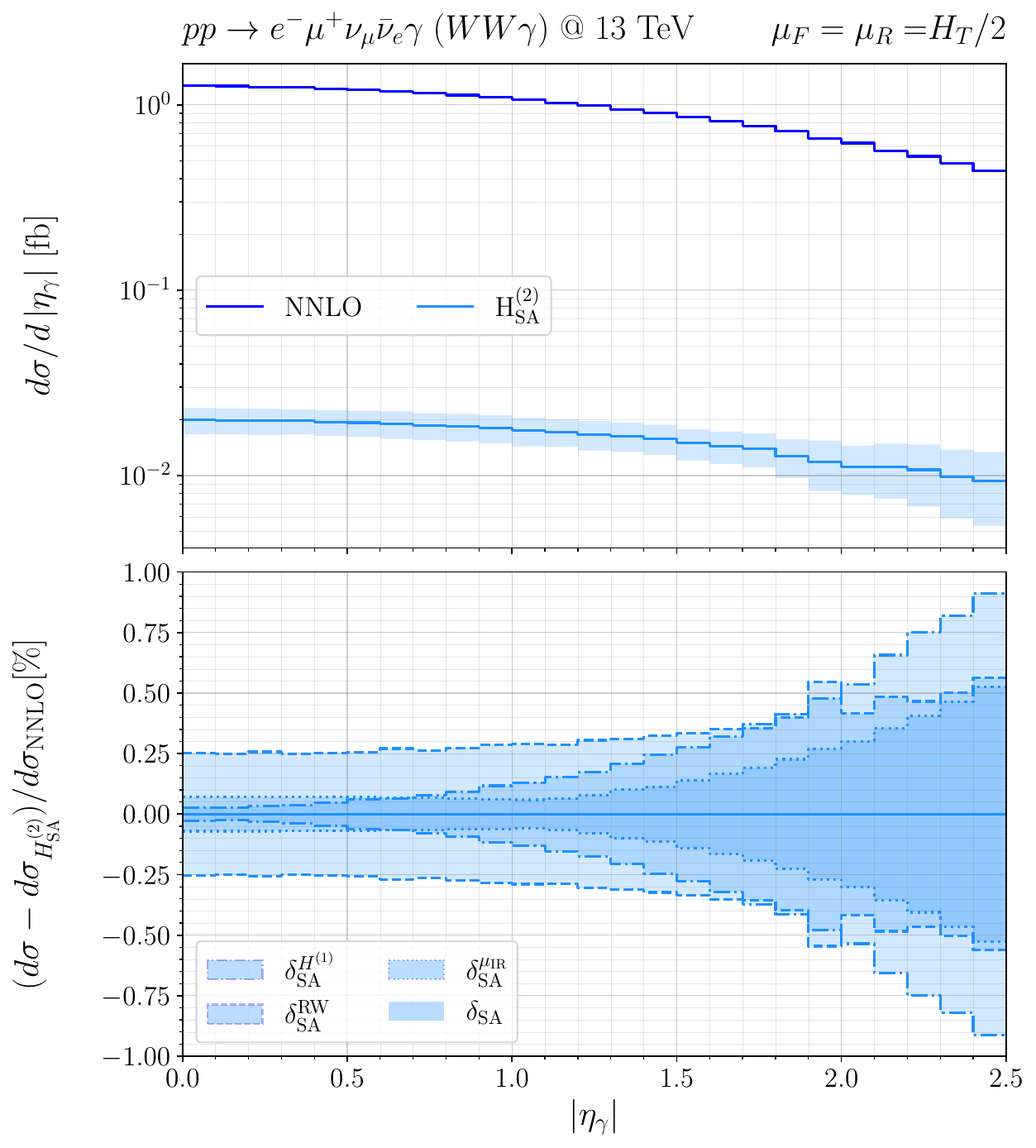}
\vspace*{1ex}
\caption{\label{fig:errorWWAcombined} Distributions to illustrate the construction of the
  combined error estimate, \DeltaSA in the photon transverse momentum,
  \pTgamma (left), and its absolute pseudo-rapidity, \absetagamma (right), for \WWgamma production.
  The upper panels show the nominal NNLO result, obtained in the \MoneMzeroRW SA,
  as well as the corresponding integrated \Htwo coefficient.
  The lower panels show
  the three individual error estimates
  \DeltaSAHone, \DeltaSARW, and \DeltaSAmuIR as bands, relative to
  the nominal NNLO prediction.
  Also depicted in both panels is \DeltaSA, which is defined as the envelope
  of the three individual error estimates and thus corresponds to the entire shaded area.
}
\end{figure}

Moving to NNLO, we apply the \textit{\MoneMzeroRW} prescription, which has
turned out to perform best for \Zgamma production, as our nominal prediction.
Following \refse{sec:errors}, we use exactly the same procedure as in \refse{sec:validationZA} to define the
three uncertainties \DeltaSAHone, \DeltaSARW, and \DeltaSAmuIR,
which are indicated separately for the \pTgamma and \absetagamma distributions in \reffi{fig:errorWWAcombined}.
From the left plot we see that for the lowest \pTgamma values the dominant uncertainty
is \DeltaSAHone, \DeltaSARW takes over at \mbox{$\pTgamma\sim50\,\GeV$},
and \DeltaSAmuIR becomes the dominant contribution for \mbox{$\pTgamma\gtrsim500\,\GeV$}.
The right plot indicates that in the low-\absetagamma region \DeltaSARW dominates,
while for \mbox{$\absetagamma\gtrsim2$} \DeltaSAHone takes over.
The final SA error estimate is constructed
as the envelope of the three individual errors. We show this combined error estimate,
\DeltaSA, in absolute terms about \dsHtwoSA in the upper panel and, together with the individual error bands
\DeltaSAHone, \DeltaSARW, and \DeltaSAmuIR, relative to the full NNLO cross section
in the lower panel.
We find that the combined error estimate remains below $1\%$ of the NNLO cross section throughout,
typically at the level of only few per mille.
When including the wider survey of kinematic distributions collected in
\reffis{WWAplots_pT_gamma}{WWAplots_dm_em_mup_dR_em_mup_dR_em_gamma} of \refapp{app:resultsWWA},
slightly larger errors of up to about $4\%$ occur,
characterised by large rapidity separations
between a charged lepton and the photon, as already observed for \Zgamma production.

Compared to \Zgamma production, the approximation error estimate is typically somewhat smaller,
despite the slightly larger impact of the two-loop virtual contribution.
To get a better understanding of this observation, we can leverage the distribution in $\pTgamma/Q$,
where $Q$ is the invariant mass of the colourless system, for \Zgamma and \WWgamma production. 
This distribution allows us to estimate how {\it soft} the photon is with respect to the underlying hard-scattering process,
and, therefore, how good our approximation of the two-loop virtual contribution may be expected to be.
We find that, on average, the photon in \WWgamma is softer than in \Zgamma.
This is a consequence of the fact that the \WWgamma process probes larger values of $Q$ compared to \Zgamma,
while the photon $p_T$ threshold used in the two analyses is similar.

\begin{figure}[t]
\centering
\includegraphics[height=\plotheightstd]{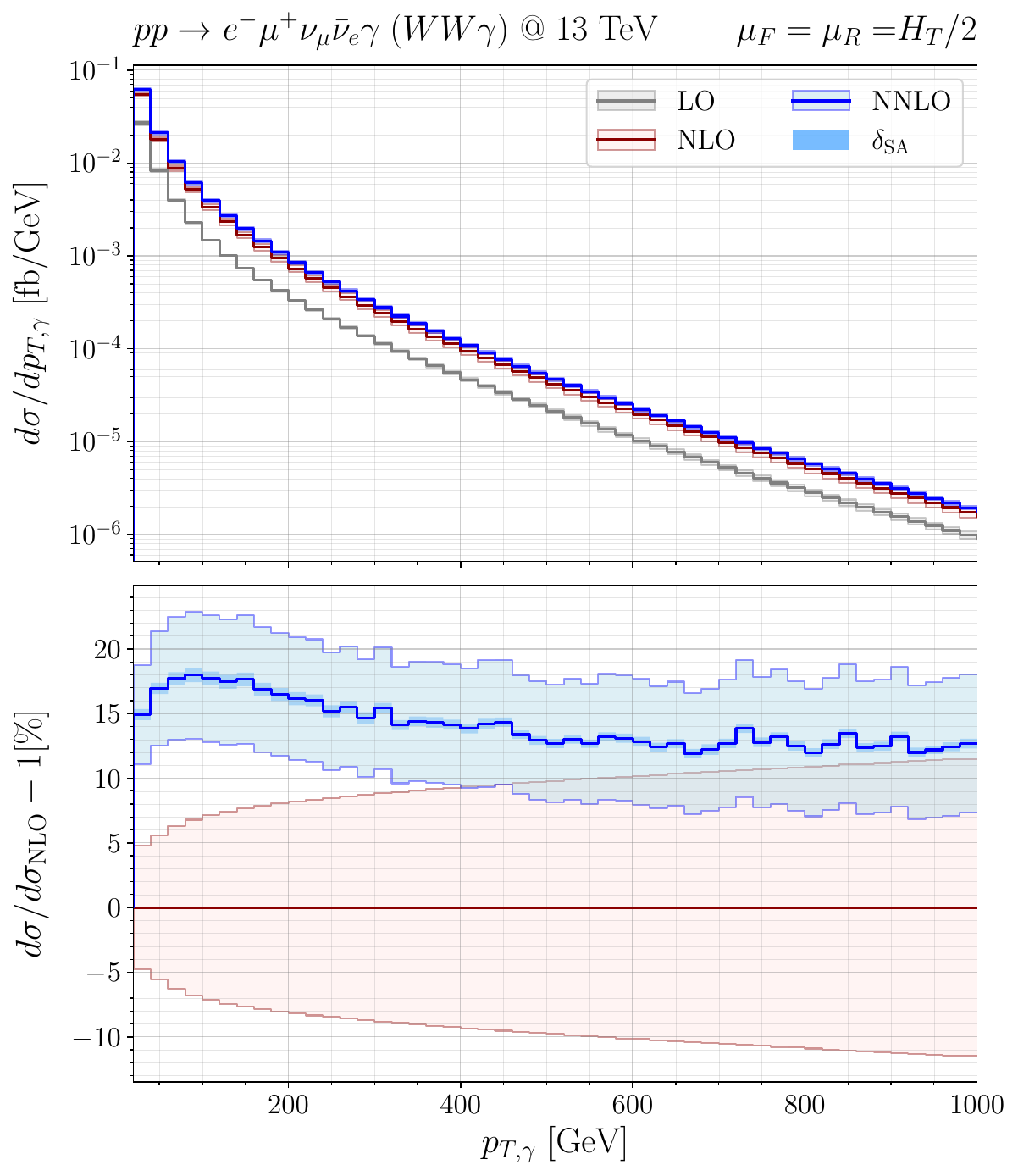}
\hfill 
\includegraphics[height=\plotheightstd]{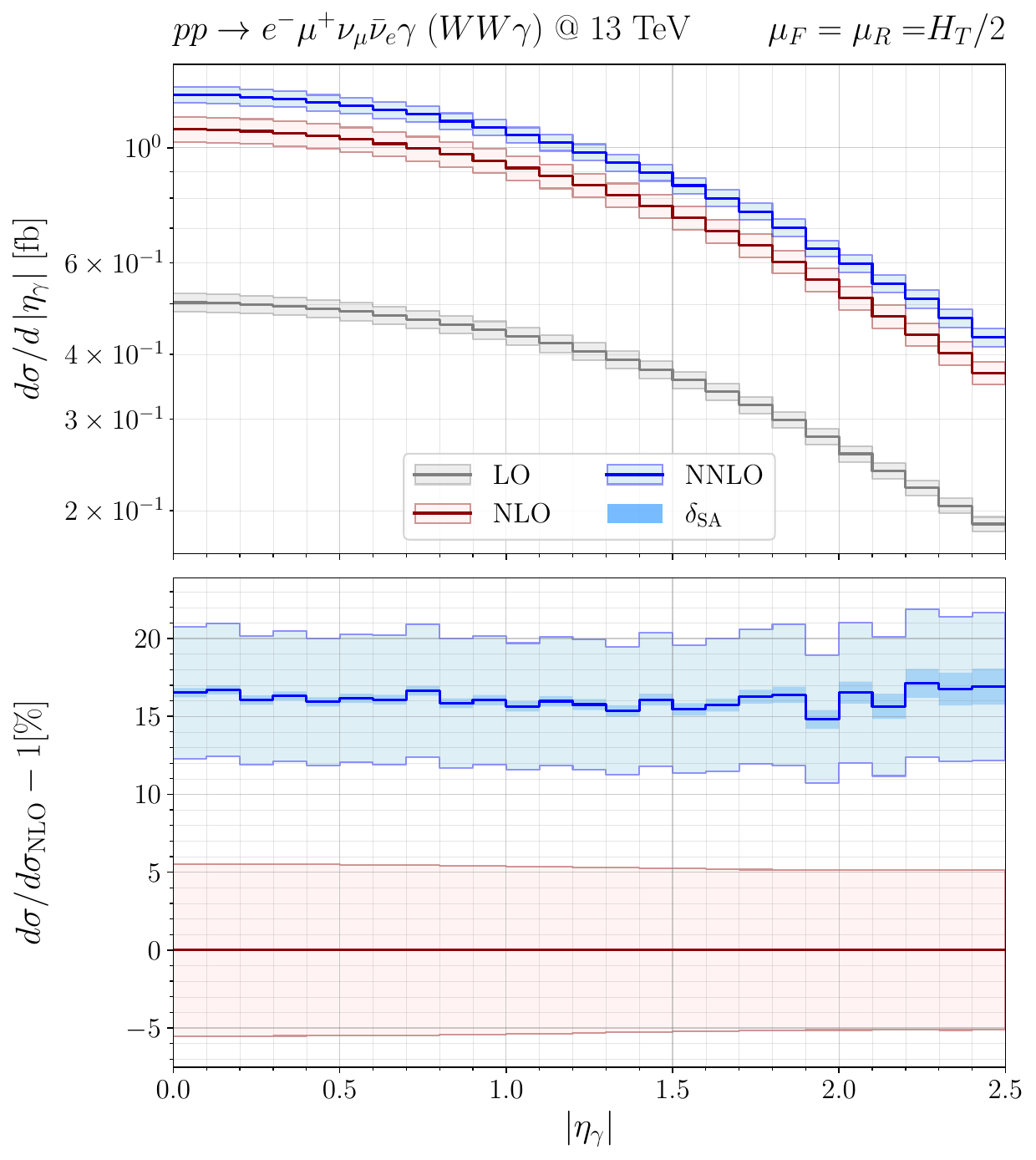}
\vspace*{1ex}
\caption{\label{fig:errorWWAKfactors} Distributions in the photon
  transverse momentum, \pTgamma (left),
  and its absolute pseudo-rapidity,
  \absetagamma (right), for \WWgamma production.
  The upper panels show the absolute predictions at LO, NLO, and NNLO accuracy,
  the latter obtained in the nominal \MoneMzeroRW SA,
  together with their conventional seven-point scale-uncertainty bands.
  The lower panels depict the NLO and NNLO results relative to the former,
  both again with their scale-uncertainty bands. For the latter,
  we additionally show the final SA error estimate, \DeltaSA.
}
\end{figure}

Finally, we compare the combined SA error estimate with the
perturbative uncertainties of our NNLO prediction, estimated through scale variations,
in \reffi{fig:errorWWAKfactors}.
We find, as previously for \Zgamma production,
that the approximation error is subdominant throughout,
i.e.\ the NNLO accuracy of the result is by no means compromised
by approximating the two-loop amplitudes.

We conclude this section with a few comments on the impact of the NNLO corrections.
For the distributions in \reffi{fig:errorWWAKfactors}, they are relatively flat,
in particular for the $|\eta_{\gamma}|$ distribution, where they remain at about $16\%$ throughout the full range.
For the \pTgamma distribution they increase from about $15\%$ at very low \pTgamma
to about $18\%$ around \mbox{$\pTgamma\sim 50-100$\,GeV}, before gradually decreasing again to about $13\%$ in the high-\pTgamma region.
For both distributions, the NLO and NNLO scale-variation bands do not overlap
in the bulk region, confirming the observation for multi-boson processes, in particular involving photons, that those bands underestimate true
perturbative uncertainties at least up to NLO.
The tail of the \pTgamma distribution draws a slightly different
picture: there, the corrections are dominated
by topologies with the \WWgamma system recoiling against a light jet,
which open up only at NLO and therefore lead to larger LO-like scale uncertainties of the NLO cross section.
The dominance of these topologies also leads to slightly larger scale uncertainties
of the NNLO prediction in the tail of the \pTgamma distribution.
Consequently, the NLO and NNLO scale-variation bands tend to overlap there.

\begin{figure}[t]
\centering
\includegraphics[height=\plotheightstd]{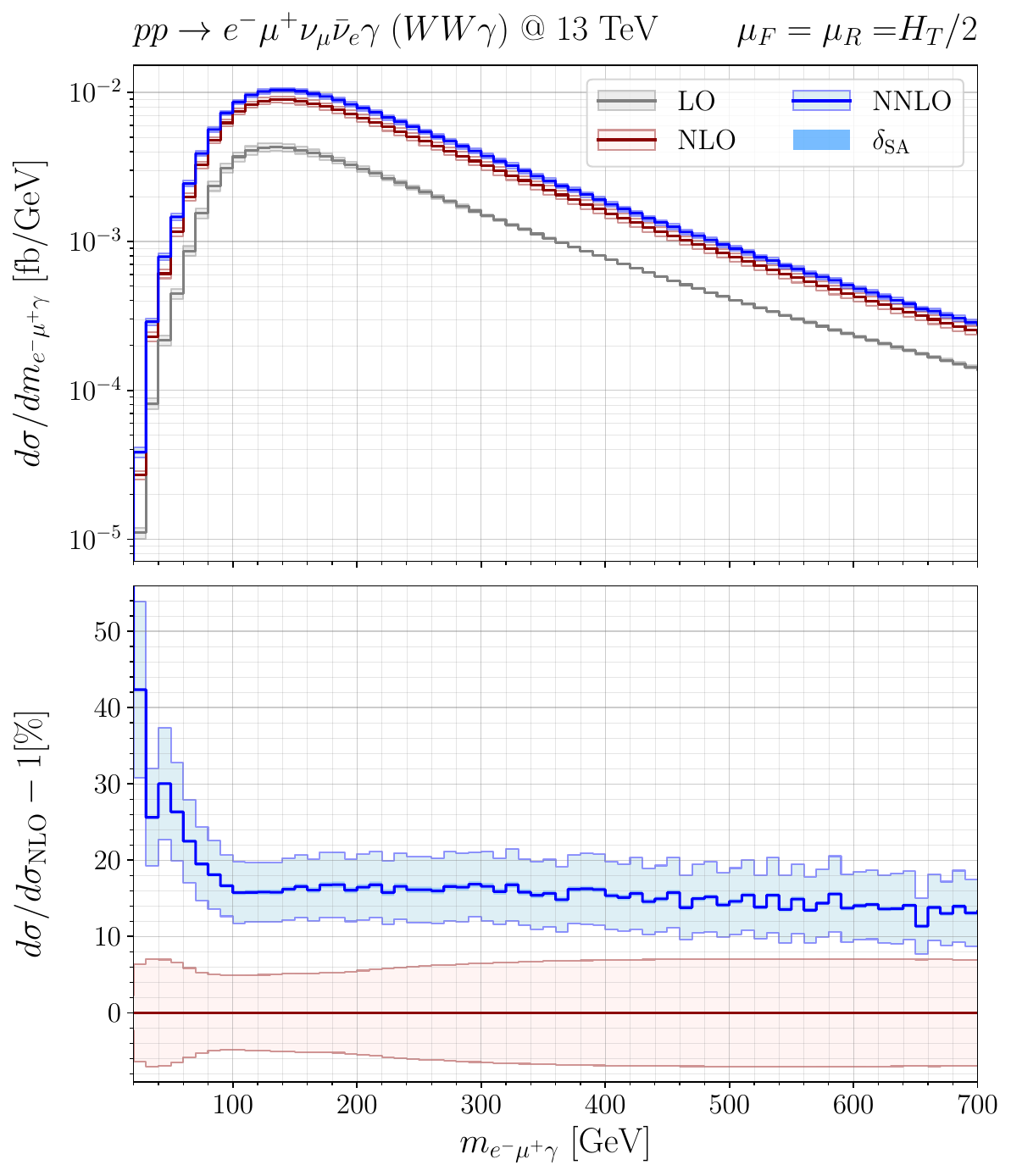}
\hfill 
\includegraphics[height=\plotheightstd]{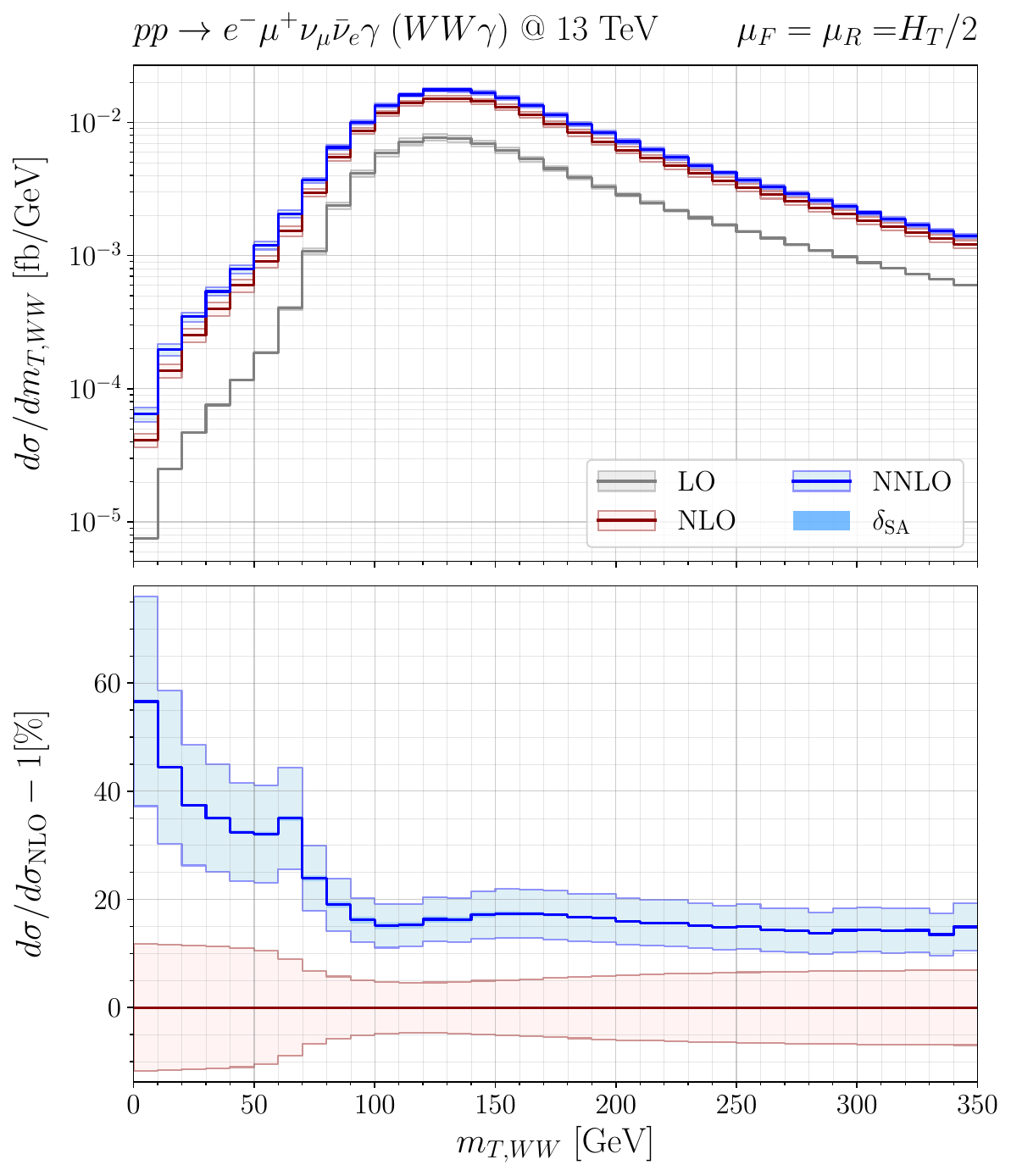}
\vspace*{1ex}
\caption{\label{fig:errorWWAKfactorsCMS} Distributions in the invariant mass
  of the dilepton--photon system, $m_{e^-\mu^+\gamma}$ (left), and
  the transverse mass of the \WW system, $m_{T,WW}$ (right), for \WWgamma production.
  Details as in \reffi{fig:errorWWAKfactors}.
}
\end{figure}

While no differential unfolded measurements have been provided so far, we
show in \reffi{fig:errorWWAKfactorsCMS} the invariant-mass distribution of the
dilepton--photon system, $m_{e^-\mu^+\gamma}$, and the transverse-mass distribution of the $W$-pair,
$m_{T,WW}$, corresponding to the observables used to establish the observation of \WWgamma production
by the CMS Collaboration~\cite{CMS:2023rcv}.
For these two distributions as well, the NNLO corrections remain relatively flat at about $16\%$ around and above the peak of
the cross section. Only in the low-$m_{e^-\mu^+\gamma}$ region, and particularly at low $m_{T,WW}$, where the cross section is suppressed,
we find larger corrections of up to about $40-60\%$.

In the wider survey of distributions presented in
\reffis{WWAplots_pT_gamma}{WWAplots_dm_em_mup_dR_em_mup_dR_em_gamma} of \refapp{app:resultsWWA},
we find a similar pattern:
NNLO $K$-factors around the fiducial value in the bulk regions, with moderate
variations typically within about $\pm10\%$,
and more pronounced effects in suppressed phase-space
regions, with non-overlapping scale-variation bands throughout.
The picture is slightly different for transverse-momentum distributions
 as exemplarily discussed for \pTgamma.

\section{Summary}
\label{sec:summary}
In this paper, we have presented first NNLO QCD predictions for the production of a $W^+W^-$ pair
in association with a photon at the LHC. The leptonic decays of the $W$ bosons are treated fully off-shell, thereby
accounting for all non-resonant effects, spin correlations, and interferences.
The calculation is exact apart from the finite part of the two-loop virtual contribution,
which is estimated by applying a soft-photon approximation.
A similar approach has been used in recent computations of
\ttH~\cite{Catani:2022mfv,Devoto:2024nhl} and \ttW~\cite{Buonocore:2023ljm} production.
We have validated our approach using \Zgamma production, for which the exact two-loop
amplitude~\cite{Gehrmann:2011ab} and complete NNLO results~\cite{Grazzini:2013bna,Grazzini:2015nwa,Campbell:2017aul}
have been long available.
Since the photon is massless, a soft approximation is certainly motivated,
and indeed works extremely well in its region of validity.
The selection cuts typically applied for these processes set a minimum transverse momentum
for the photon of the order of $15-20$ GeV.
The quality of the approximation therefore largely depends on the typical energy scales of the partonic processes.
Our SA is defined through a reweighting procedure, with an error estimate constructed from several ingredients,
including the quality of its performance at NLO and the intrinsic ambiguities of the approximation procedure.

For our validation process, \Zgamma production, we find that the NNLO result based on the SA
is compatible with the exact NNLO result
within the error estimate we have derived, across the entire phase-space and for the wide range of observables studied.
Moreover, we find that this error estimate is always much smaller than the residual perturbative uncertainties.

The main reason for the success of this approach is the small size of the two-loop virtual contribution.
In the region of photon transverse momenta close to the selection cut, the photon is typically relatively soft,
and the SA provides a reasonable approximation of the small two-loop contribution. When the photon is hard,
such that the SA is not well justified, the reweighting procedure still corrects the kinematics reasonably well,
while the two-loop contribution is further suppressed since the cross section is dominated by real emissions.
Therefore, even if the quality of the SA deteriorates at high $\pTgamma$,
the impact on the NNLO cross section is essentially negligible.

We have then applied the same procedure to \WWgamma production.
Due to the similarities in the production mechanisms between diboson and triboson processes,
we do not expect significant changes in the performance of the SA.
The pattern of QCD radiative corrections at NLO is quite similar to
\Zgamma production, and we find about $+16\%$ NNLO corrections, with the
two-loop finite remainder in SA affecting the NNLO cross section by less than $2\%$.
The residual perturbative uncertainties are estimated to be at the $4\%$ level.
On the other hand, the error estimate from approximating the two-loop amplitude is found to be
at the few per mille level, even slightly smaller than for \Zgamma production.
This is not completely unexpected, given that the photon radiated in \WWgamma production is, on average, slightly softer.
With this error estimate and its validation in \Zgamma production at hand,
our predictions can safely be regarded as NNLO accurate.
Comparing our results against the measurements of the CMS~\cite{CMS:2023rcv} and ATLAS~\cite{ATLAS:2025yxf} Collaborations,
we find agreement at the fiducial level within the experimentally dominated uncertainties. 

Thanks to the continuous progress in the field of two-loop computations, explicit results for the
two-loop amplitudes for triboson processes might become available in the foreseeable future.
In the meantime, our approach can be employed to obtain accurate predictions for the remaining
processes in this class, helping to fully exploit upcoming experimental measurements of triboson processes
at the LHC, which are of utmost importance in testing the gauge structure of the Standard Model.

\section*{Acknowledgments}
We are grateful to Chiara Savoini for several useful discussions on many
aspects of the soft-photon approximation during the completion of this work and valuable comments on this manuscript.
This work is supported in part by the Swiss National Science Foundation (SNF) under contract $200020\_219367$.

\appendix

\section[Differential \Zgamma results in soft-photon approximation]{Differential $\boldsymbol{Z\gamma}$ results in soft-photon approximation}
\label{app:validationZA}
In this appendix, we provide a survey of differential cross sections
for \Zgamma production to illustrate the performance of the soft-photon approximation
and its associated error estimate across a broader range of phase-space regions.
All plots follow the descriptions in \reffi{fig:errorZAH1based} (left columns),
\reffi{fig:errorZAcombined} (central columns) and
\reffi{fig:errorZAKfactors} (right columns), respectively. For reference, we additionally include
the \dsHtwoSAnoRW contribution in the plots of the central column.
Some of the distributions presented in this appendix have already been shown in the main text.
They are included to provide a consistent overview of the full survey of observables considered here.

\begin{figure}[t]
\centering
\includegraphics[height=\plotheightapp]{figures/ppeexa03_LHC13_error_estimate_NLO_H1_pT_gamma.pdf}
\hfill 
\includegraphics[height=\plotheightapp]{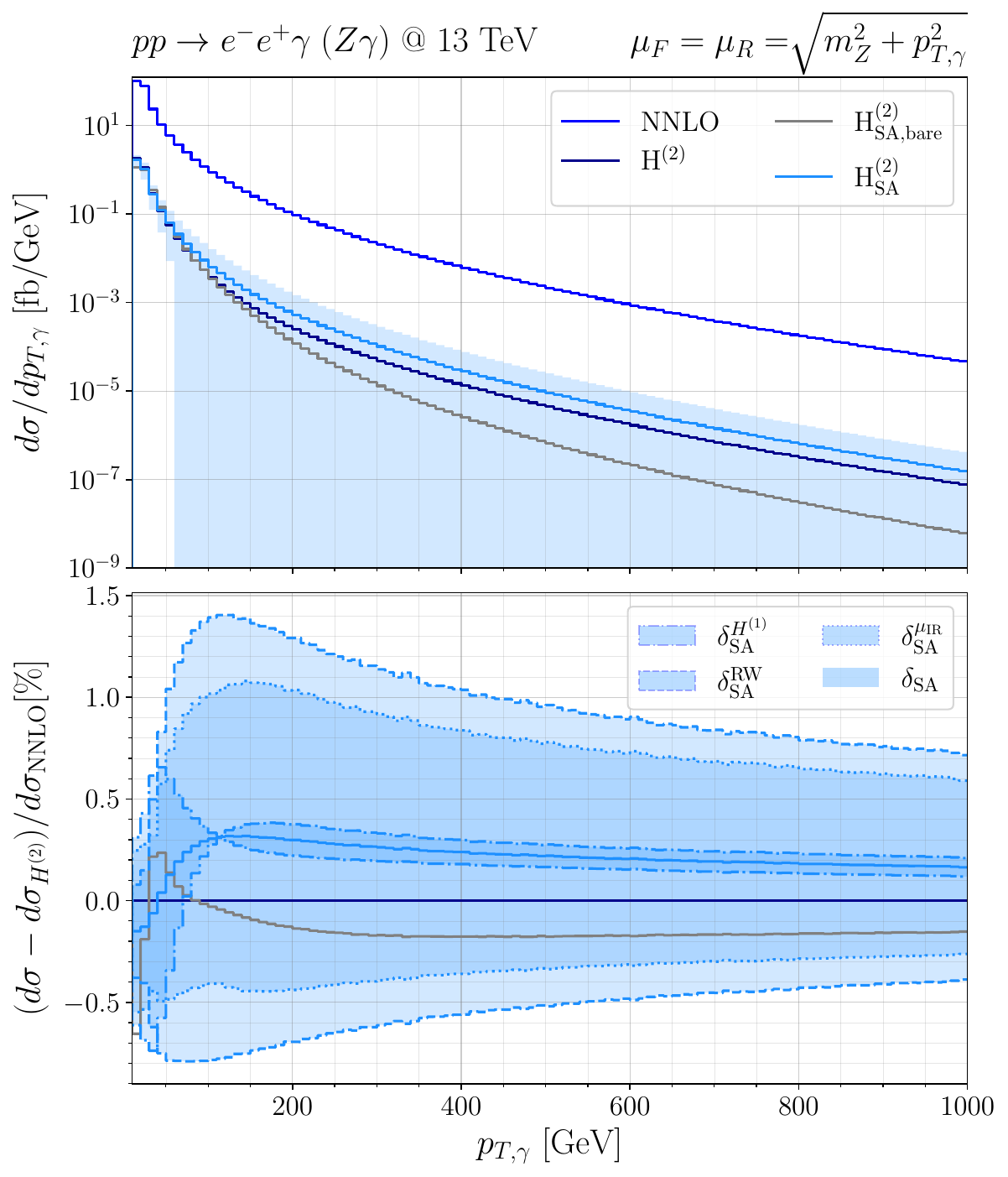}
\hfill 
\includegraphics[height=\plotheightapp]{figures/ppeexa03_LHC13_Kfactors_NNLO_pT_gamma.pdf}\\[2ex]

\caption{\label{ZAplots_pT_gamma} Distribution
  in the photon transverse momentum, \pTgamma, for \Zgamma production. The
  plots follow the descriptions in
  \reffi{fig:errorZAH1based} (left column),
  \reffi{fig:errorZAcombined} (central column) and
  \reffi{fig:errorZAKfactors} (right column), respectively.
  For reference, the \dsHtwoSAnoRW contribution is added in the plot of the central column.
}
\vspace*{5ex}
\end{figure}

\begin{figure}[p]
\centering
\includegraphics[height=\plotheightapp]{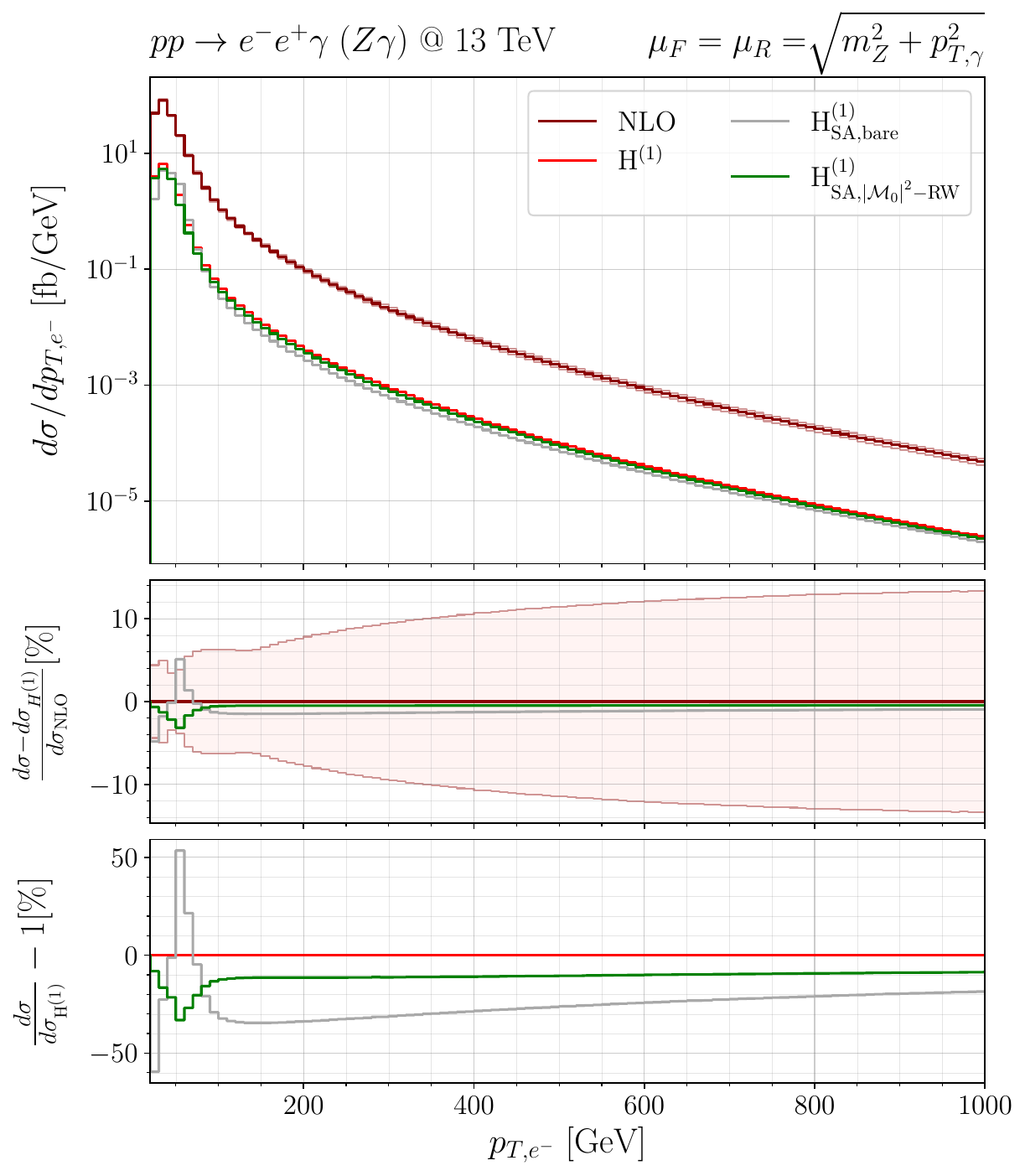}
\hfill 
\includegraphics[height=\plotheightapp]{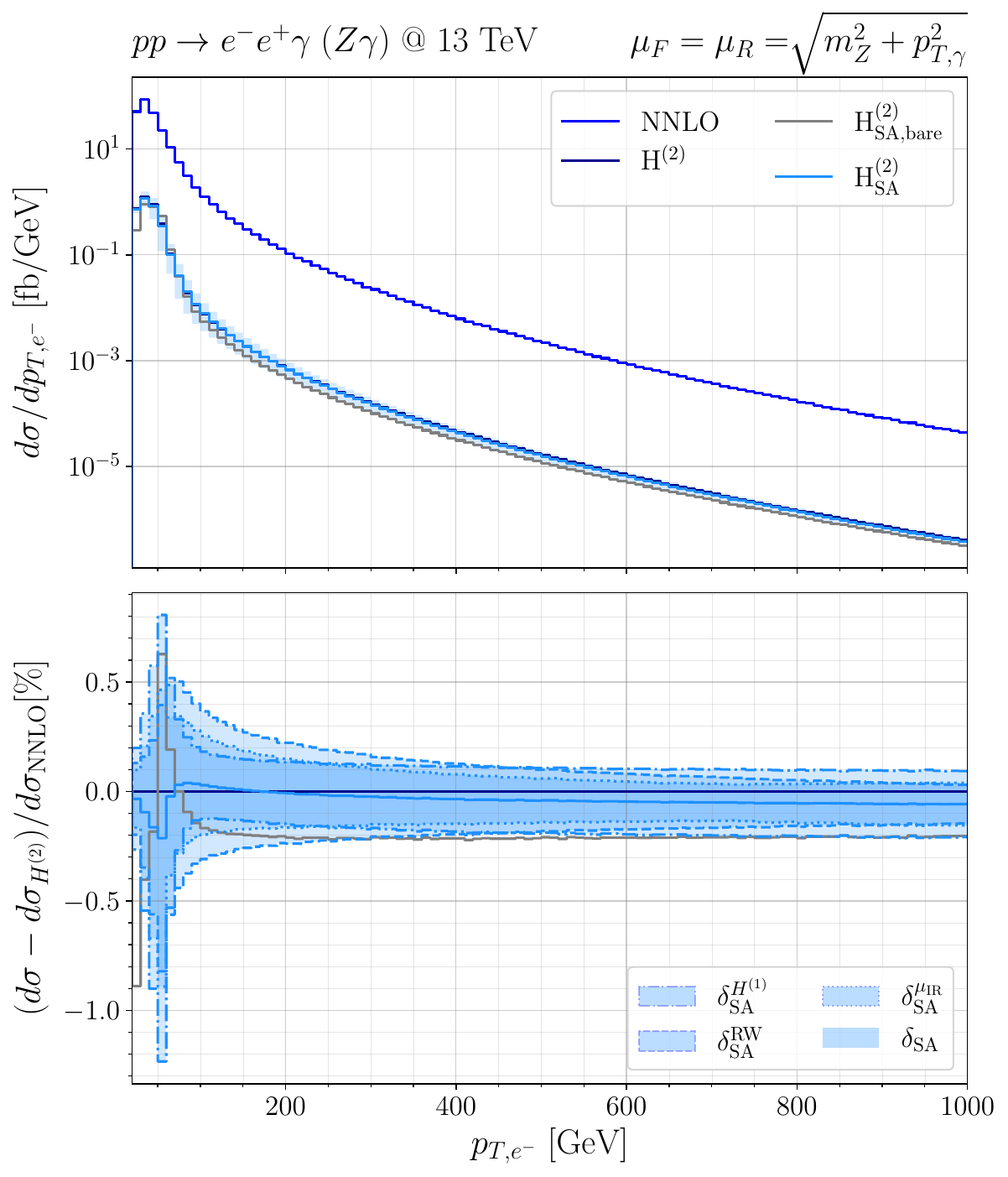}
\hfill 
\includegraphics[height=\plotheightapp]{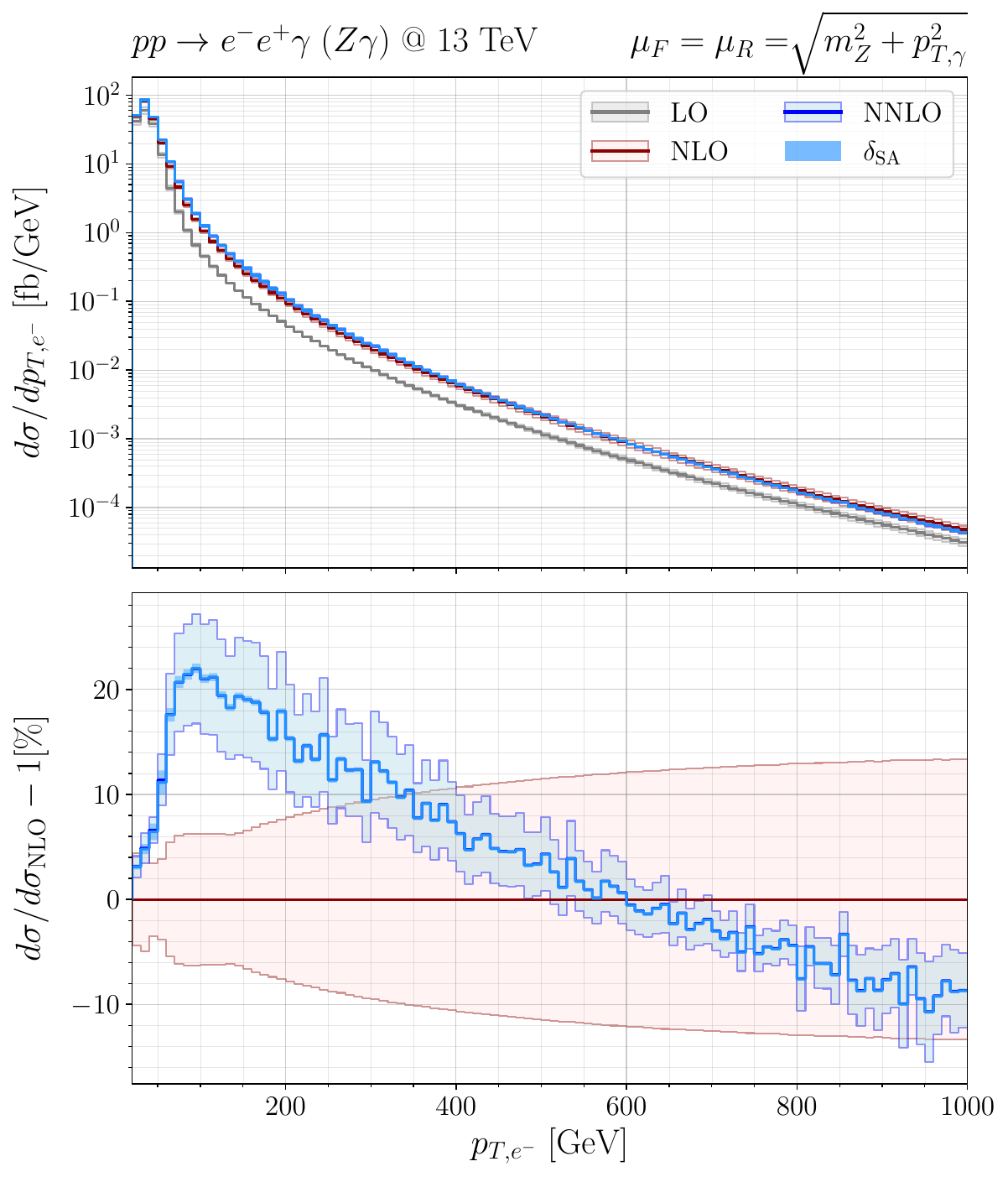}\\[2ex]

\includegraphics[height=\plotheightapp]{figures/ppeexa03_LHC13_error_estimate_NLO_H1_eta_gamma.pdf}
\hfill 
\includegraphics[height=\plotheightapp]{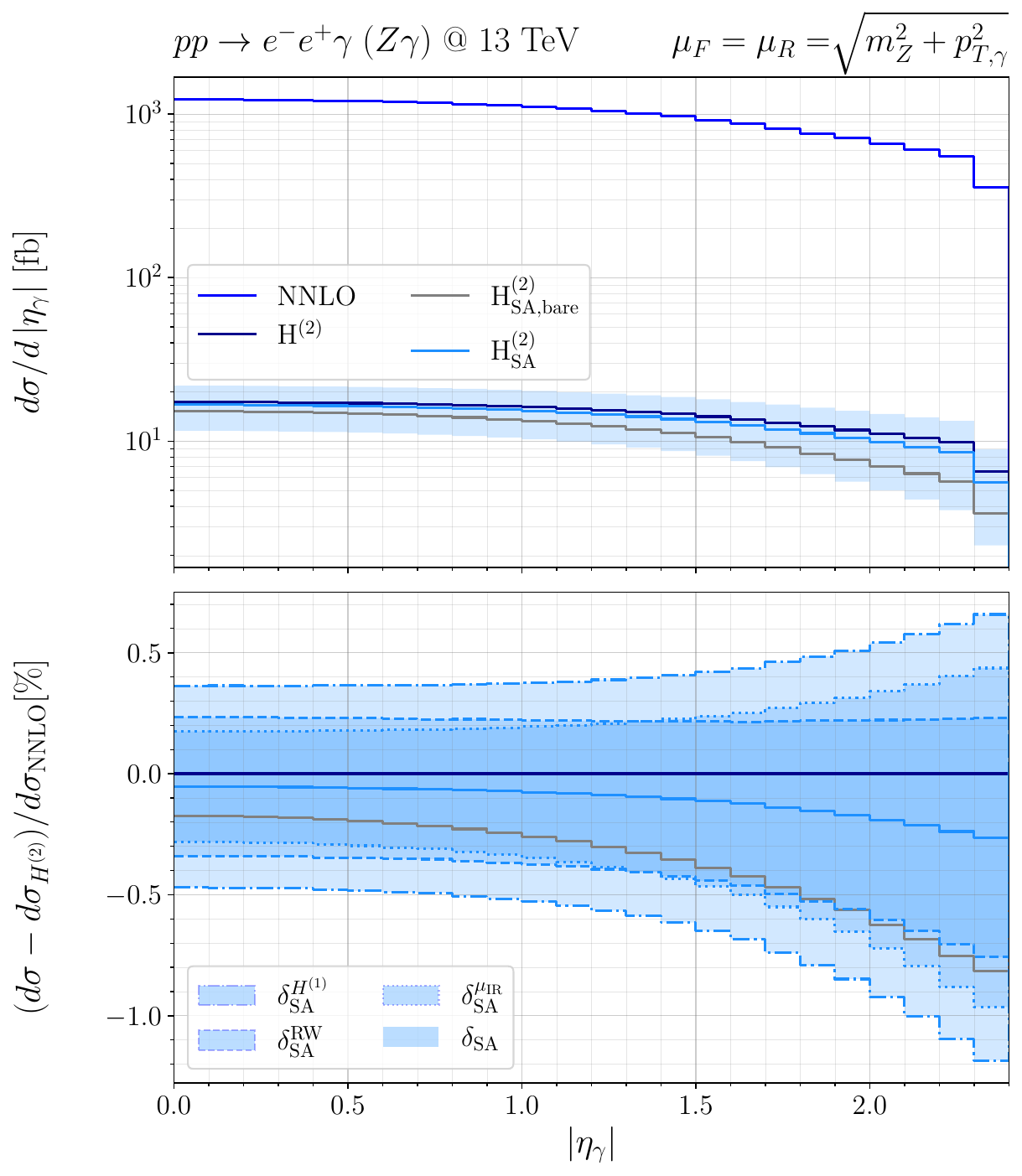}
\hfill 
\includegraphics[height=\plotheightapp]{figures/ppeexa03_LHC13_Kfactors_NNLO_eta_gamma.pdf}\\[2ex]

\includegraphics[height=\plotheightapp]{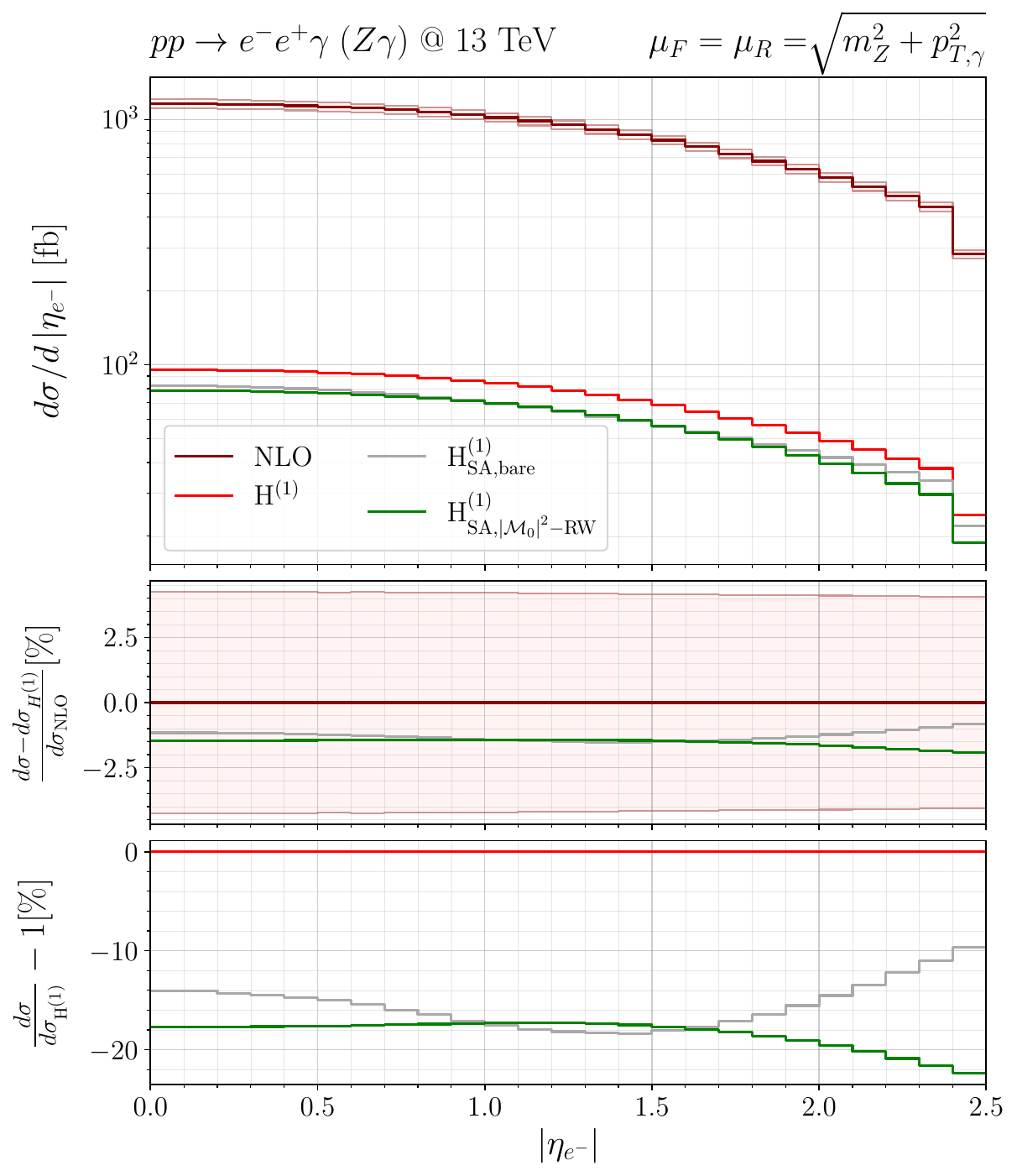}
\hfill 
\includegraphics[height=\plotheightapp]{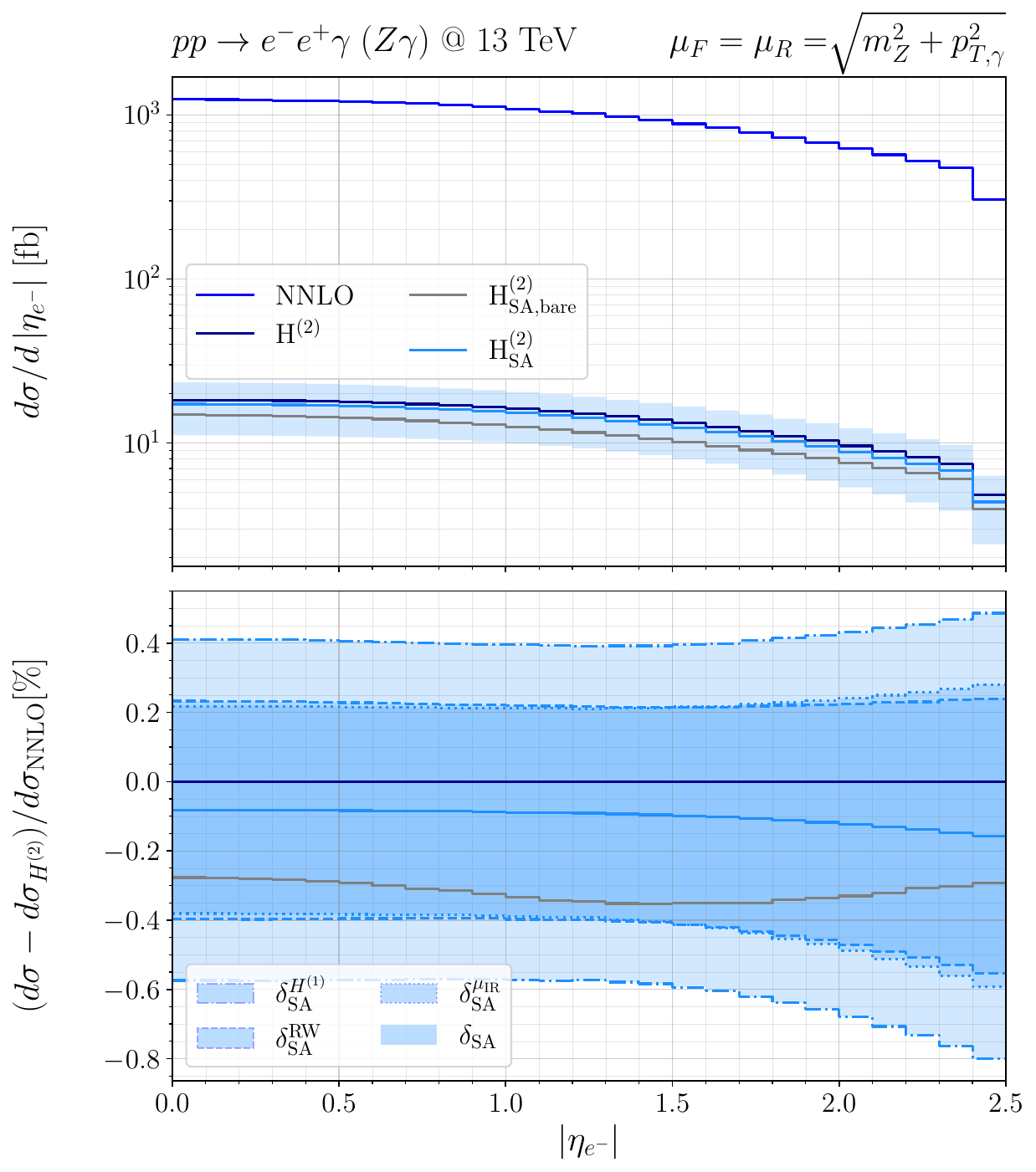}
\hfill 
\includegraphics[height=\plotheightapp]{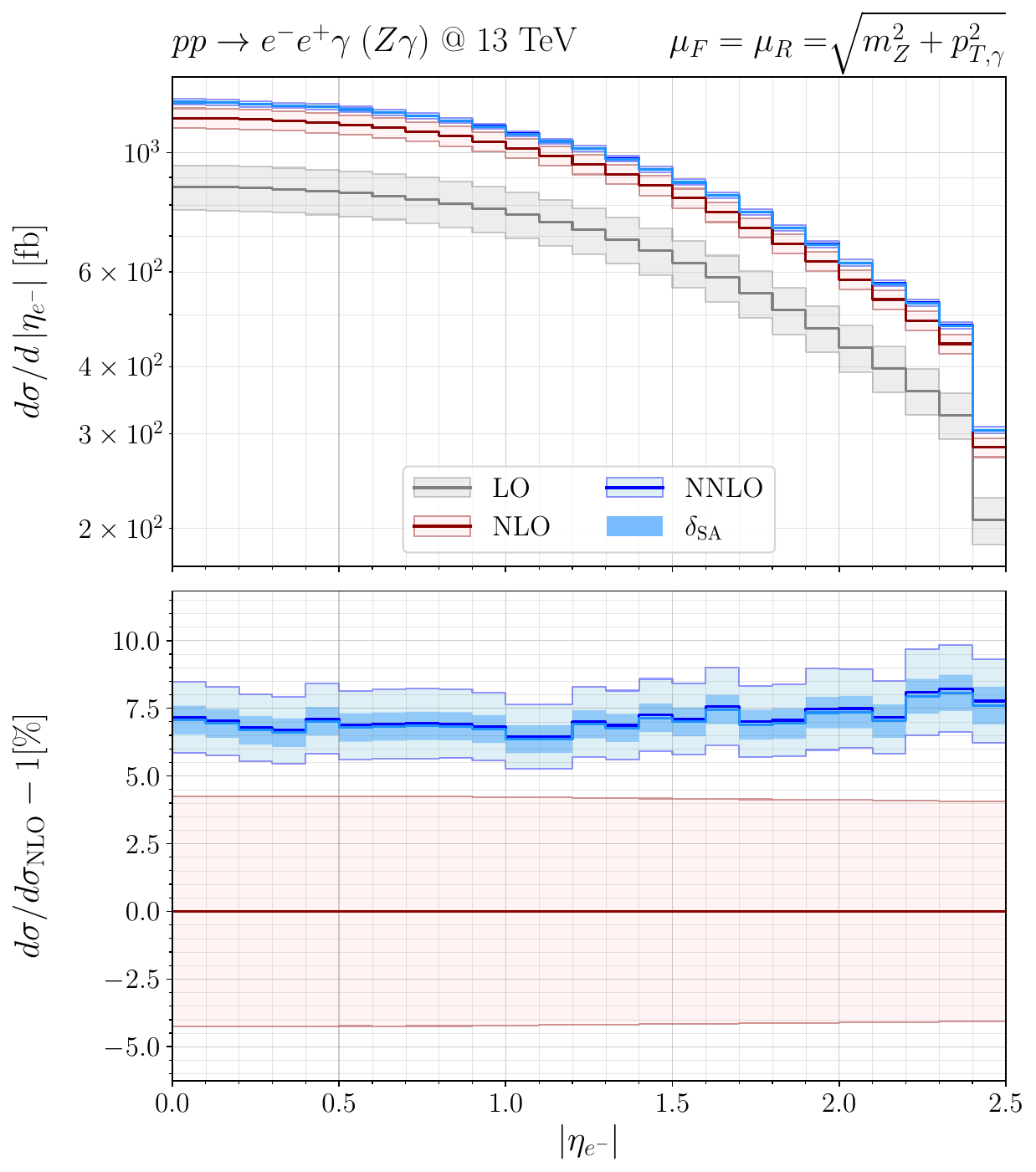}\\[2ex]

\caption{\label{ZAplots_pT_em_eta_gamma_em} Distributions in the electron transverse momentum, \pTem (first row),
  and in the absolute pseudo-rapidities of the photon, $|\eta_{\gamma}|$ (second row), and
  the electron, $|\eta_{e^-}|$ (third row) for \Zgamma production. The
  plots follow the descriptions in
  \reffi{fig:errorZAH1based} (left column),
  \reffi{fig:errorZAcombined} (central column) and
  \reffi{fig:errorZAKfactors} (right column), respectively.
  For reference, the \dsHtwoSAnoRW contribution is added in the plots of the central column.
}
\end{figure}

\begin{figure}[p]
\centering
\includegraphics[height=\plotheightapp]{figures/ppeexa03_LHC13_error_estimate_NLO_H1_m_em_ep_gamma.pdf}
\hfill 
\includegraphics[height=\plotheightapp]{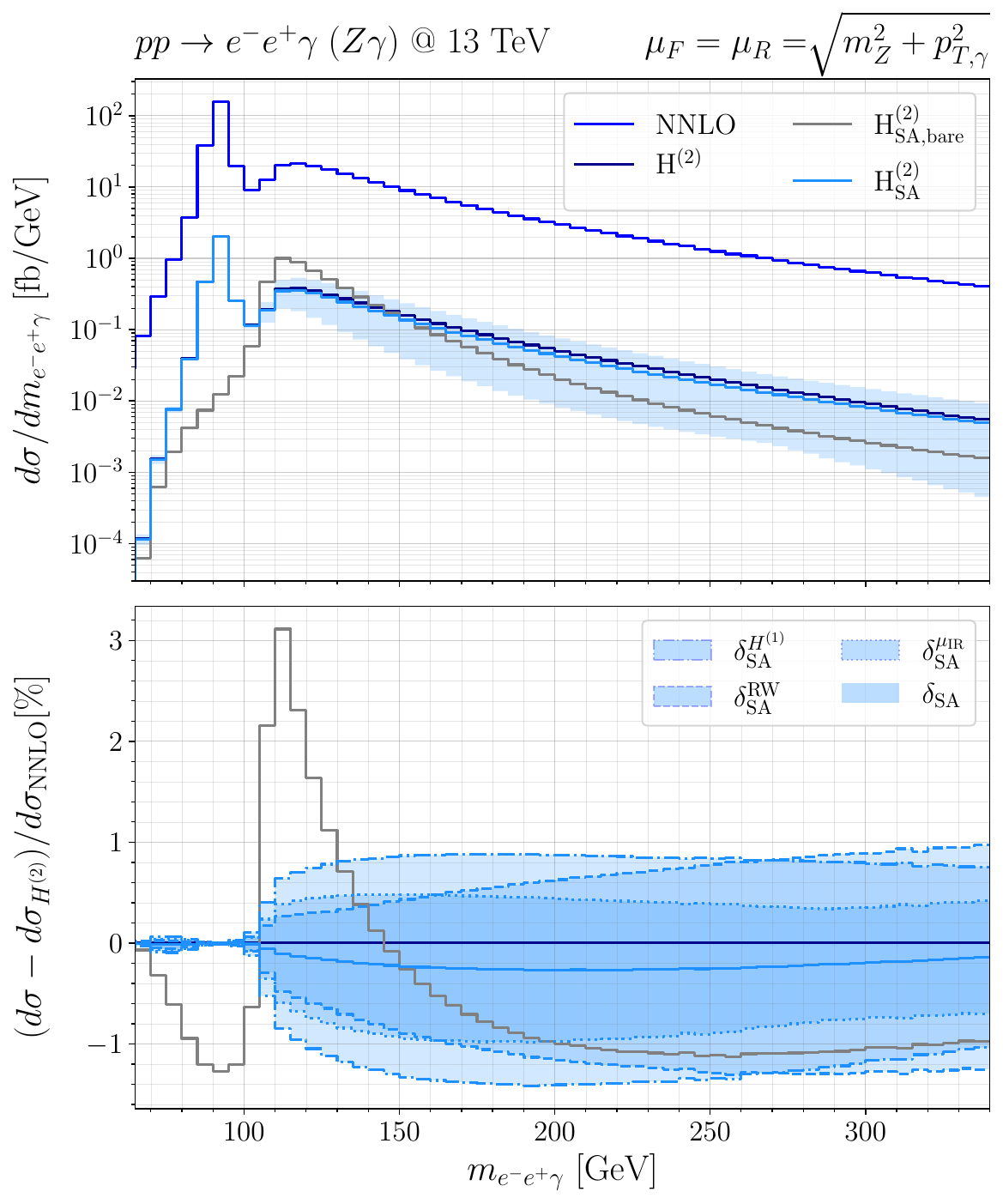}
\hfill 
\includegraphics[height=\plotheightapp]{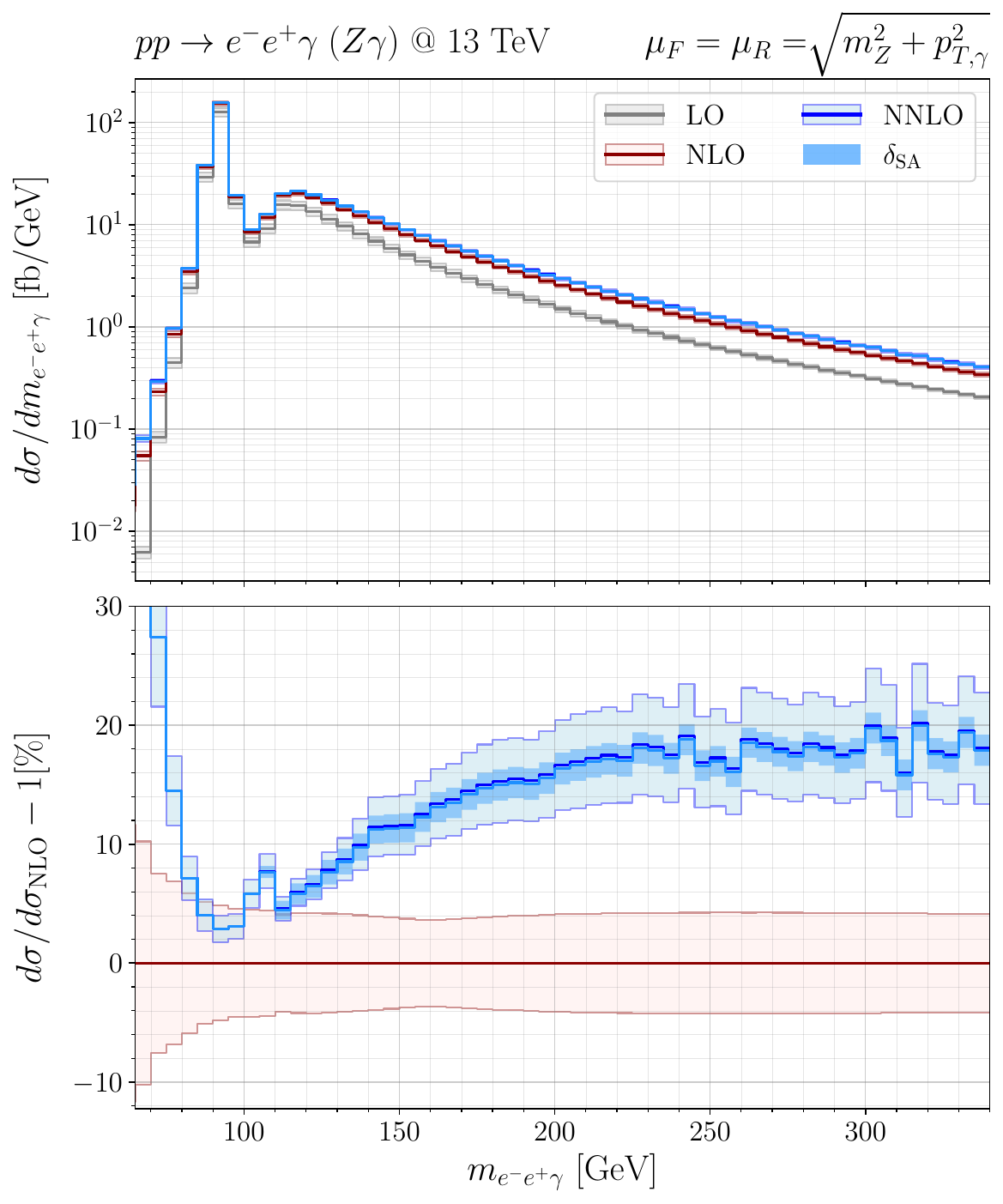}\\[2ex]

\includegraphics[height=\plotheightapp]{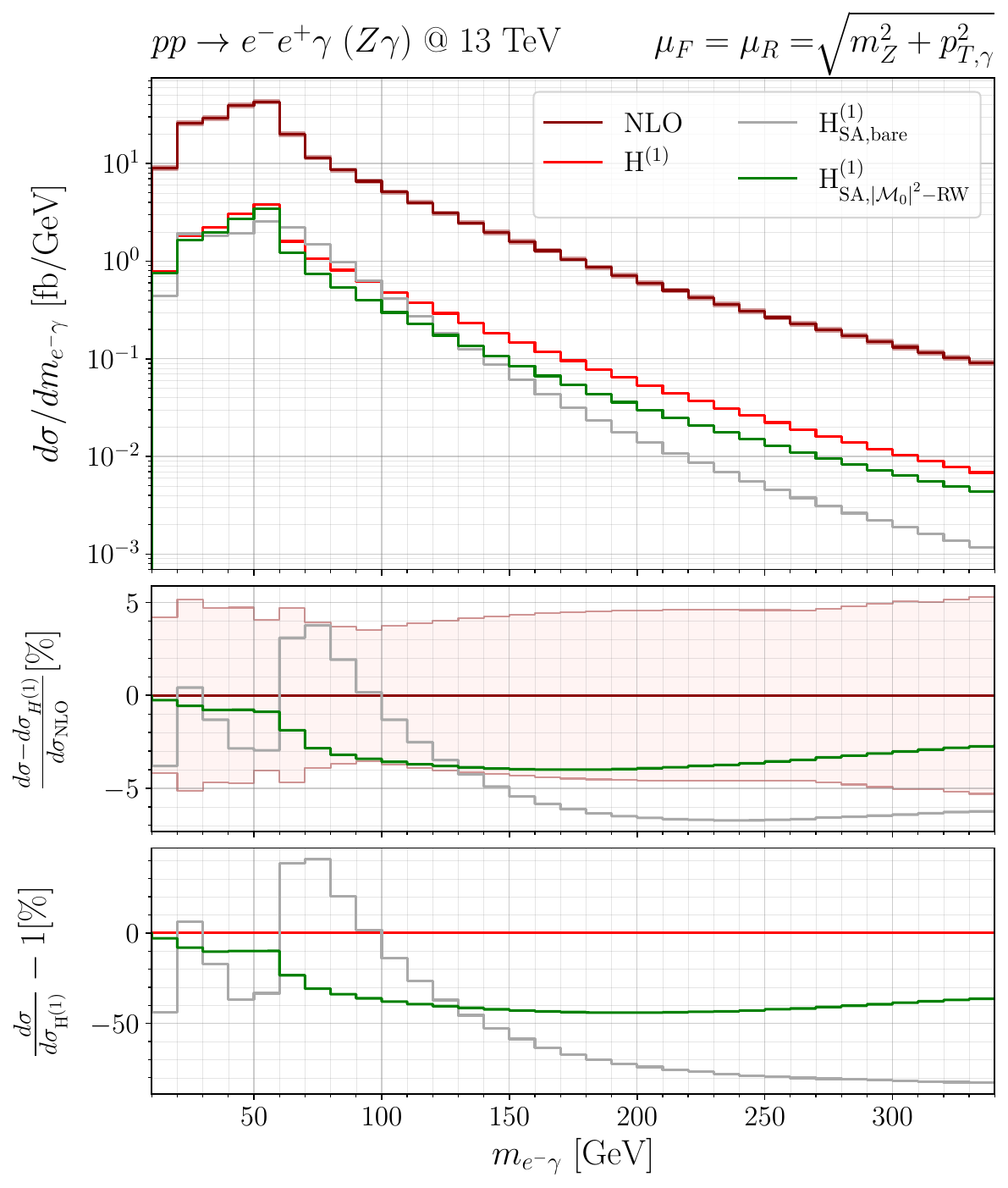}
\hfill 
\includegraphics[height=\plotheightapp]{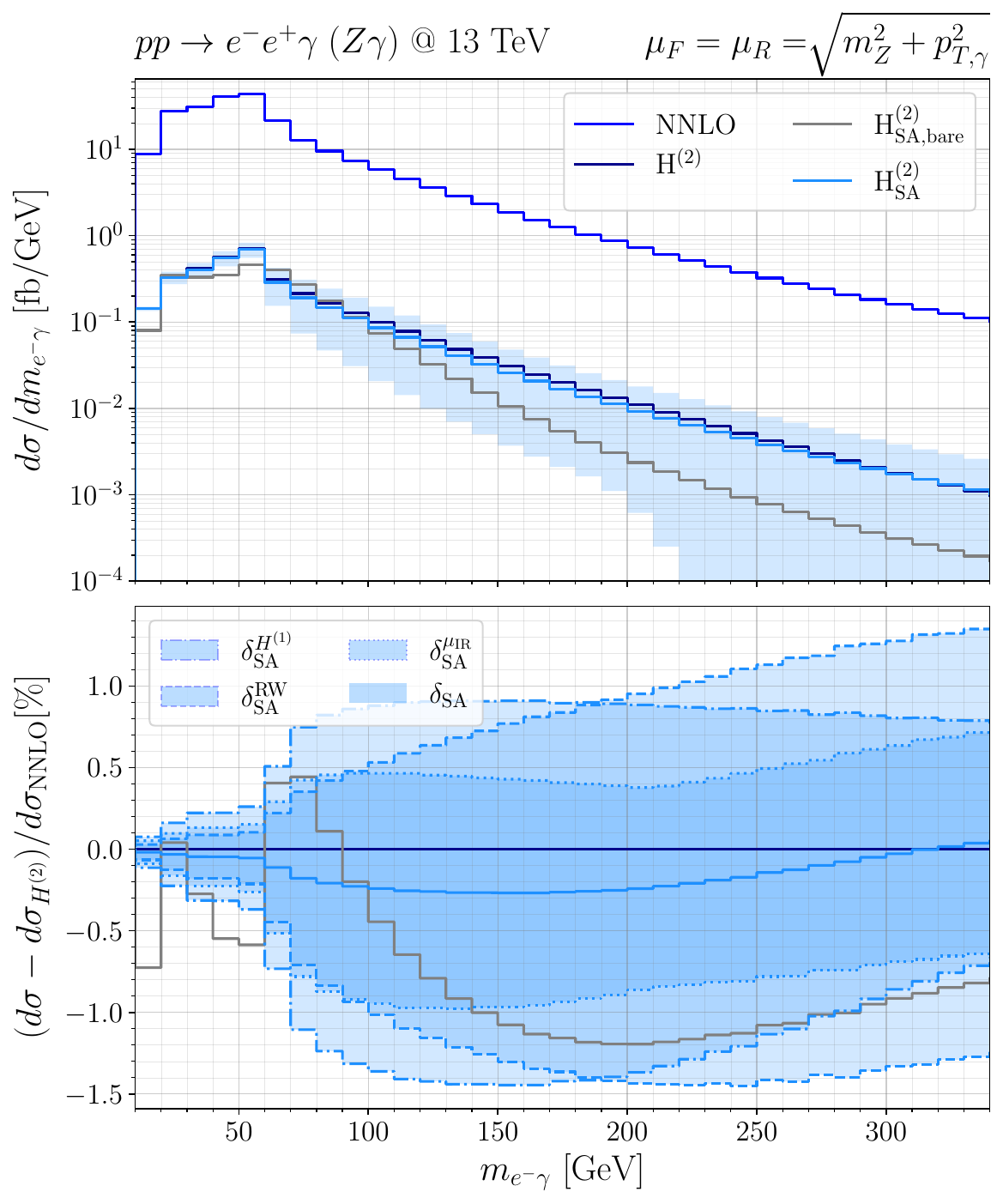}
\hfill 
\includegraphics[height=\plotheightapp]{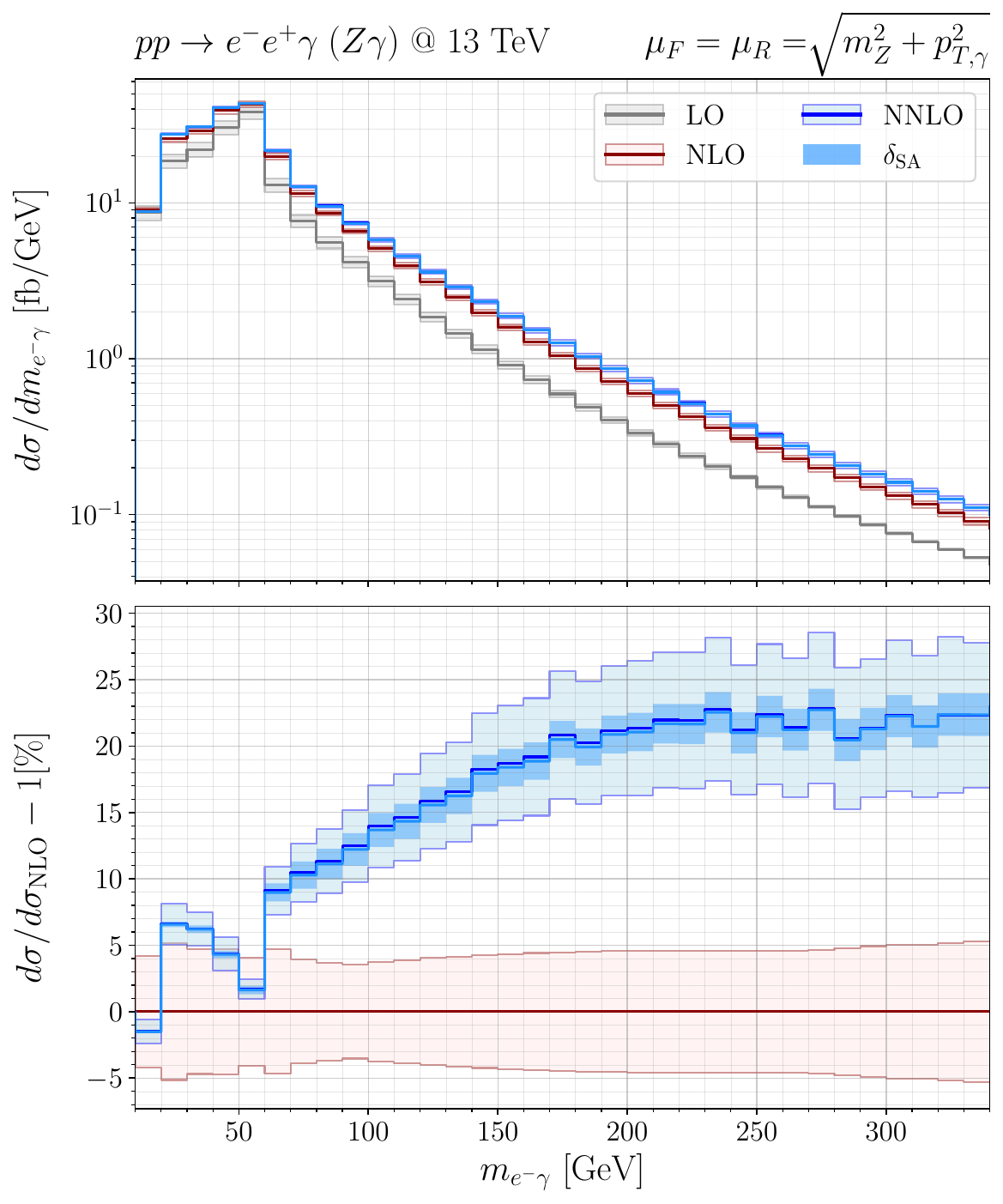}\\[2ex]

\includegraphics[height=\plotheightapp]{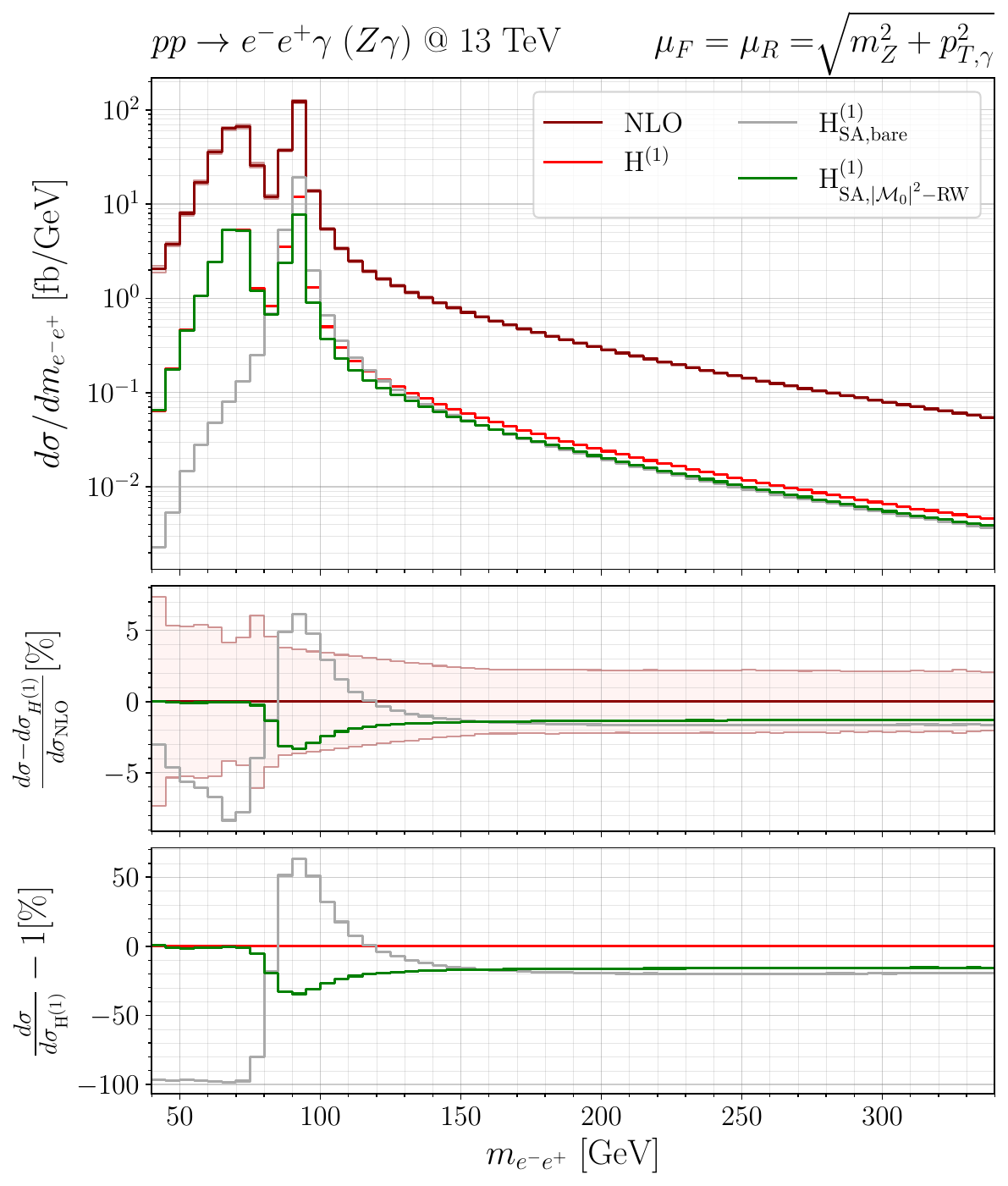}
\hfill 
\includegraphics[height=\plotheightapp]{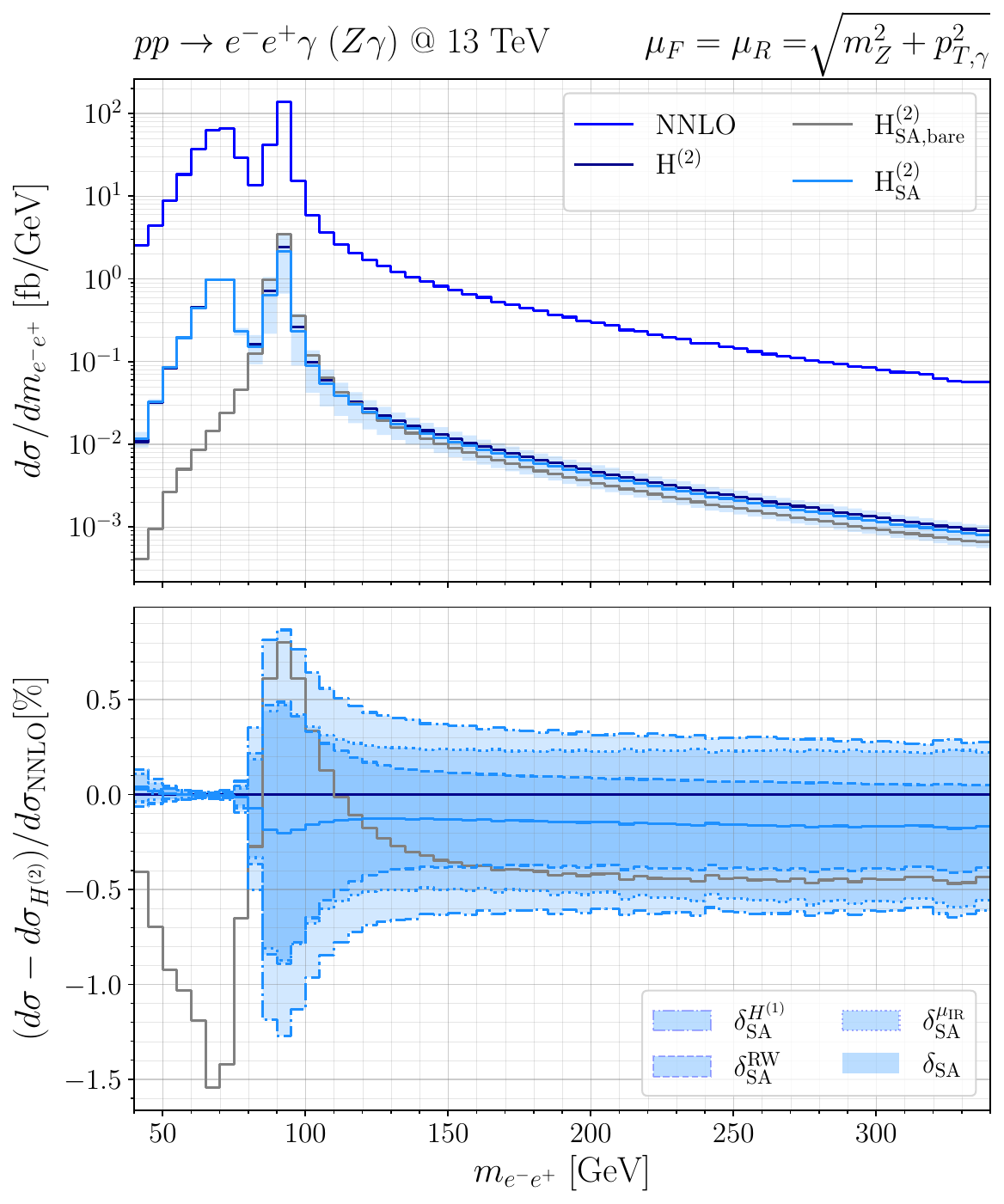}
\hfill 
\includegraphics[height=\plotheightapp]{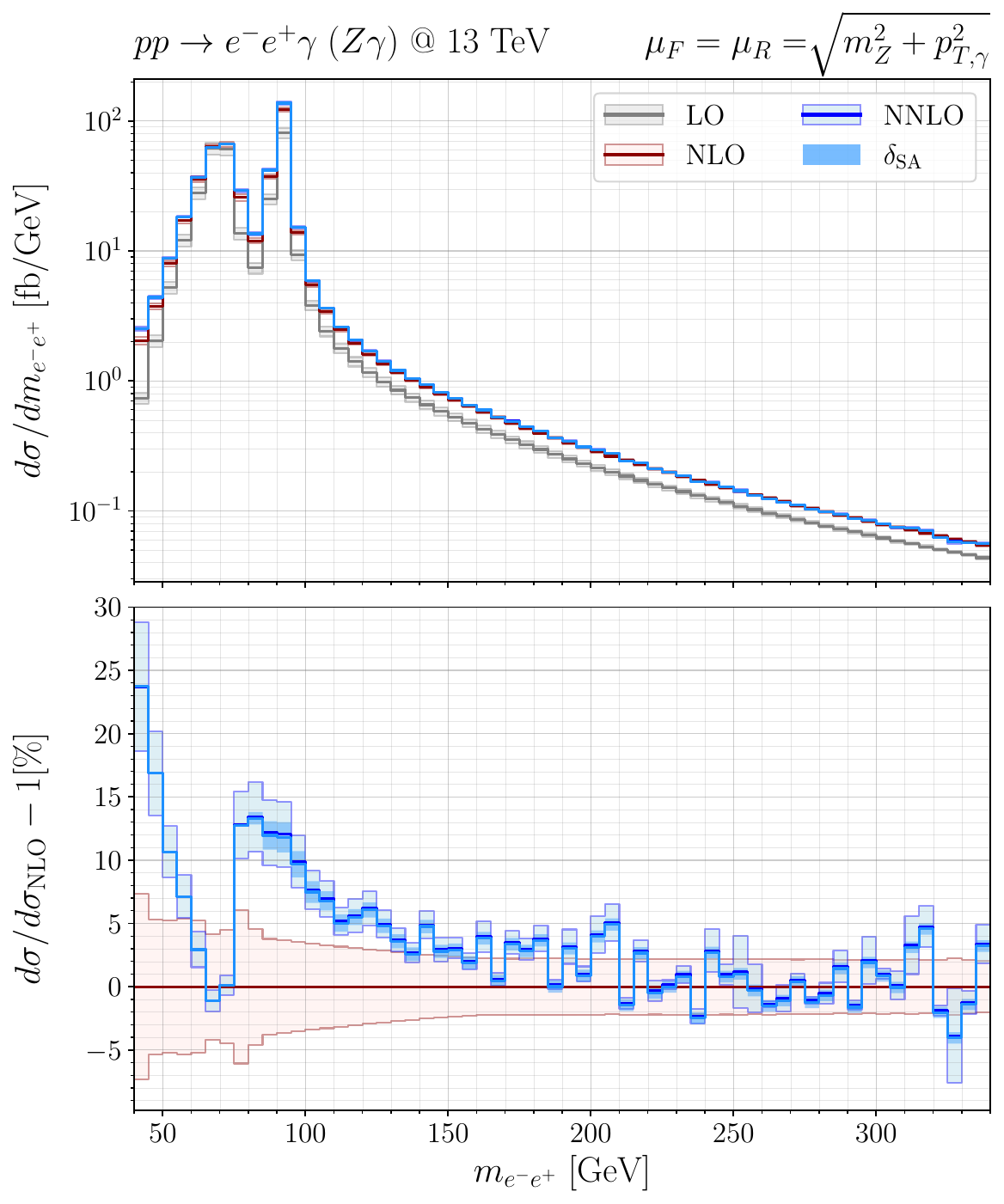}\\[2ex]

\caption{\label{ZAplots_m_em_ep_gamma_m_em_gamma_m_em_ep} Distributions in the invariant masses of the dilepton--photon system, $m_{e^-e^+\gamma}$ (first row),
  the electron--photon system, $m_{e^-\gamma}$ (second row),
  and the electron--positron system, $m_{e^-e^+}$ (third row) for \Zgamma production. The plots follow the descriptions in
  \reffi{fig:errorZAH1based} (left column),
  \reffi{fig:errorZAcombined} (central column) and
  \reffi{fig:errorZAKfactors} (right column), respectively.
  For reference, the \dsHtwoSAnoRW contribution is added in the plots of the central column.
}
\end{figure}

\begin{figure}[p]
\centering
\includegraphics[height=\plotheightapp]{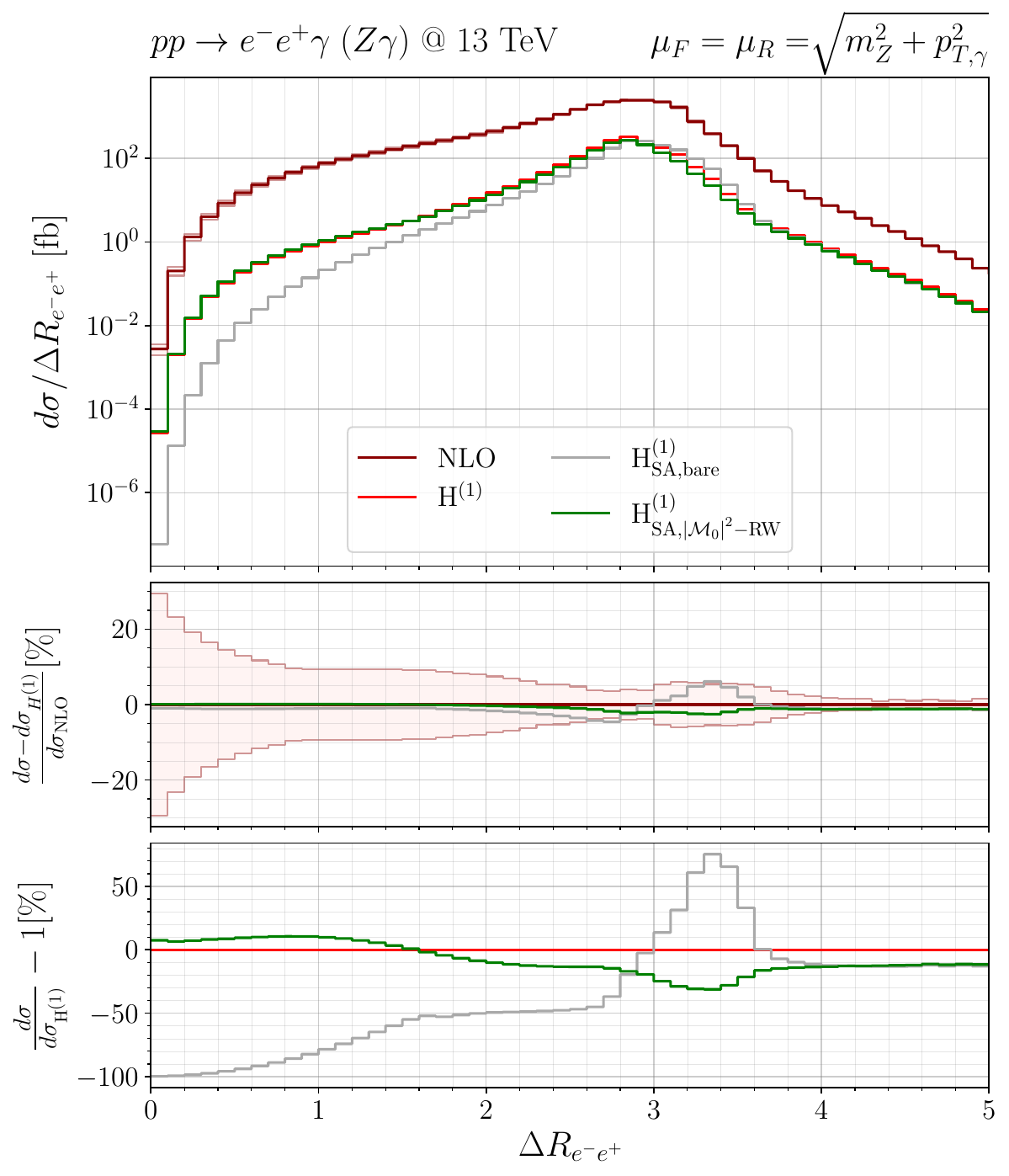}
\hfill 
\includegraphics[height=\plotheightapp]{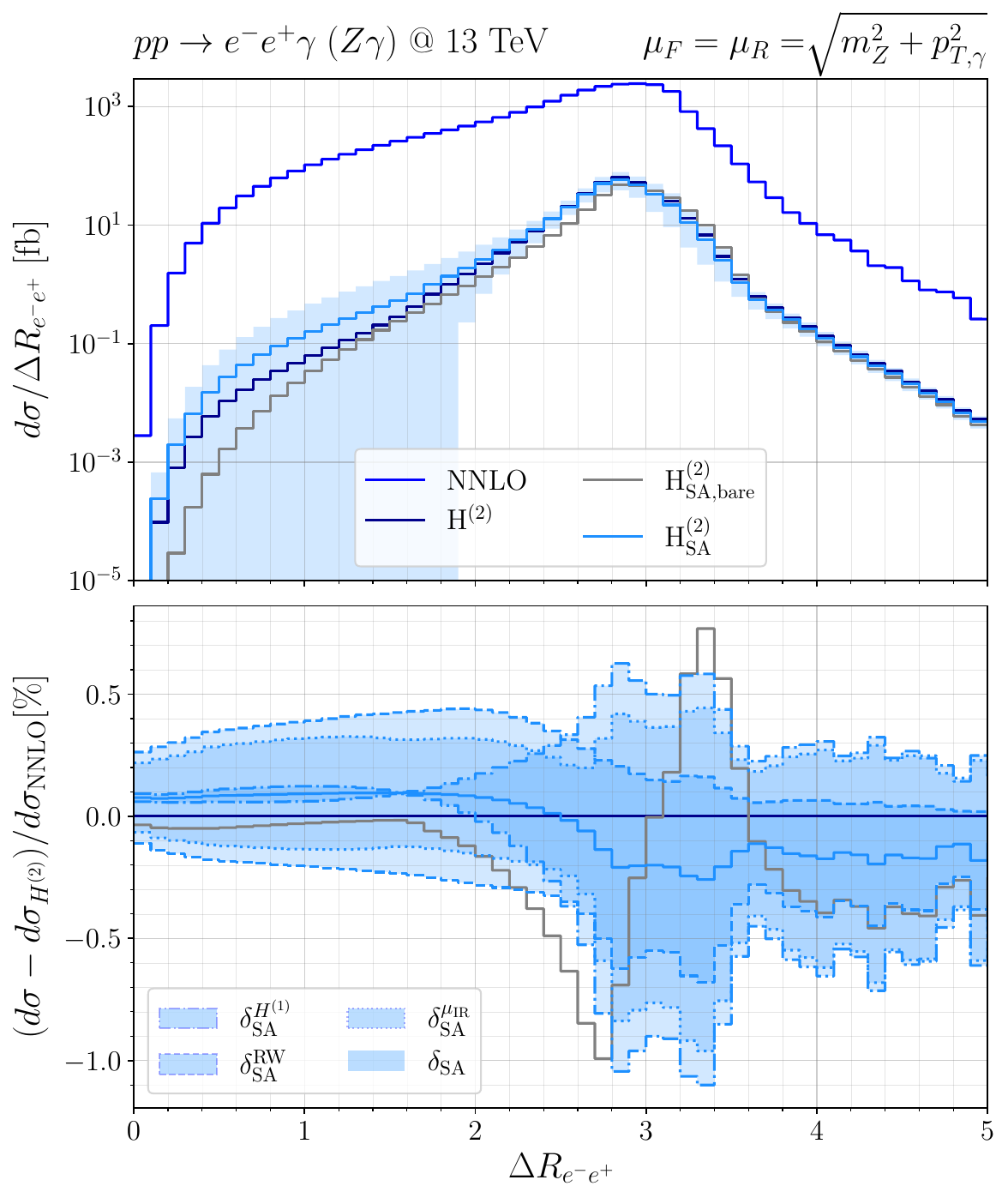}
\hfill 
\includegraphics[height=\plotheightapp]{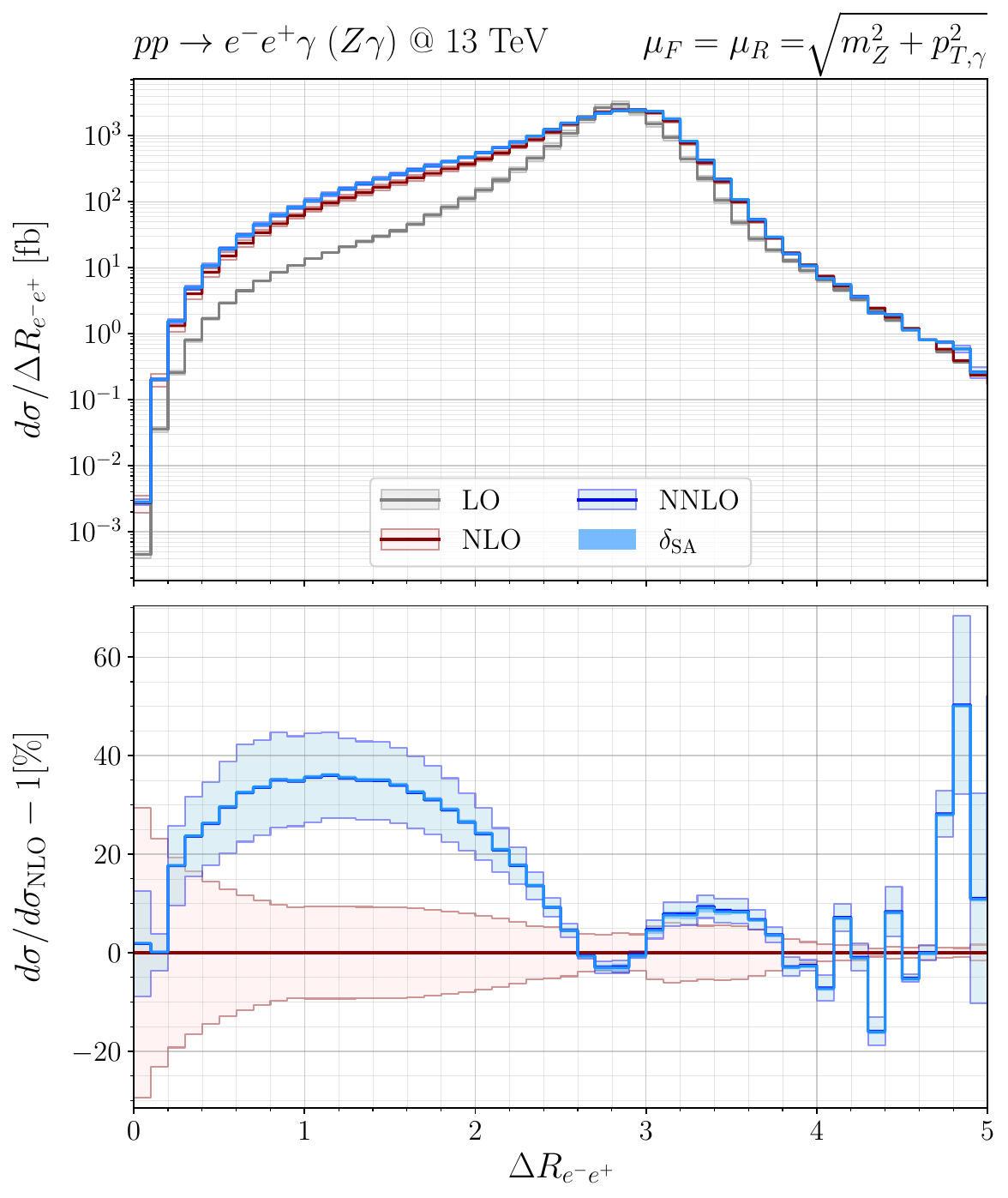}\\[2ex]

\includegraphics[height=\plotheightapp]{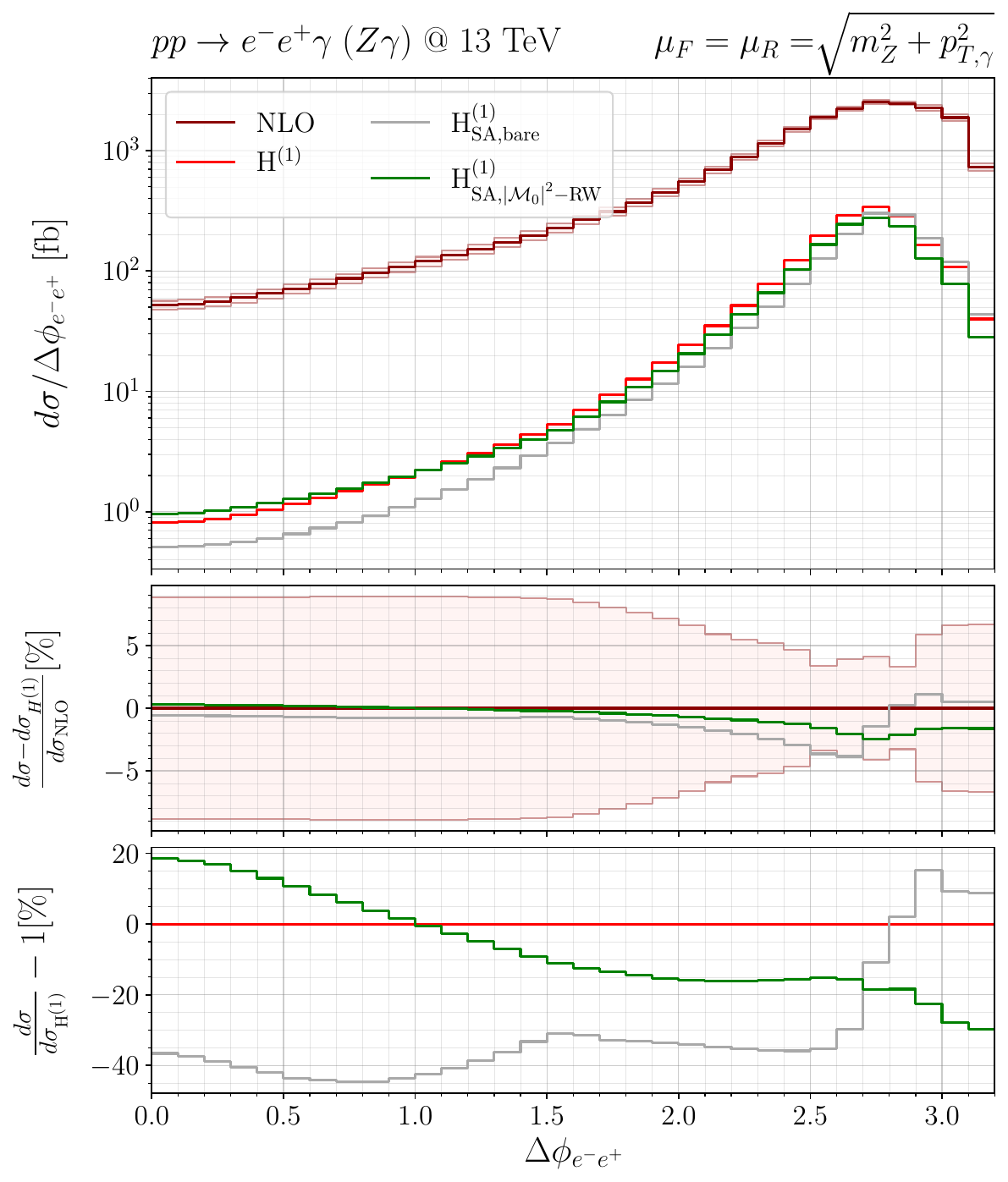}
\hfill 
\includegraphics[height=\plotheightapp]{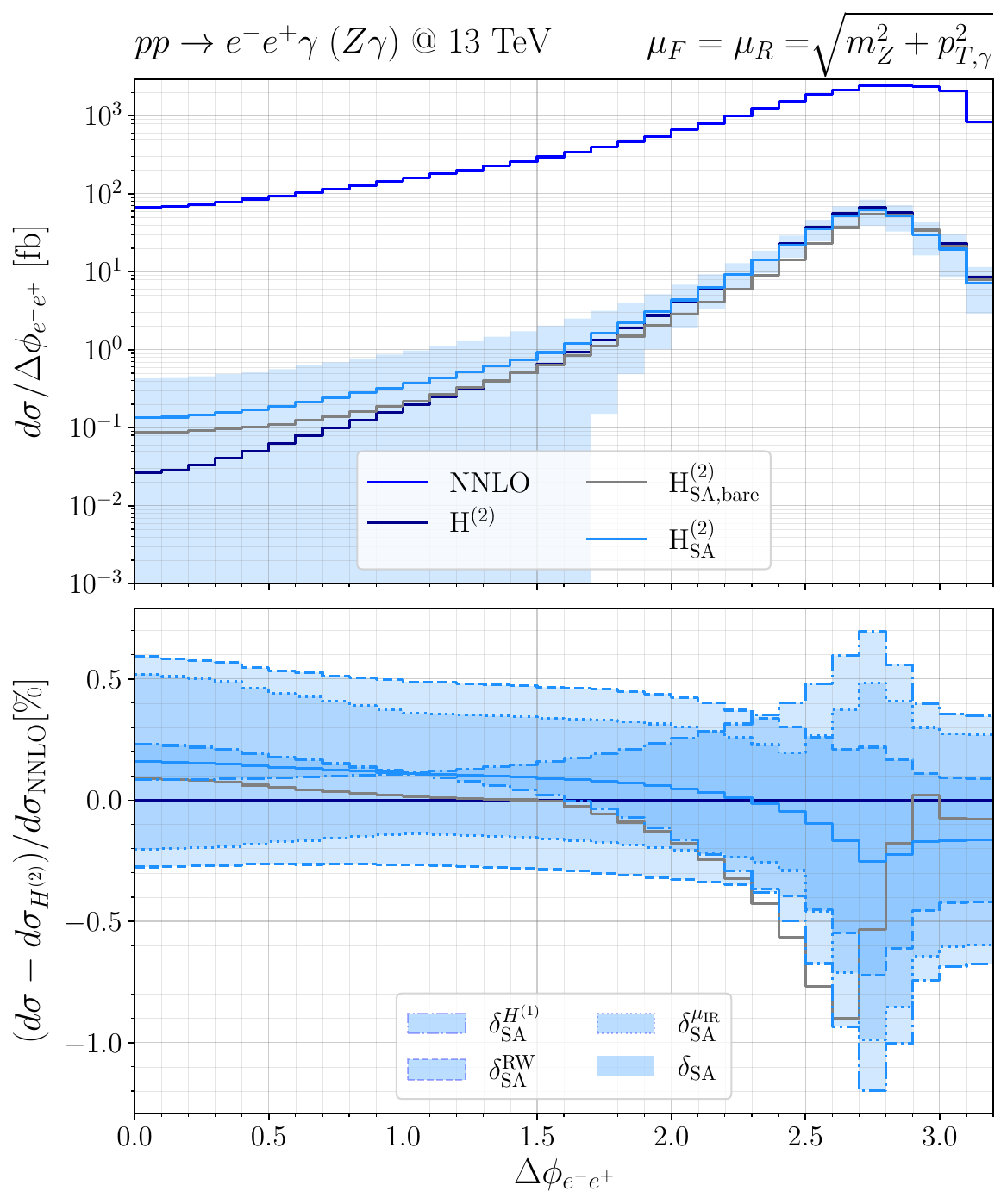}
\hfill 
\includegraphics[height=\plotheightapp]{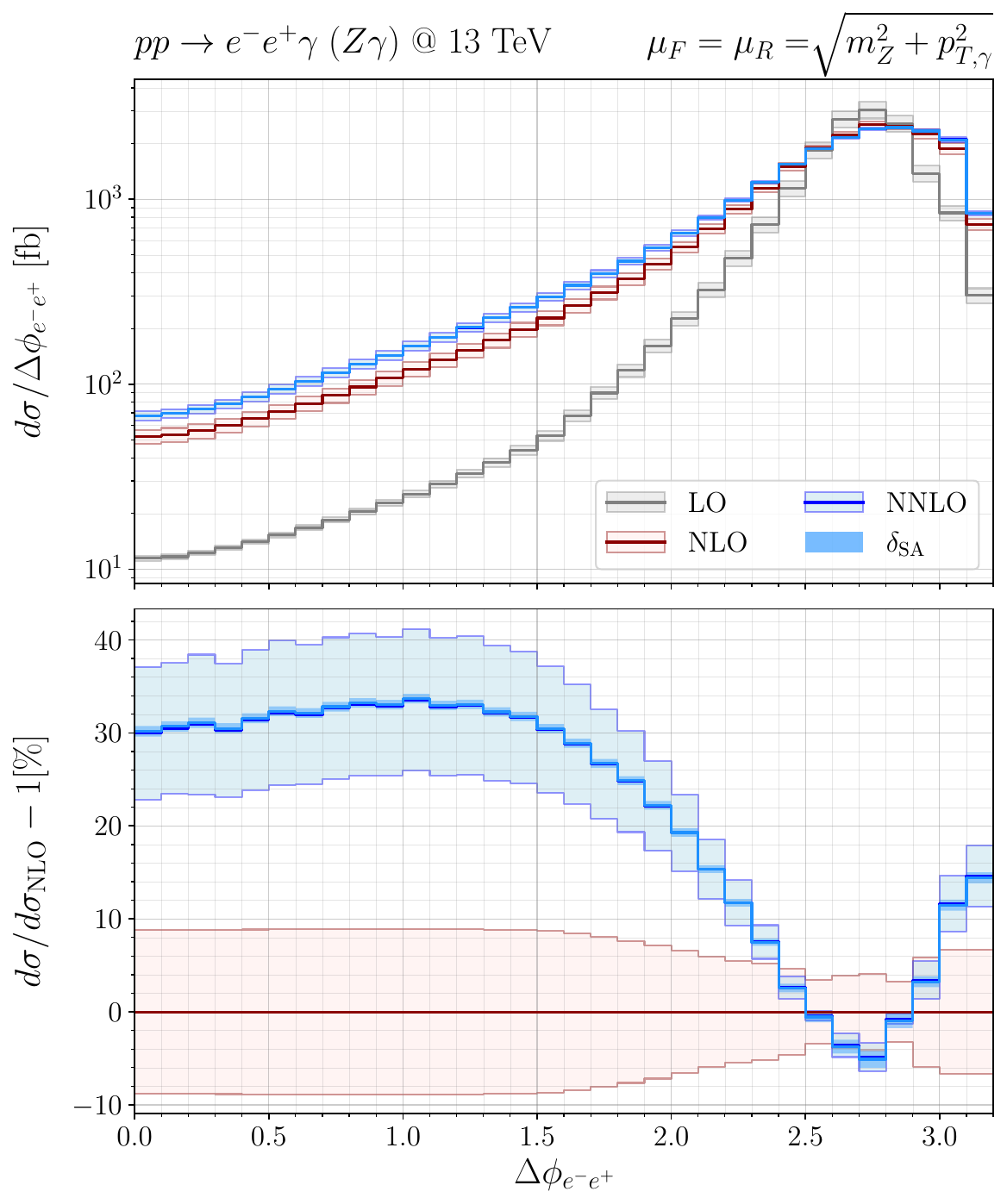}\\[2ex]

\includegraphics[height=\plotheightapp]{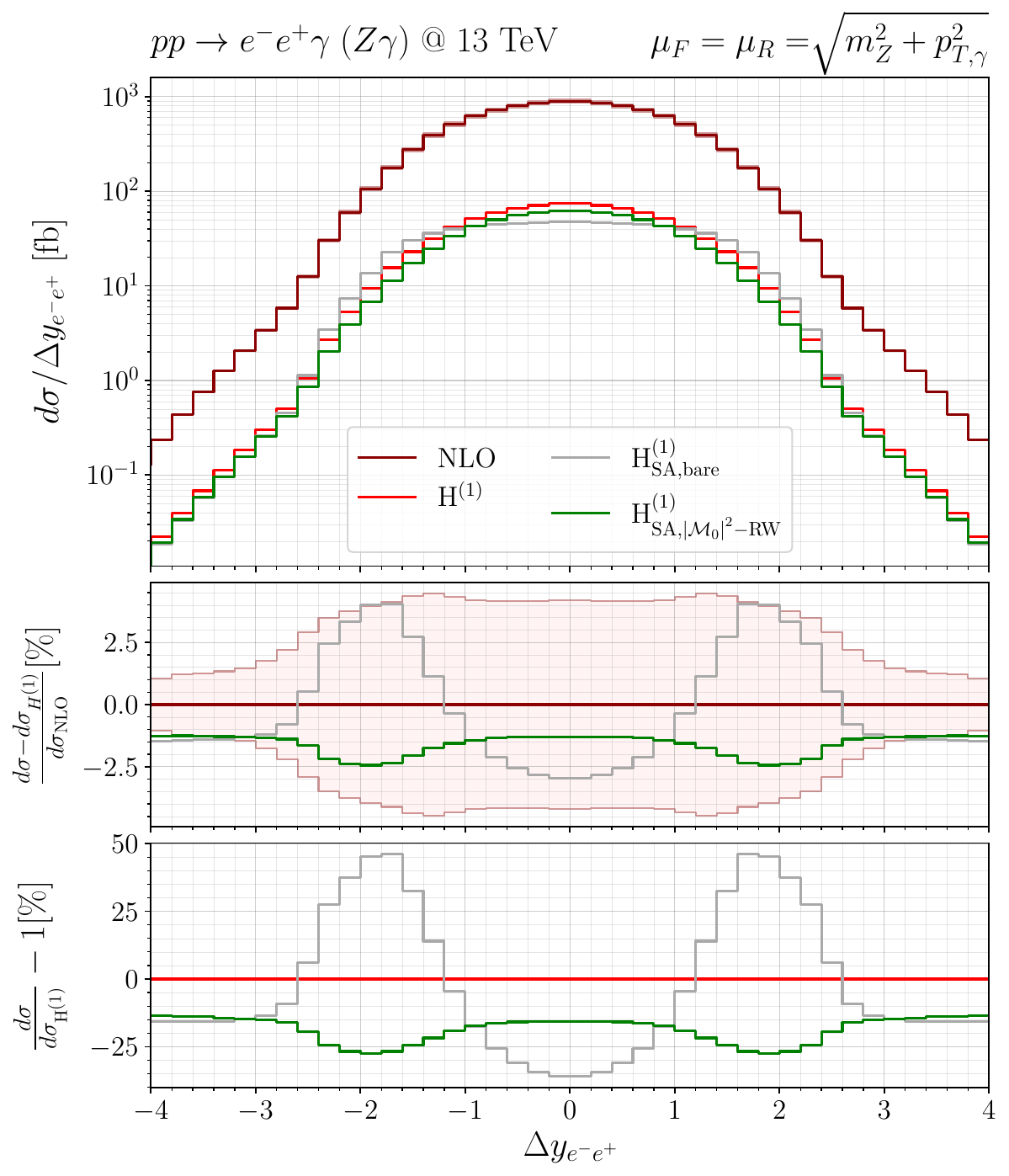}
\hfill 
\includegraphics[height=\plotheightapp]{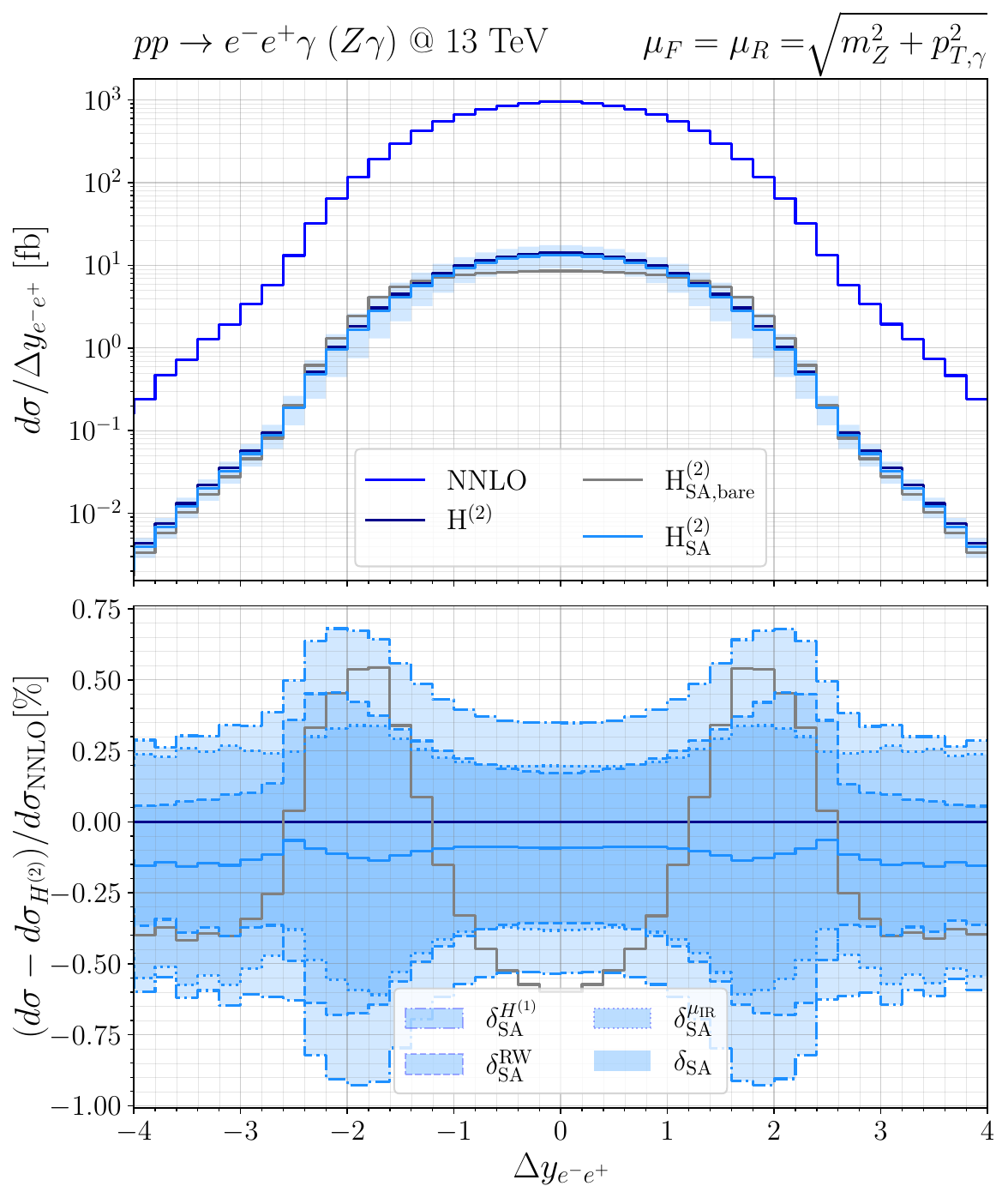}
\hfill 
\includegraphics[height=\plotheightapp]{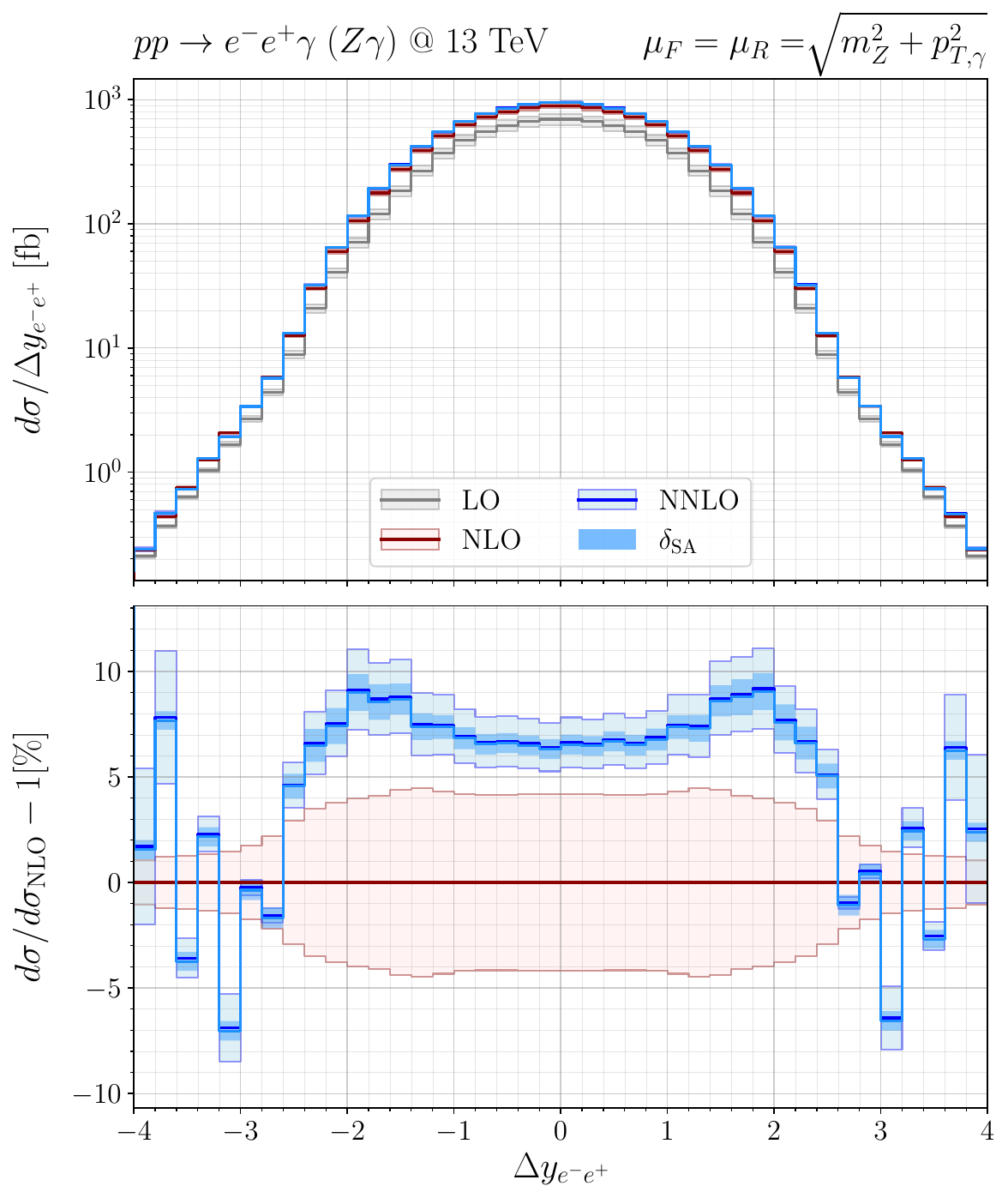}\\[2ex]

\caption{\label{ZAplots_dR_em_ep_dphi_em_ep_dy_em_ep} Distributions
  in the distance between the leptons in the $\phi$--$\eta$ plane, $\Delta R_{e^-e^+}$ (first row),
  as well as in the azimuthal-angle separation, $\Delta\phi_{e^-e^+}$ (second row),
  and the pseudo-rapidity separation, $\Delta\eta_{e^-e^+}$ (third row), for \Zgamma production. The plots follow the descriptions in
  \reffi{fig:errorZAH1based} (left column),
  \reffi{fig:errorZAcombined} (central column) and
  \reffi{fig:errorZAKfactors} (right column), respectively.
  For reference, the \dsHtwoSAnoRW contribution is added in the plots of the central column.
}
\end{figure}

\begin{figure}[p]
\centering
\includegraphics[height=\plotheightapp]{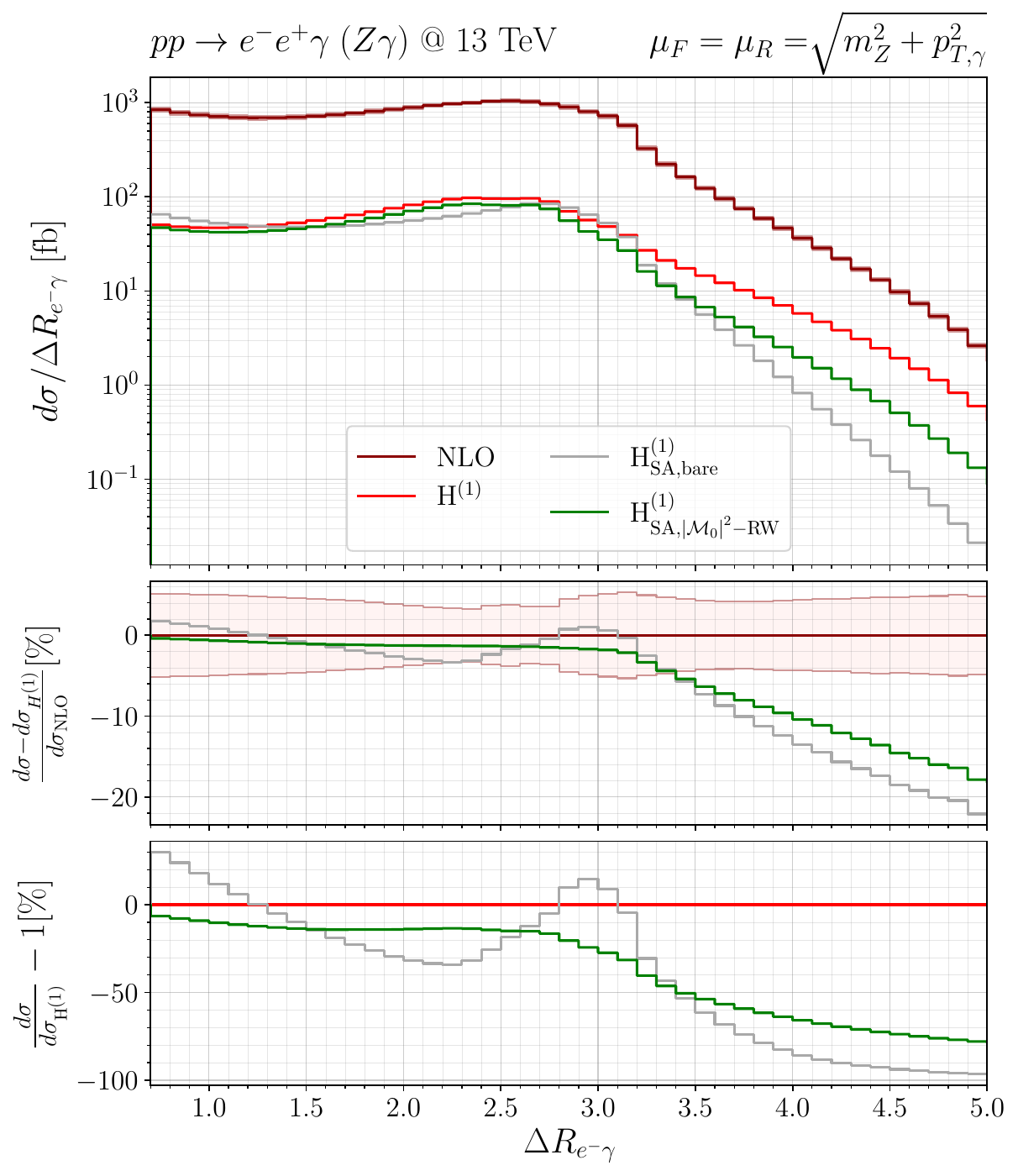}
\hfill 
\includegraphics[height=\plotheightapp]{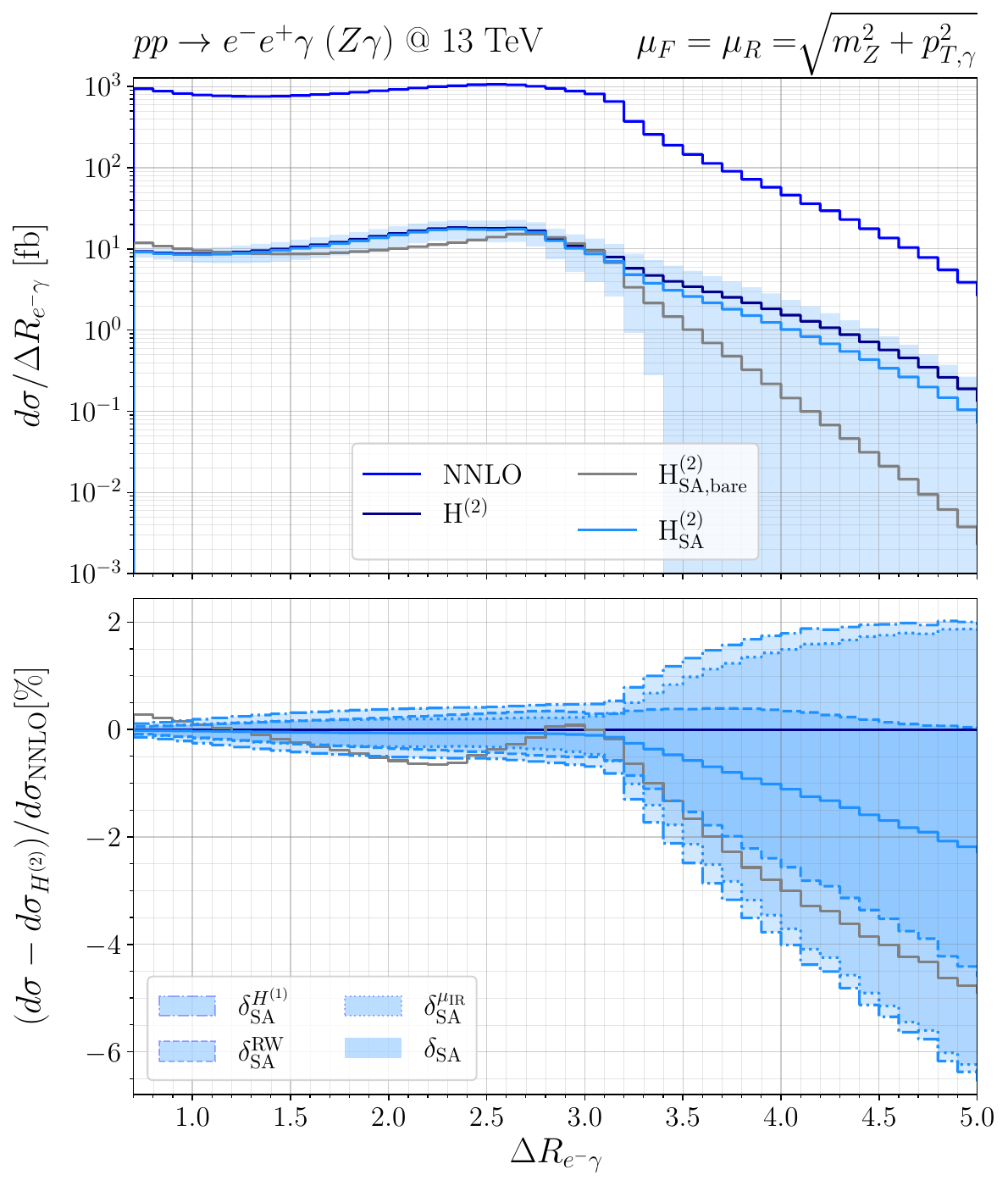}
\hfill 
\includegraphics[height=\plotheightapp]{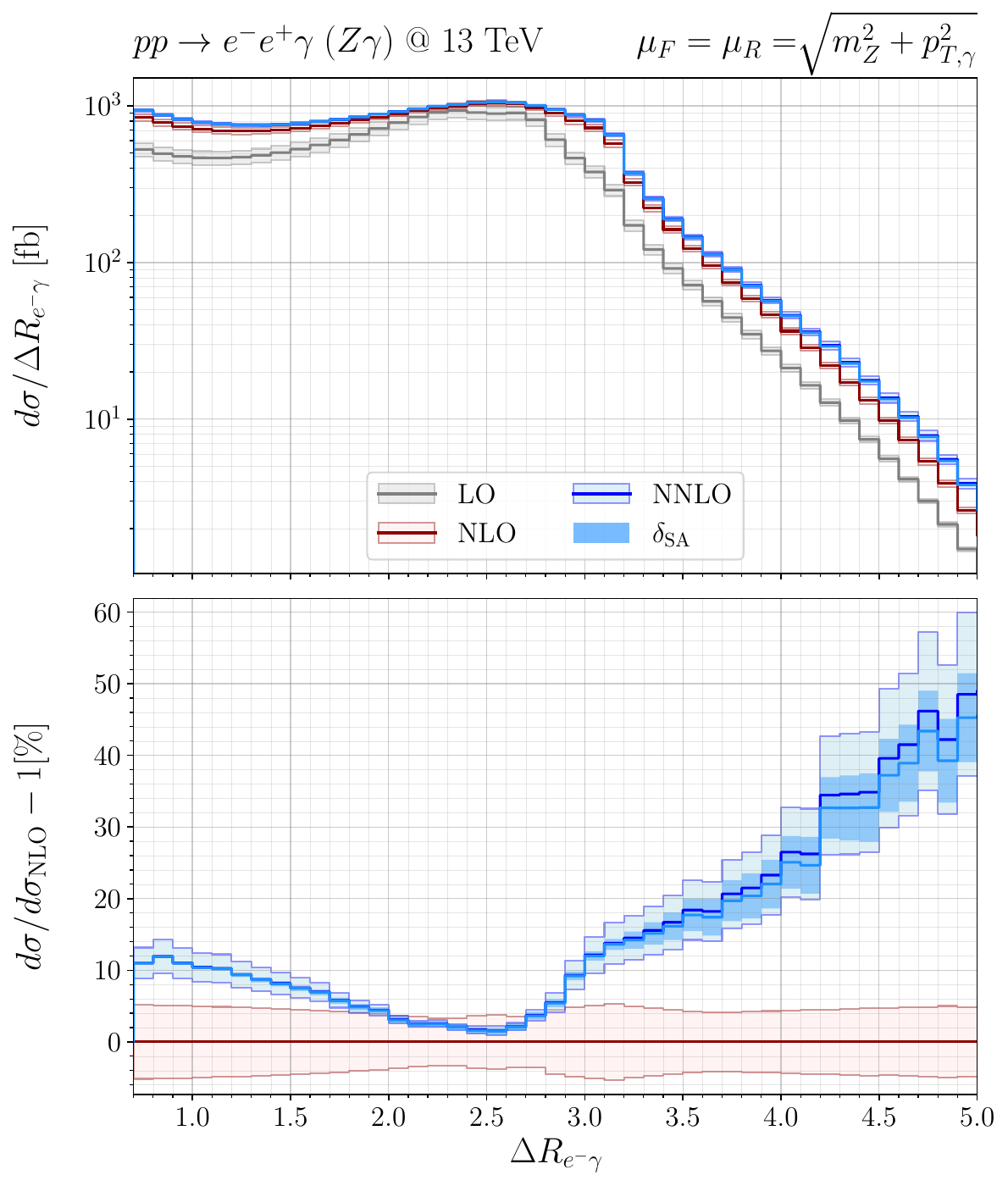}\\[2ex]

\includegraphics[height=\plotheightapp]{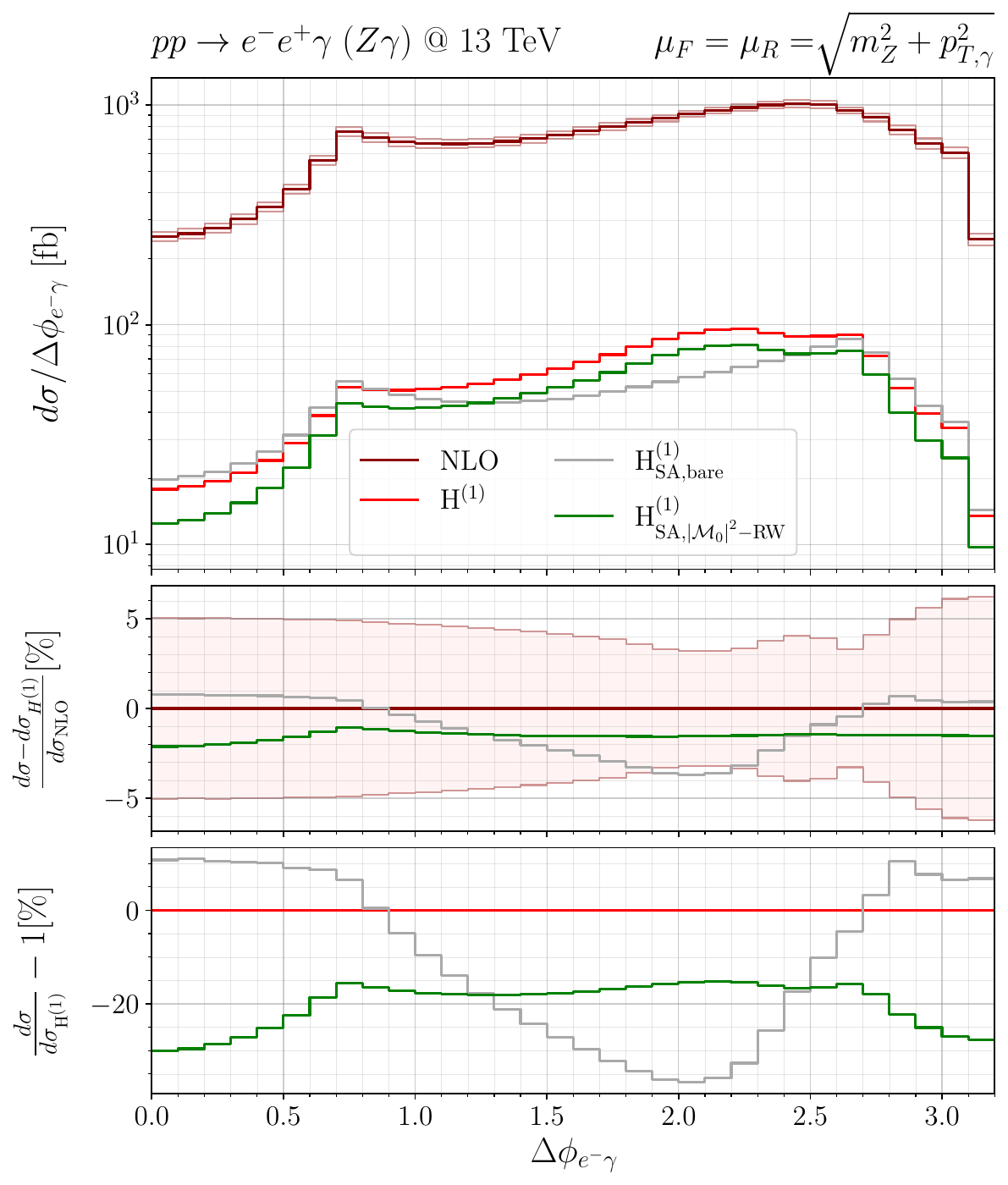}
\hfill 
\includegraphics[height=\plotheightapp]{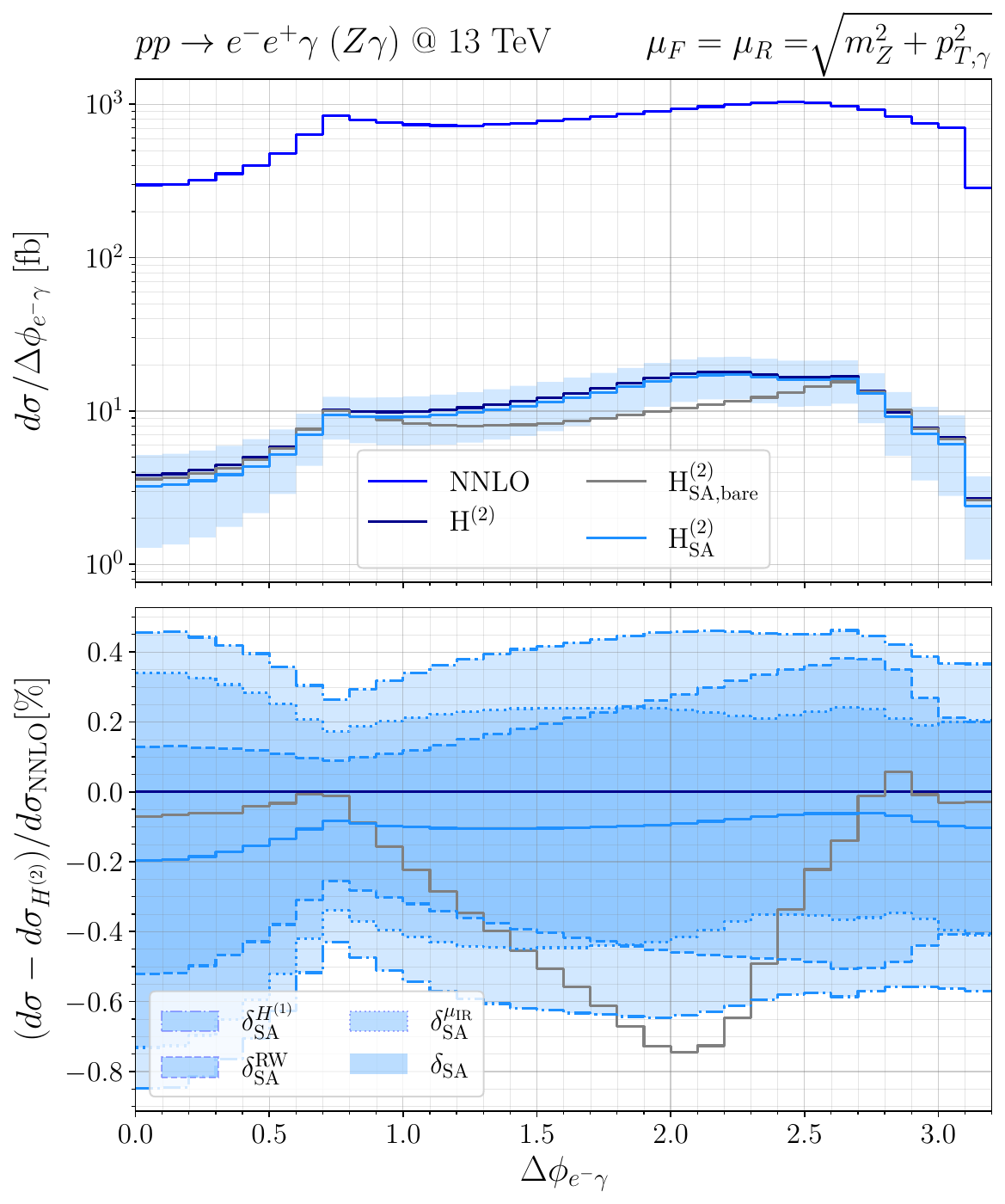}
\hfill 
\includegraphics[height=\plotheightapp]{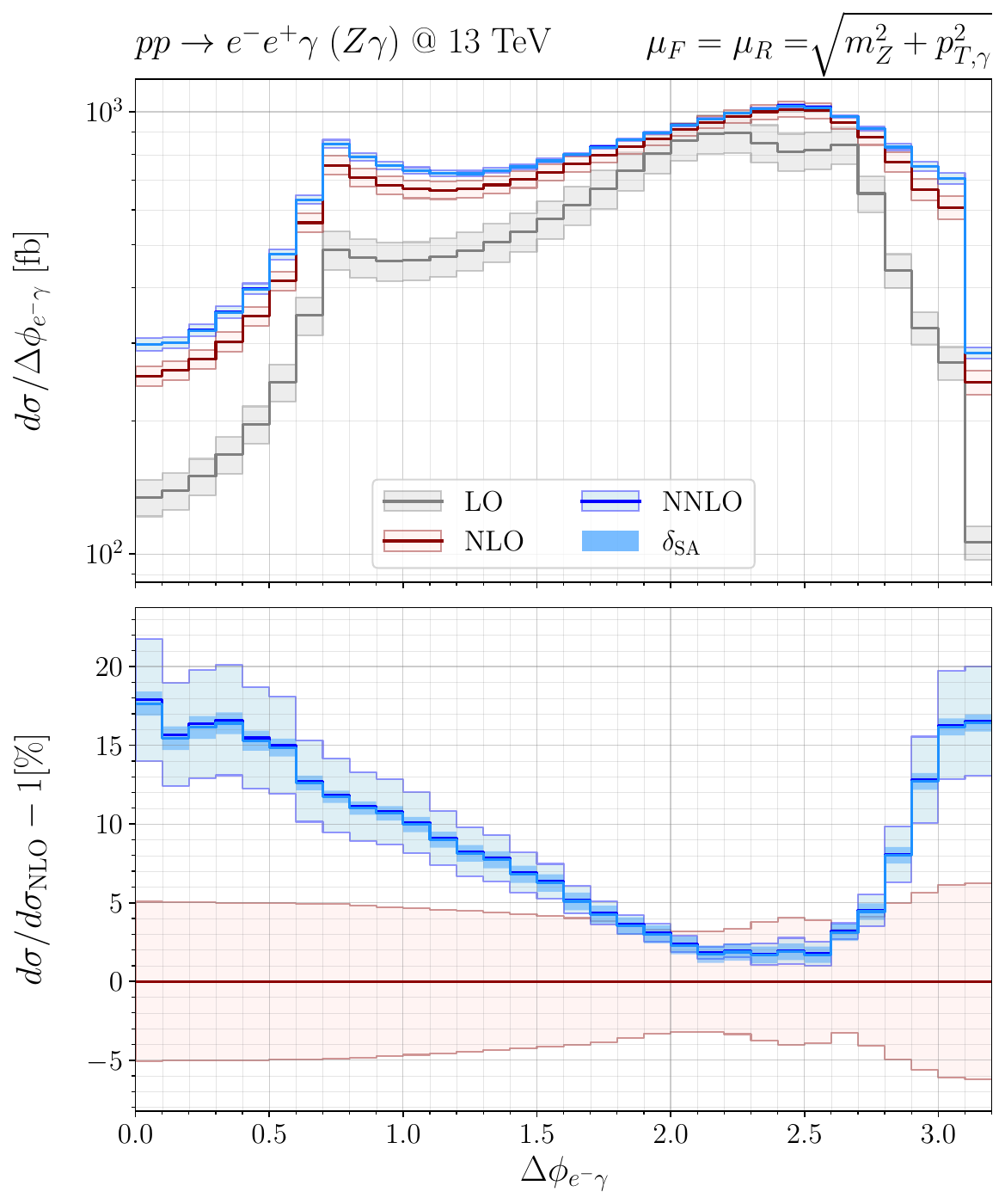}\\[2ex]

\includegraphics[height=\plotheightapp]{figures/ppeexa03_LHC13_error_estimate_NLO_H1_dy_em_gamma.pdf}
\hfill 
\includegraphics[height=\plotheightapp]{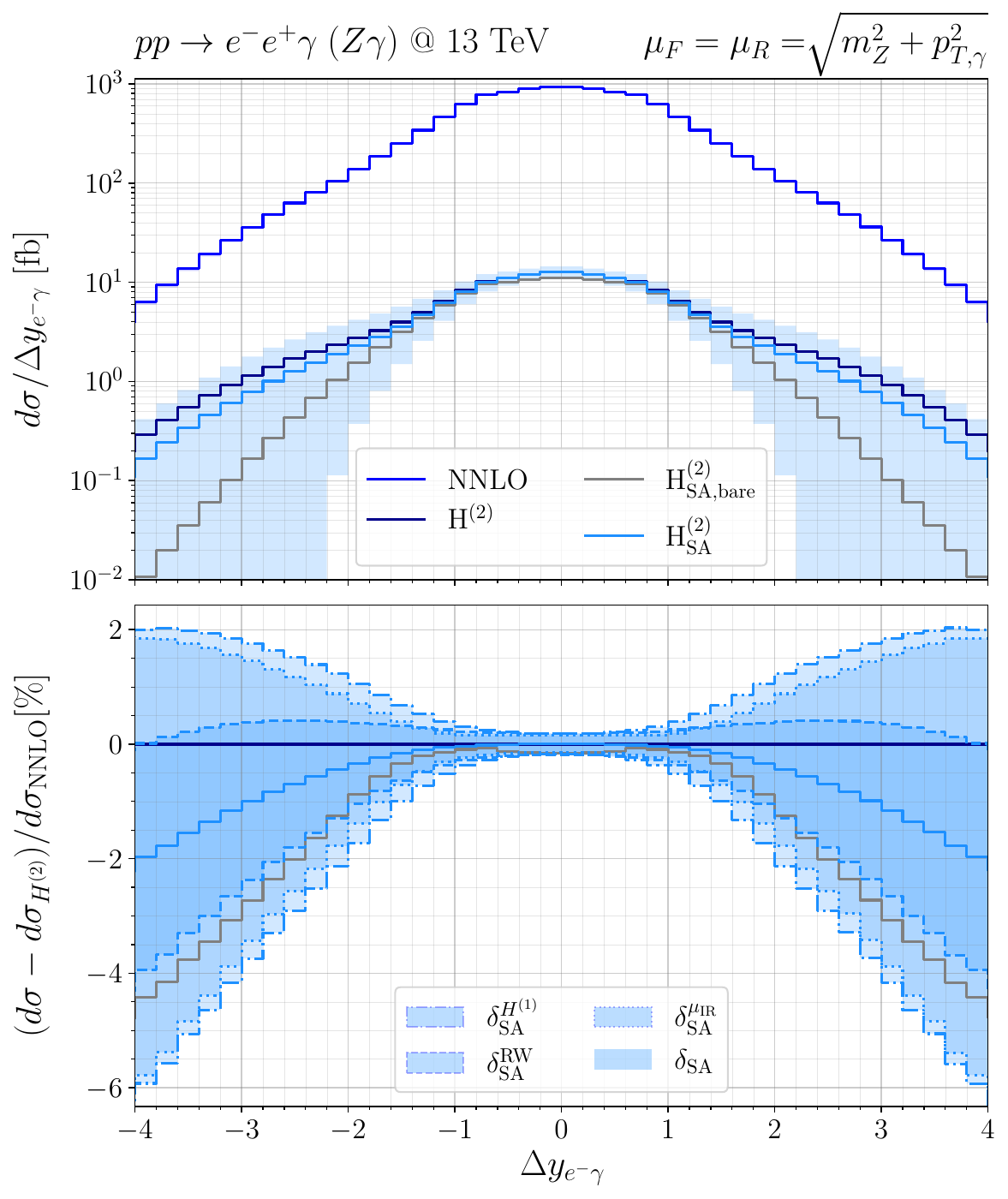}
\hfill 
\includegraphics[height=\plotheightapp]{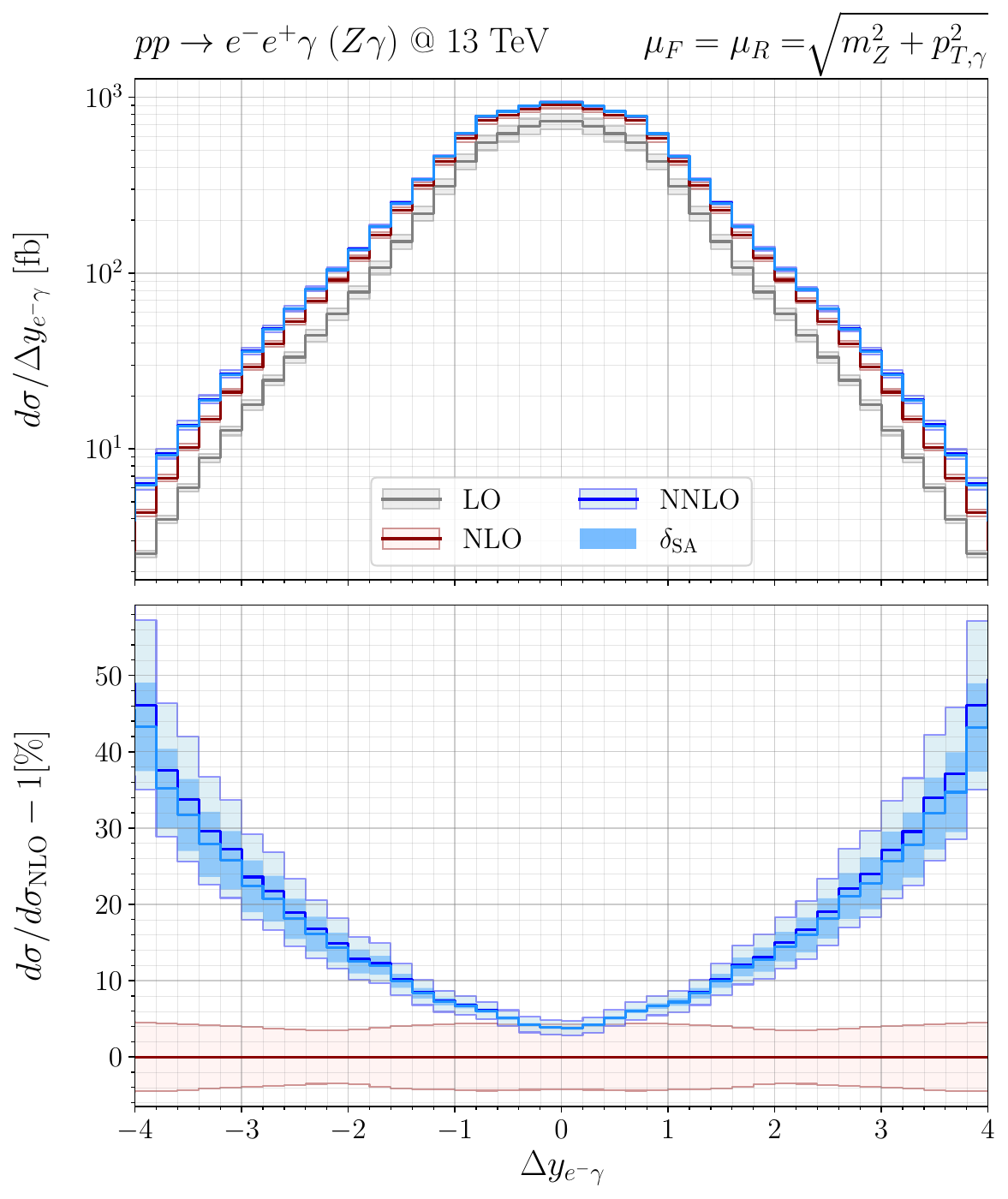}\\[2ex]

\caption{\label{ZAplots_dR_em_gamma_dphi_em_gamma_dy_em_gamma} Distributions
  in the distance between the electron and the photon in the $\phi$--$\eta$ plane,
  $\Delta R_{e^-\gamma}$ (first row), as well as in the azimuthal-angle separation, $\Delta\phi_{e^-\gamma}$ (second row),
  and the pseudo-rapidity separation, $\Delta\eta_{e^-\gamma}$ (third row), for \Zgamma production. The plots follow the descriptions in
  \reffi{fig:errorZAH1based} (left column),
  \reffi{fig:errorZAcombined} (central column) and
  \reffi{fig:errorZAKfactors} (right column), respectively.
  For reference, the \dsHtwoSAnoRW contribution is added in the plots of the central column.
}
\end{figure}

In detail, we provide the photon transverse-momentum distribution, \pTgamma, in \reffi{ZAplots_pT_gamma},
and the electron transverse-momentum distribution, \pTem,
together with the distributions in the absolute pseudo-rapidities of the photon, $|\eta_{\gamma}|$,
and the electron, $|\eta_{e^-}|$, in \reffi{ZAplots_pT_em_eta_gamma_em}.
Furthermore, we present invariant-mass distributions of the dilepton--photon system, $m_{e^-e^+\gamma}$,
the electron--photon system, $m_{e^-\gamma}$, and the electron--positron system, $m_{e^-e^+}$,
in \reffi{ZAplots_m_em_ep_gamma_m_em_gamma_m_em_ep}.
Finally, we show distributions in the distances in the $\phi$--$\eta$ plane, together with their constituents,
the azimuthal-angle and pseudo-rapidity separations,
between the electron and the positron, $\Delta R_{e^-e^+}$, $\Delta\phi_{e^-e^+}$, and $\Delta\eta_{e^-e^+}$,
in \reffi{ZAplots_dR_em_ep_dphi_em_ep_dy_em_ep},
as well as between the electron and the photon, $\Delta R_{e^-\gamma}$, $\Delta\phi_{e^-\gamma}$, and $\Delta\eta_{e^-\gamma}$,
in \reffi{ZAplots_dR_em_gamma_dphi_em_gamma_dy_em_gamma}.

\clearpage

\section[Differential \WWgamma results in soft-photon approximation]{Differential $\boldsymbol{WW\gamma}$ results in soft-photon approximation}
\label{app:resultsWWA}
In this appendix, we provide a survey of differential cross sections
for \WWgamma production to illustrate the performance of the soft-photon approximation
and its associated error estimate across a broader range of phase-space regions.
All plots follow the descriptions in \reffi{fig:errorWWAH1based} (left columns),
\reffi{fig:errorWWAcombined} (central columns) and
\reffi{fig:errorWWAKfactors} (right column), respectively. For reference, we additionally include
the \dsHtwoSAnoRW contribution in the plots of the central column.
Some of the distributions presented in this appendix have already been shown in the main text.
They are included to provide a consistent overview of the full survey of observables considered here.

\begin{figure}[t]
\centering
\includegraphics[height=\plotheightapp]{figures/ppemxnmnexa05_LHC13_CMS_error_estimate_NLO_H1_pT_gamma.pdf}
\hfill 
\includegraphics[height=\plotheightapp]{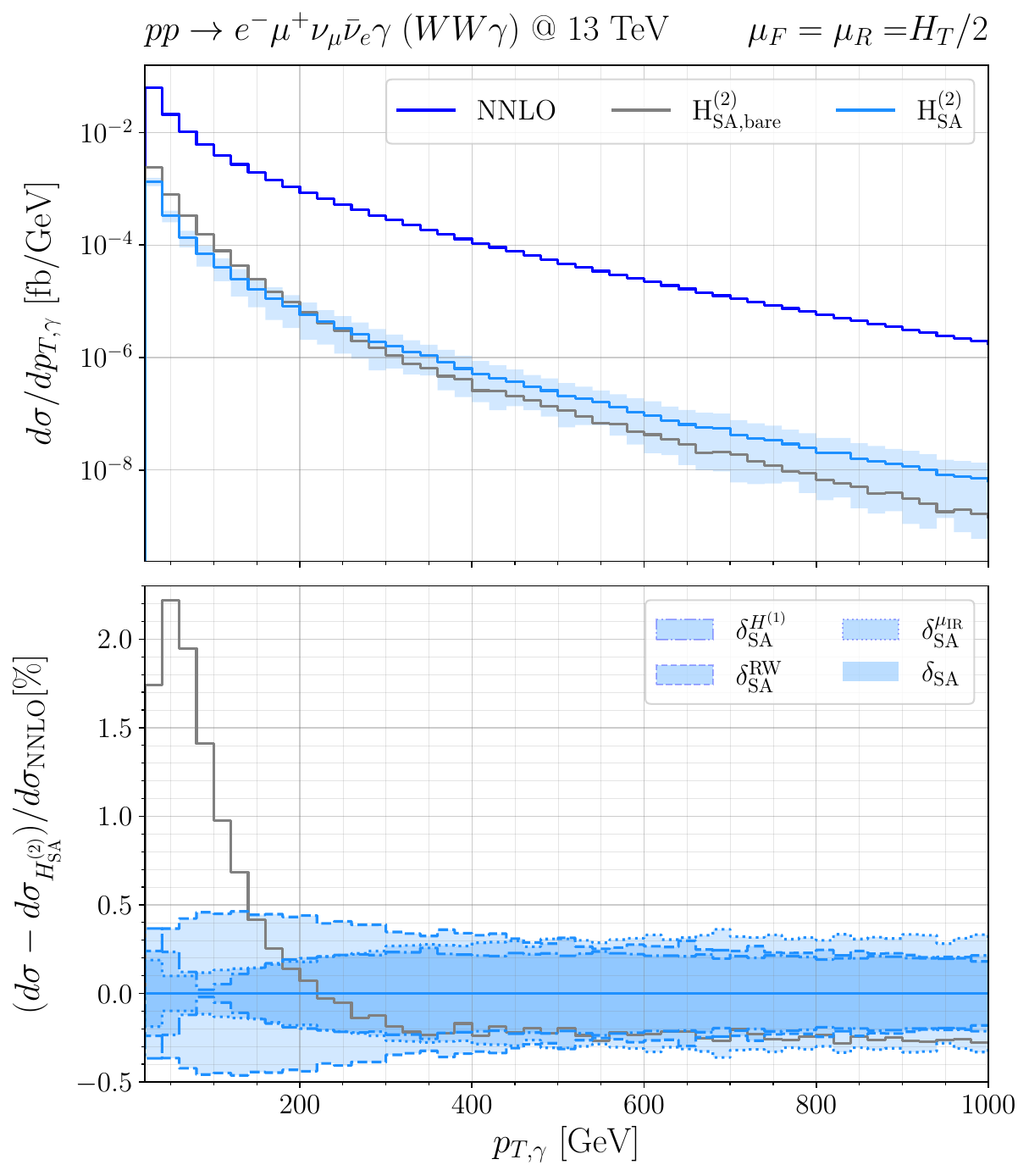}
\hfill 
\includegraphics[height=\plotheightapp]{figures/ppemxnmnexa05_LHC13_CMS_Kfactors_NNLO_pT_gamma.pdf}\\[2ex]

\caption{\label{WWAplots_pT_gamma} Distribution
  in the photon transverse momentum, \pTgamma, for \WWgamma production. The
  plots follow the descriptions in
  \reffi{fig:errorWWAH1based} (left column),
  \reffi{fig:errorWWAcombined} (central column) and
  \reffi{fig:errorWWAKfactors} (right column), respectively.
  For reference, the \dsHtwoSAnoRW contribution is added in the plot of the central column.
}
\vspace*{5ex}
\end{figure}

\begin{figure}[p]
\centering
\includegraphics[height=\plotheightapp]{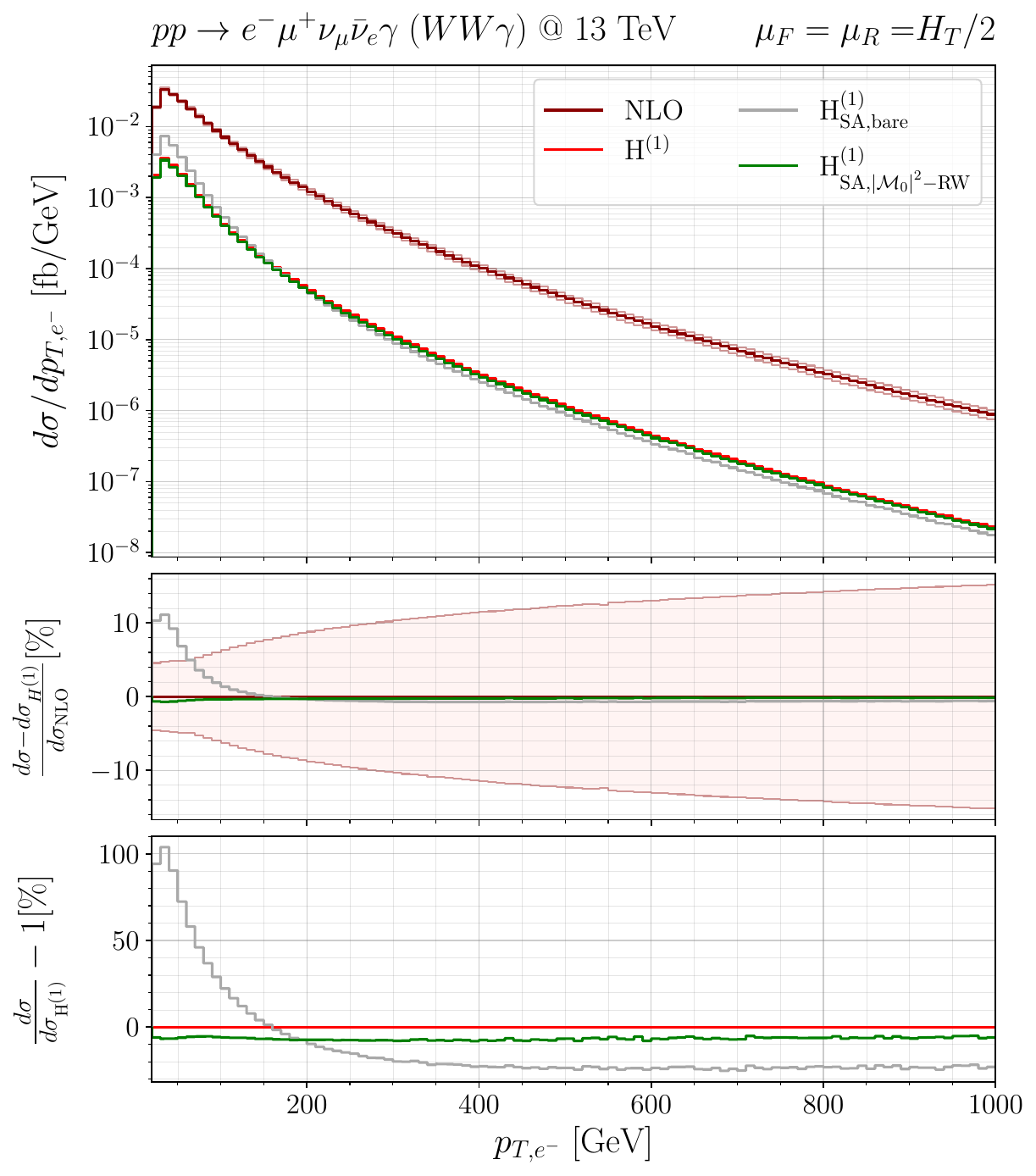}
\hfill 
\includegraphics[height=\plotheightapp]{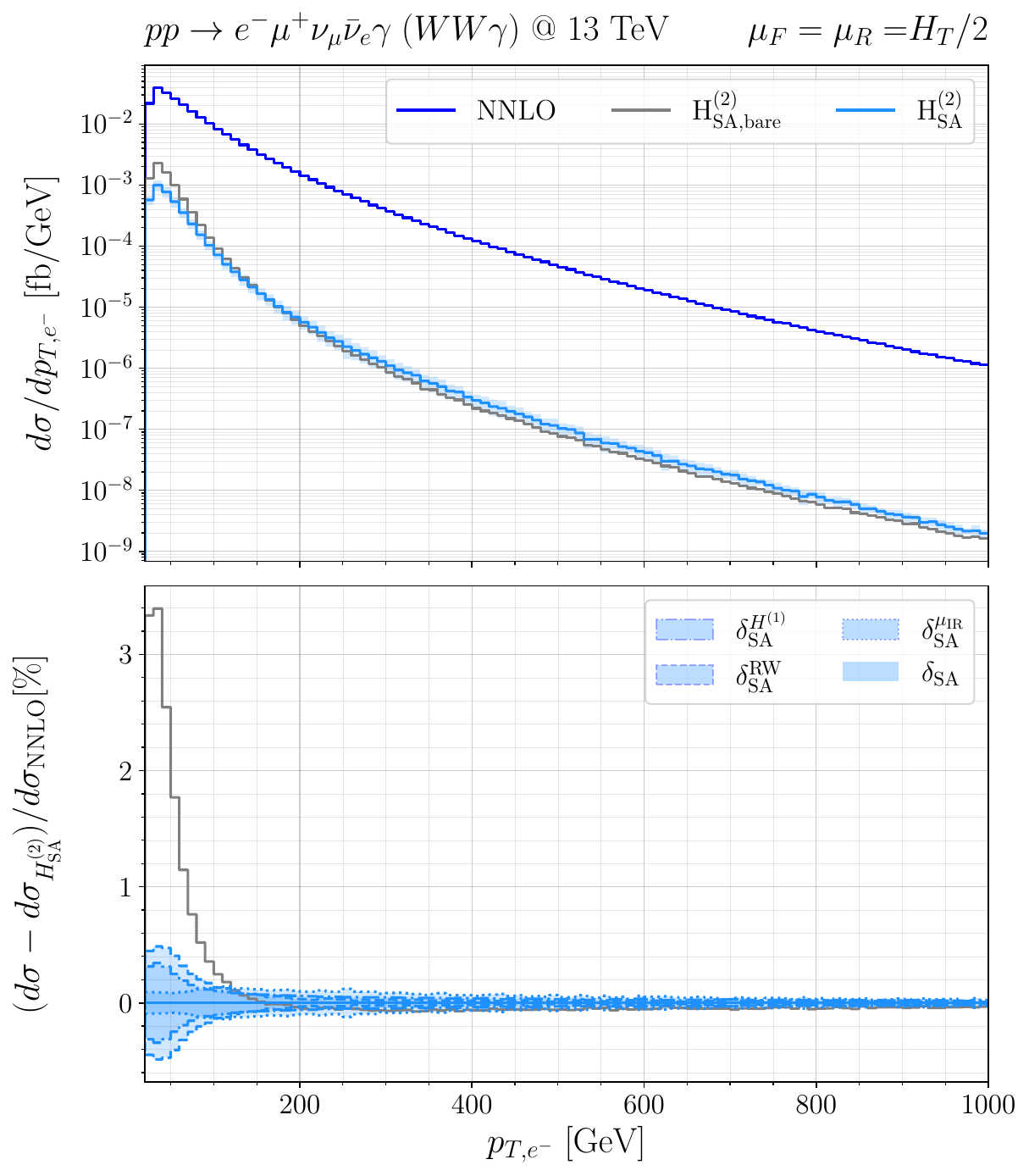}
\hfill 
\includegraphics[height=\plotheightapp]{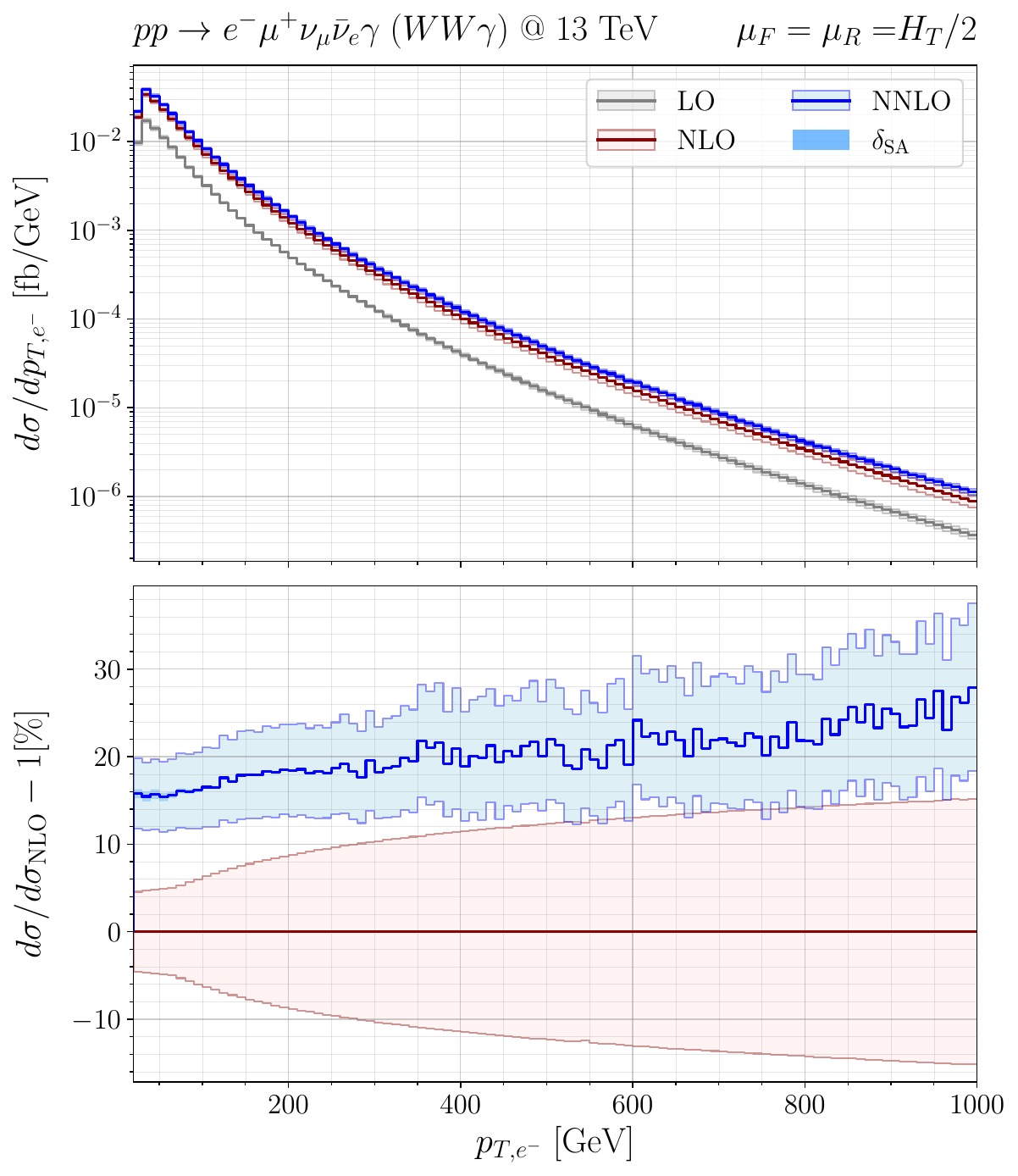}\\[2ex]

\includegraphics[height=\plotheightapp]{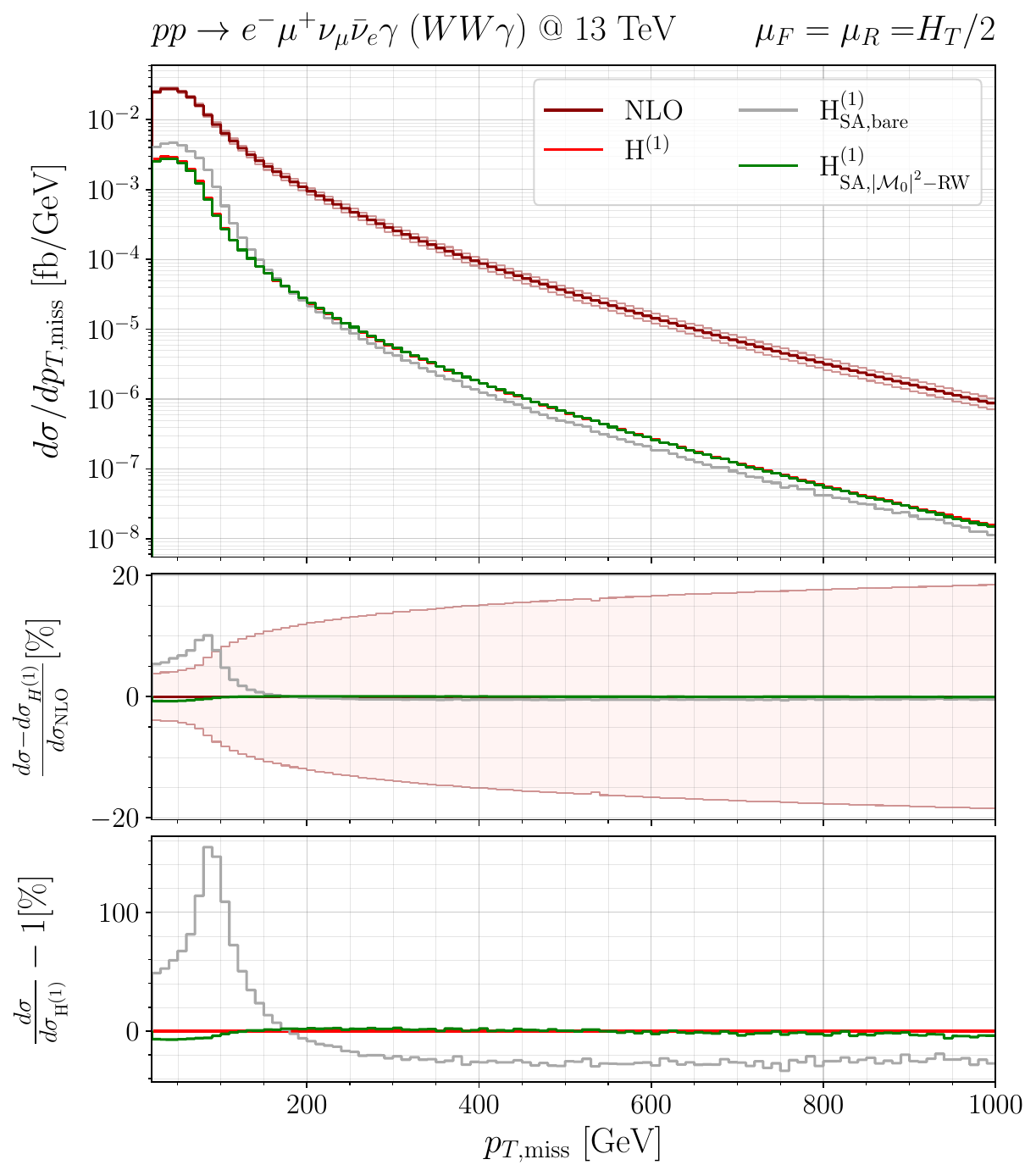}
\hfill 
\includegraphics[height=\plotheightapp]{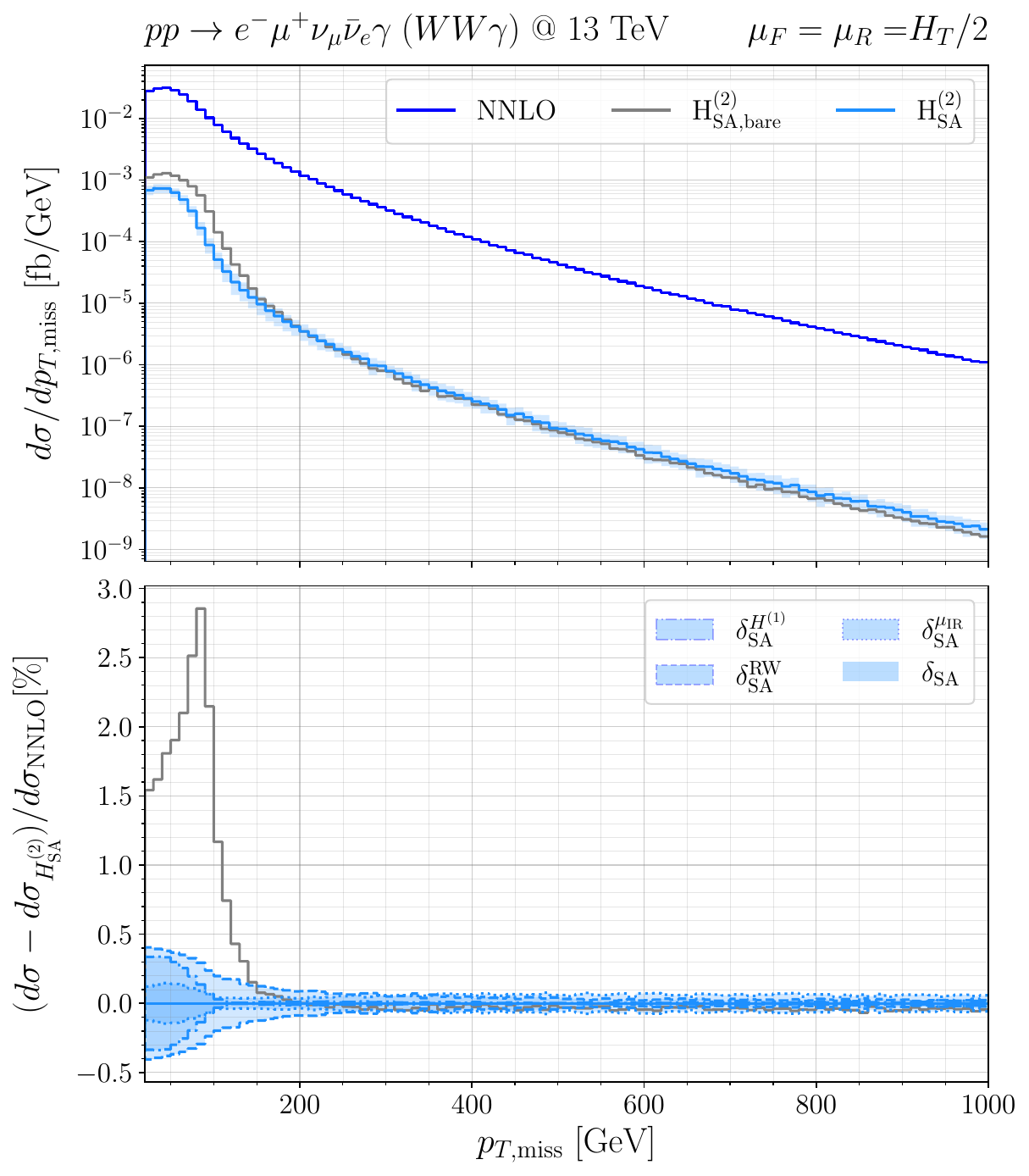}
\hfill 
\includegraphics[height=\plotheightapp]{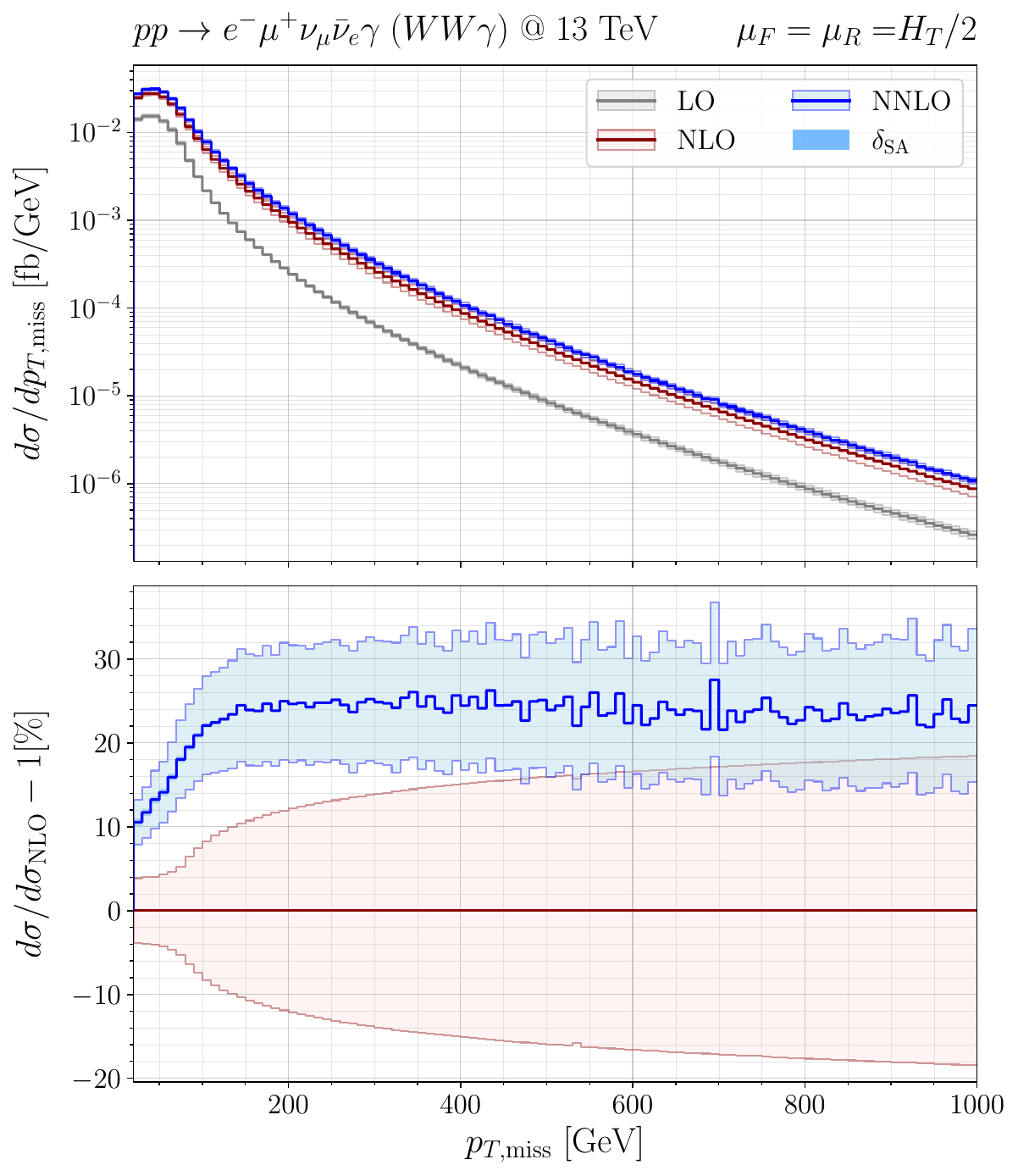}\\[2ex]

\includegraphics[height=\plotheightapp]{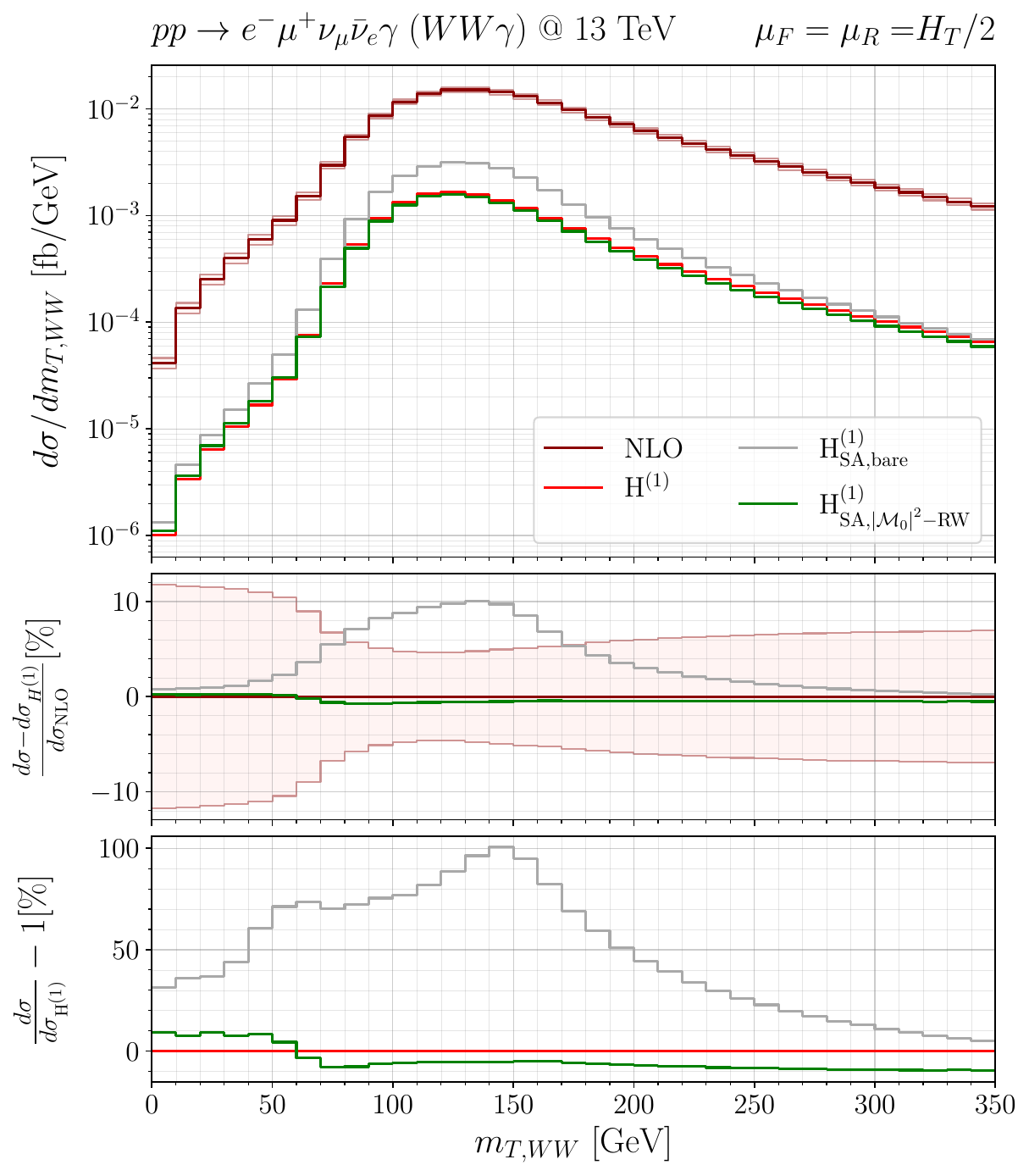}
\hfill 
\includegraphics[height=\plotheightapp]{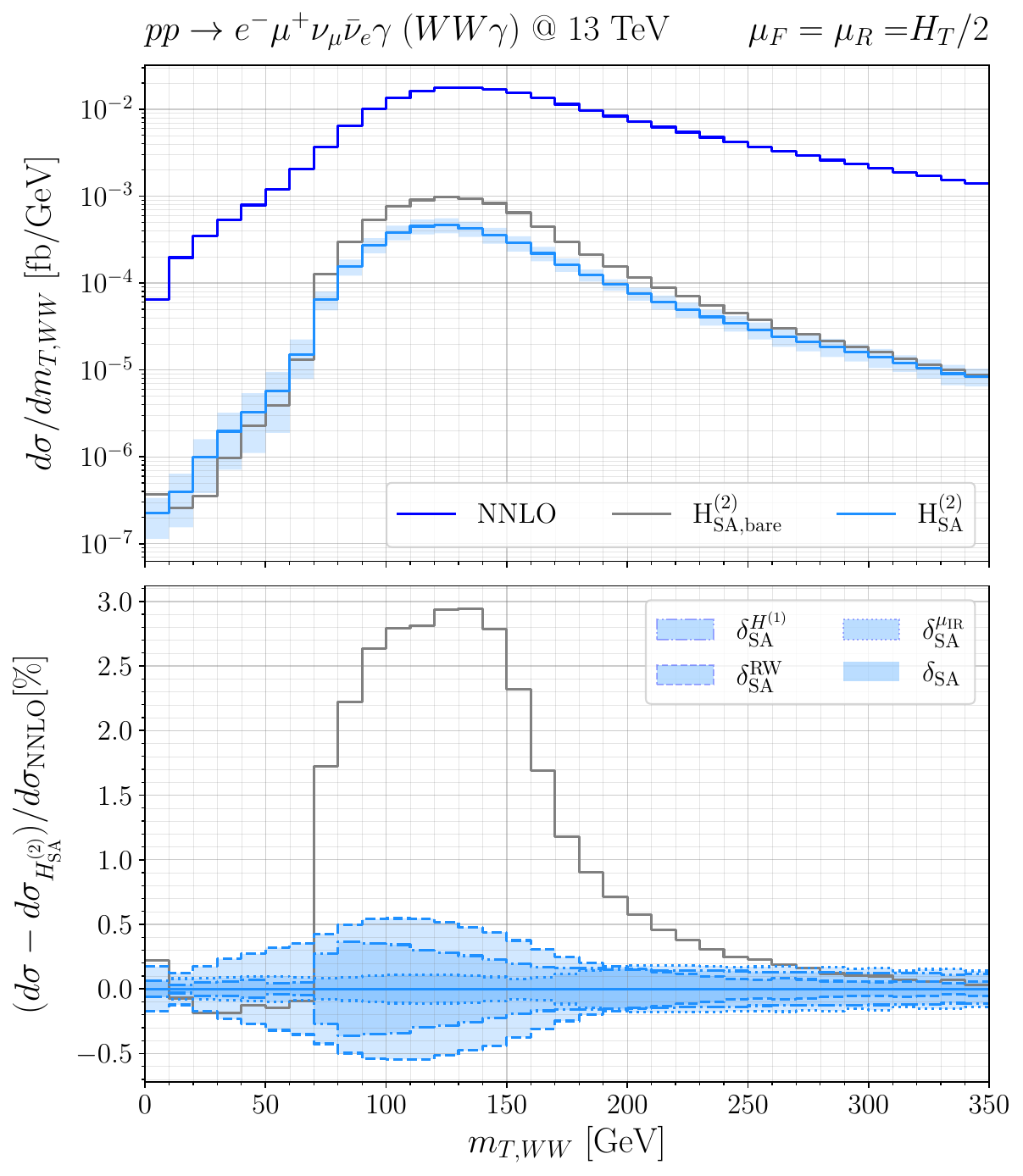}
\hfill 
\includegraphics[height=\plotheightapp]{figures/ppemxnmnexa05_LHC13_CMS_Kfactors_NNLO_mT_WW.pdf}\\[2ex]

\caption{\label{WWAplots_pT_e_pT_miss_mT_WW} Distributions in the electron transverse momentum, $p_{T,e^-}$ (first row), the missing transverse momentum, $\pTmiss$ (second row),
  and the transverse mass of the \WW system, $m_{T,WW}$ (third row), for \WWgamma production. The plots follow the descriptions in
  \reffi{fig:errorWWAH1based} (left column),
  \reffi{fig:errorWWAcombined} (central column) and
  \reffi{fig:errorWWAKfactors} (right column), respectively.
  For reference, the \dsHtwoSAnoRW contribution is added in the plots of the central column.
}
\end{figure}

\begin{figure}[p]
\centering
\includegraphics[height=\plotheightapp]{figures/ppemxnmnexa05_LHC13_CMS_error_estimate_NLO_H1_eta_gamma.pdf}
\hfill 
\includegraphics[height=\plotheightapp]{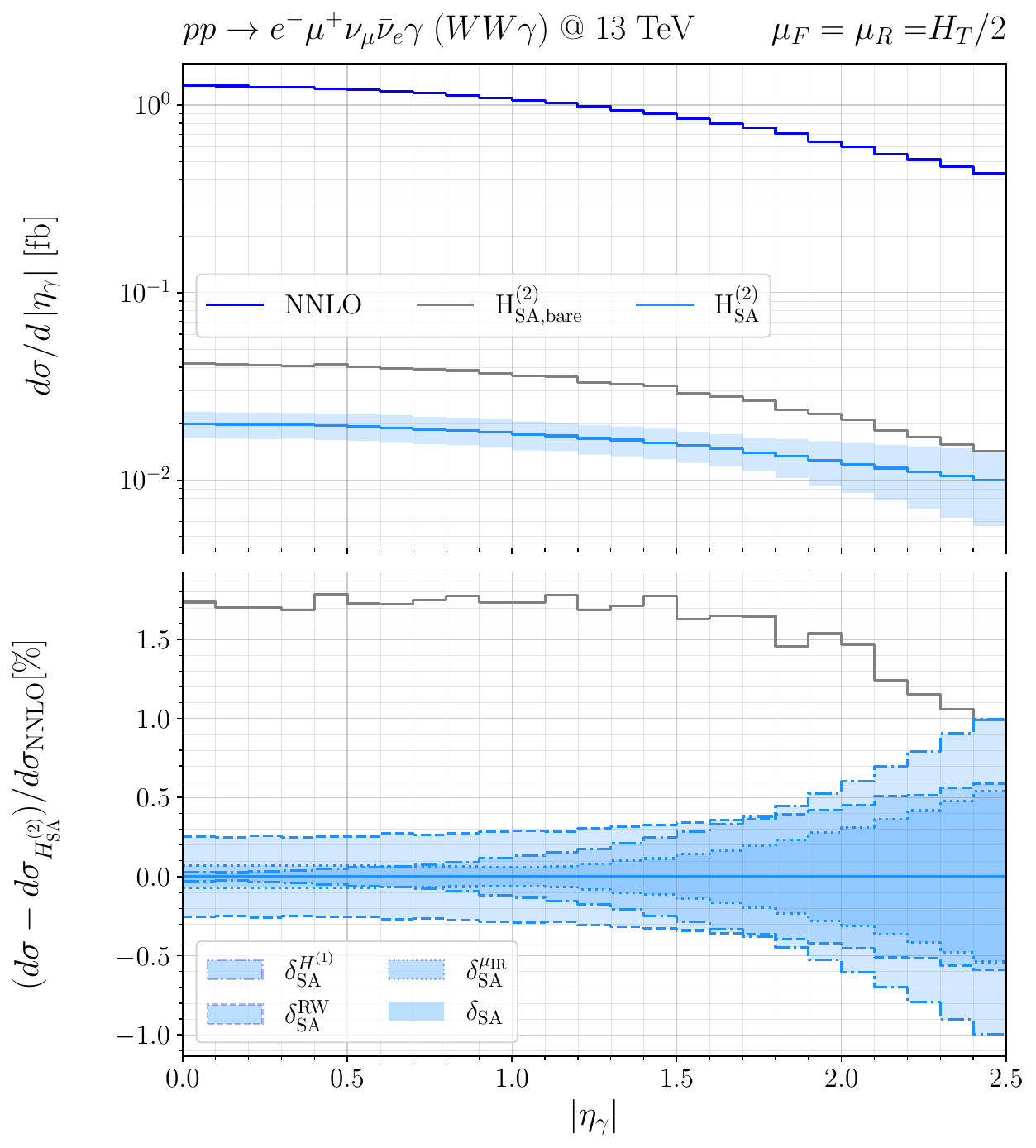}
\hfill 
\includegraphics[height=\plotheightapp]{figures/ppemxnmnexa05_LHC13_CMS_Kfactors_NNLO_eta_gamma.pdf}\\[2ex]

\includegraphics[height=\plotheightapp]{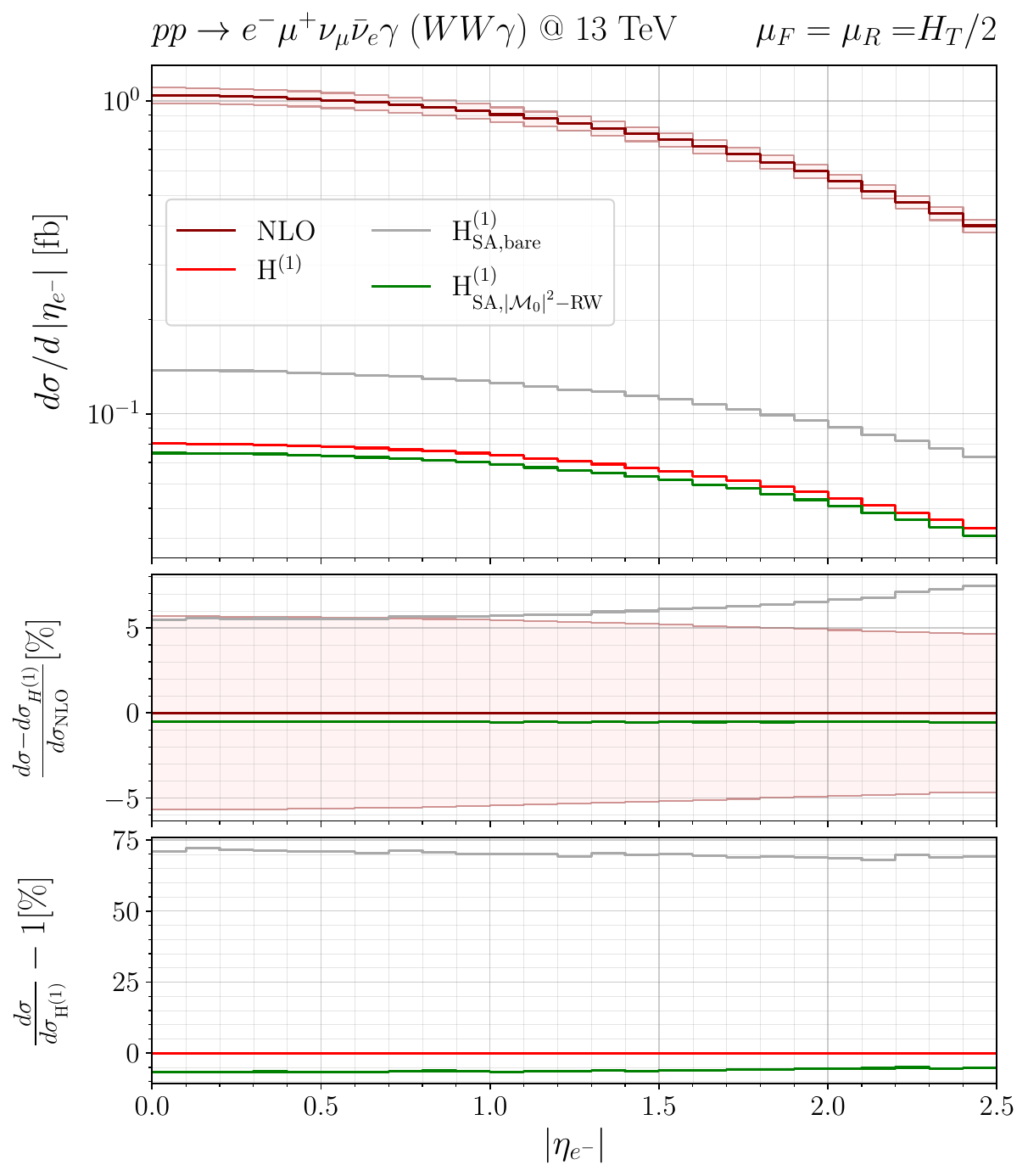}
\hfill 
\includegraphics[height=\plotheightapp]{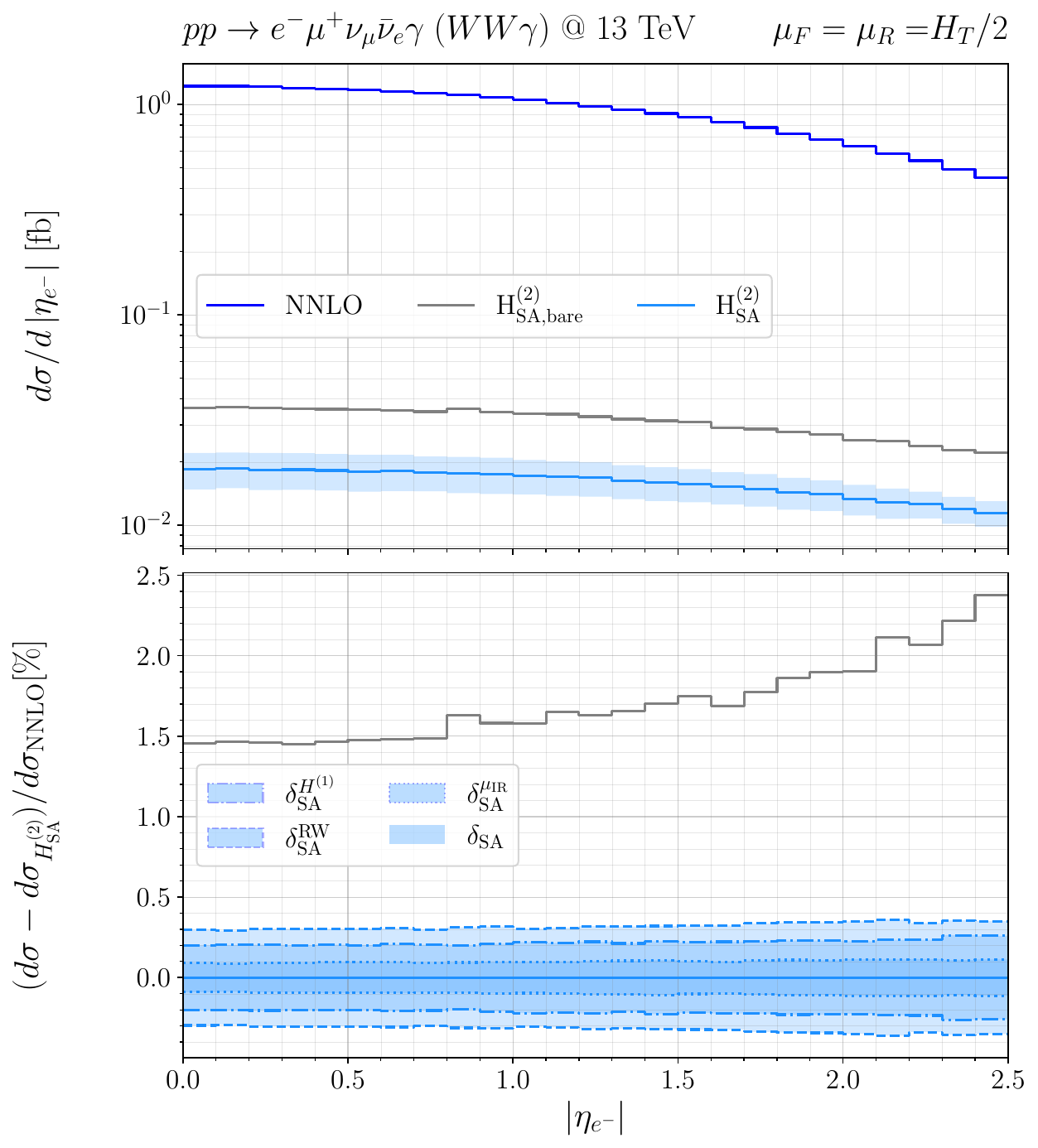}
\hfill 
\includegraphics[height=\plotheightapp]{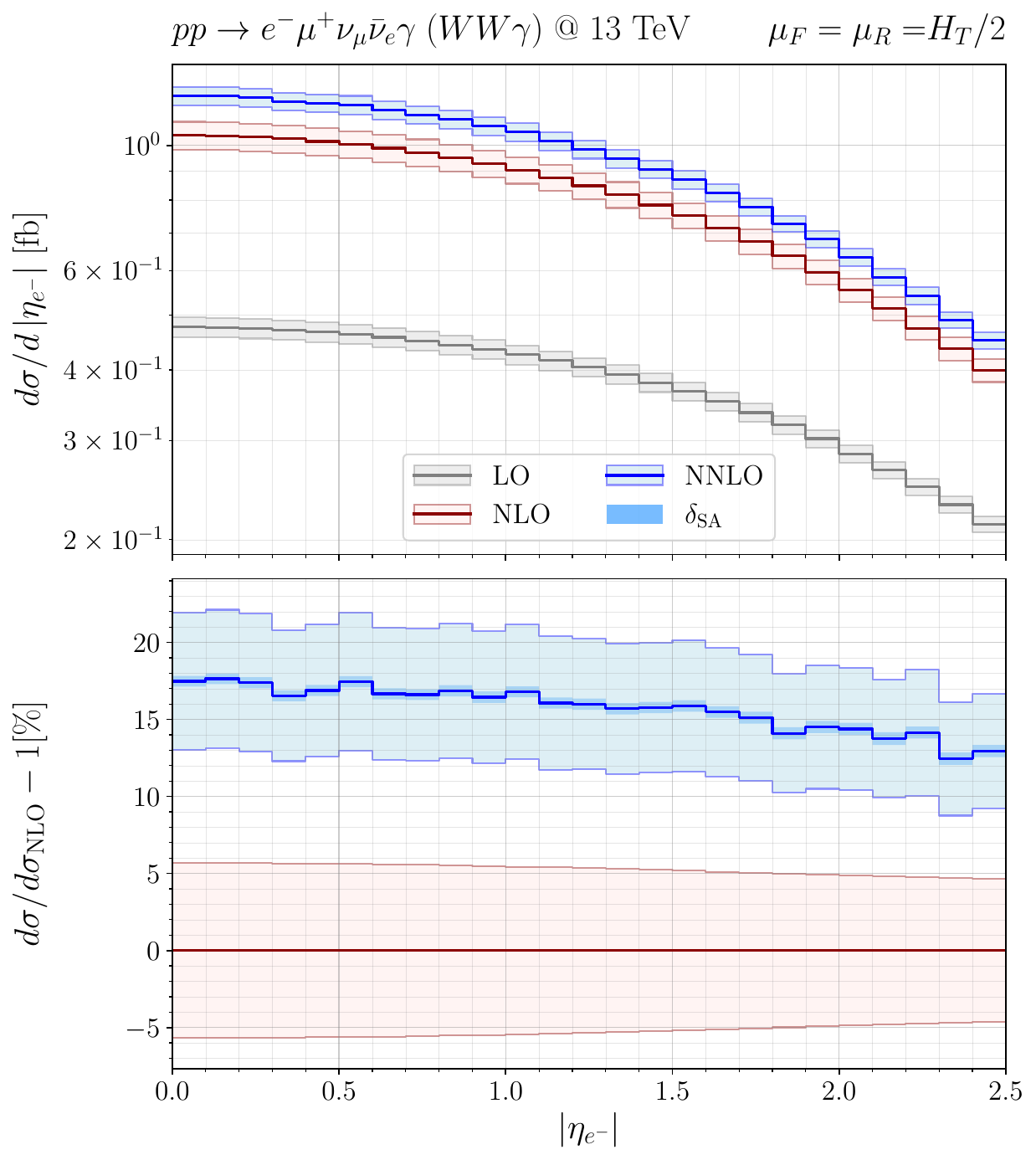}\\[2ex]

\includegraphics[height=\plotheightapp]{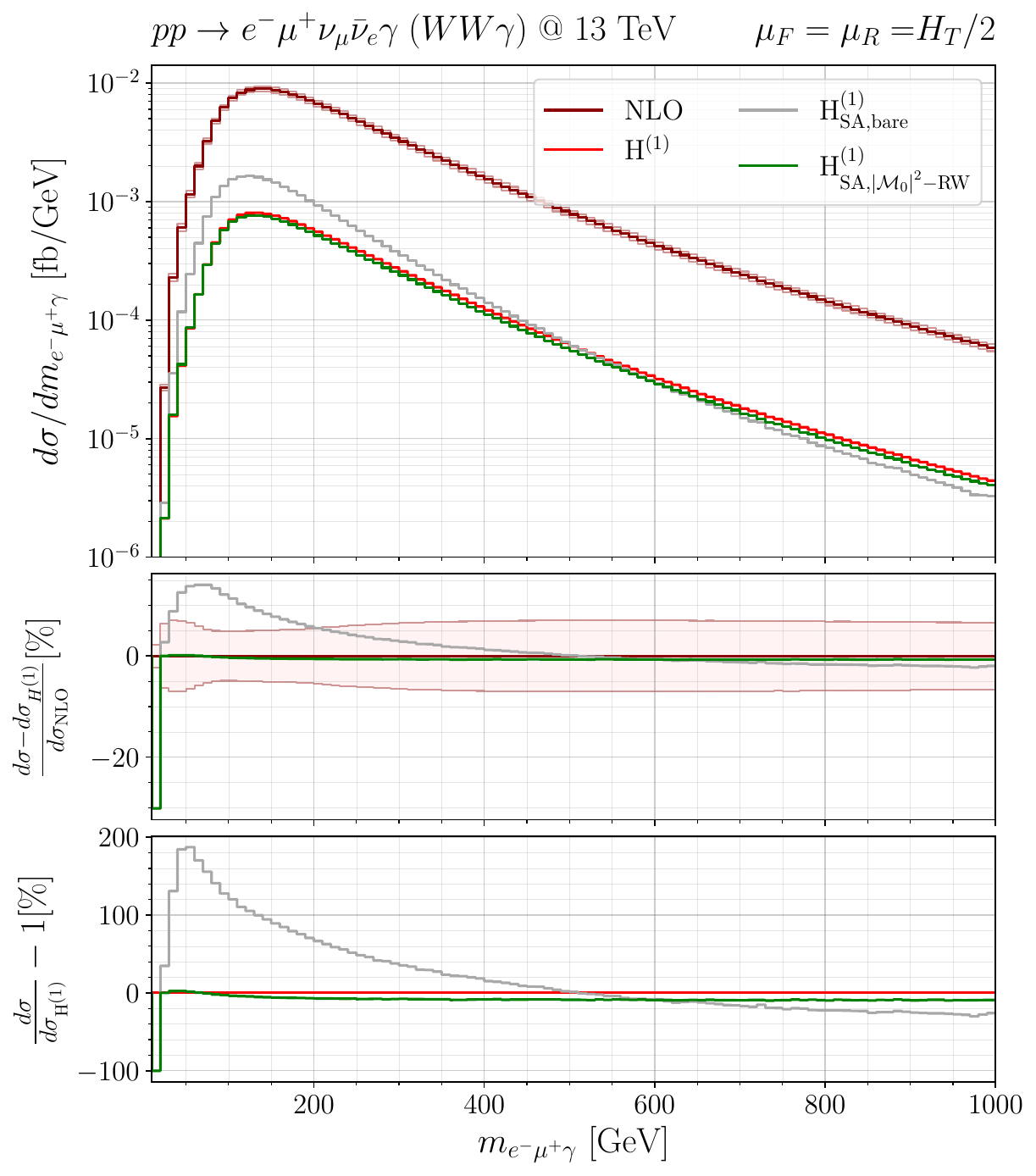}
\hfill 
\includegraphics[height=\plotheightapp]{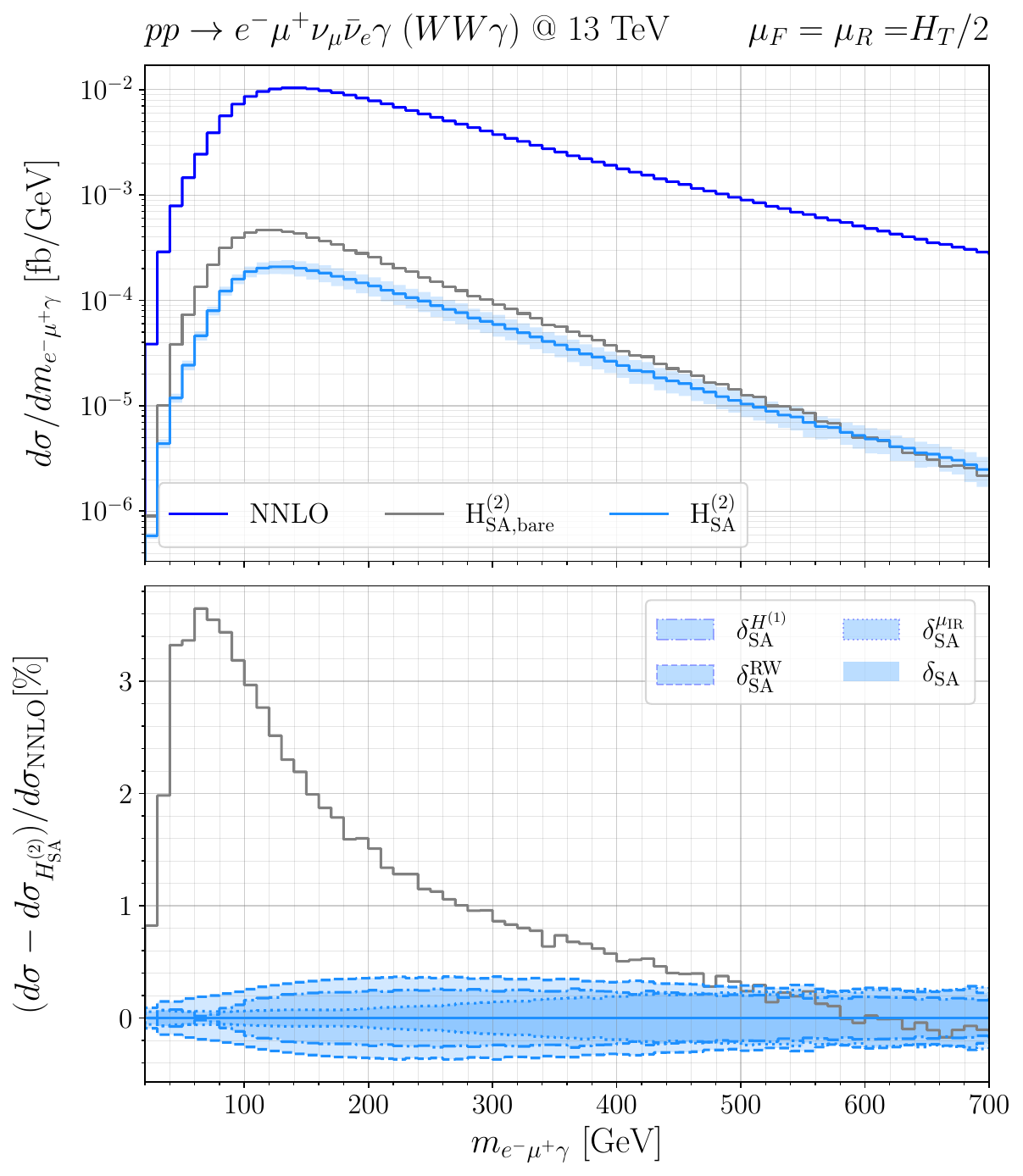}
\hfill 
\includegraphics[height=\plotheightapp]{figures/ppemxnmnexa05_LHC13_CMS_Kfactors_NNLO_m_llgamma.pdf}\\[2ex]

\caption{\label{WWAplots_eta_gamma_eta_em_m_lep_lep_gamma}Distributions
  in the absolute pseudo-rapidities of the photon, $\eta_{\gamma}$ (first row), and
  the electron, $\eta_{e^-}$ (second row), 
  and in the invariant mass of the dilepton--photon system, $m_{e^-\mu^+\gamma}$ (third row), for \WWgamma production. The plots follow the descriptions in
  \reffi{fig:errorWWAH1based} (left column),
  \reffi{fig:errorWWAcombined} (central column) and
  \reffi{fig:errorWWAKfactors} (right column), respectively.
  For reference, the \dsHtwoSAnoRW contribution is added in the plots of the central column.
}
\end{figure}

\begin{figure}[p]
\centering
\includegraphics[height=\plotheightapp]{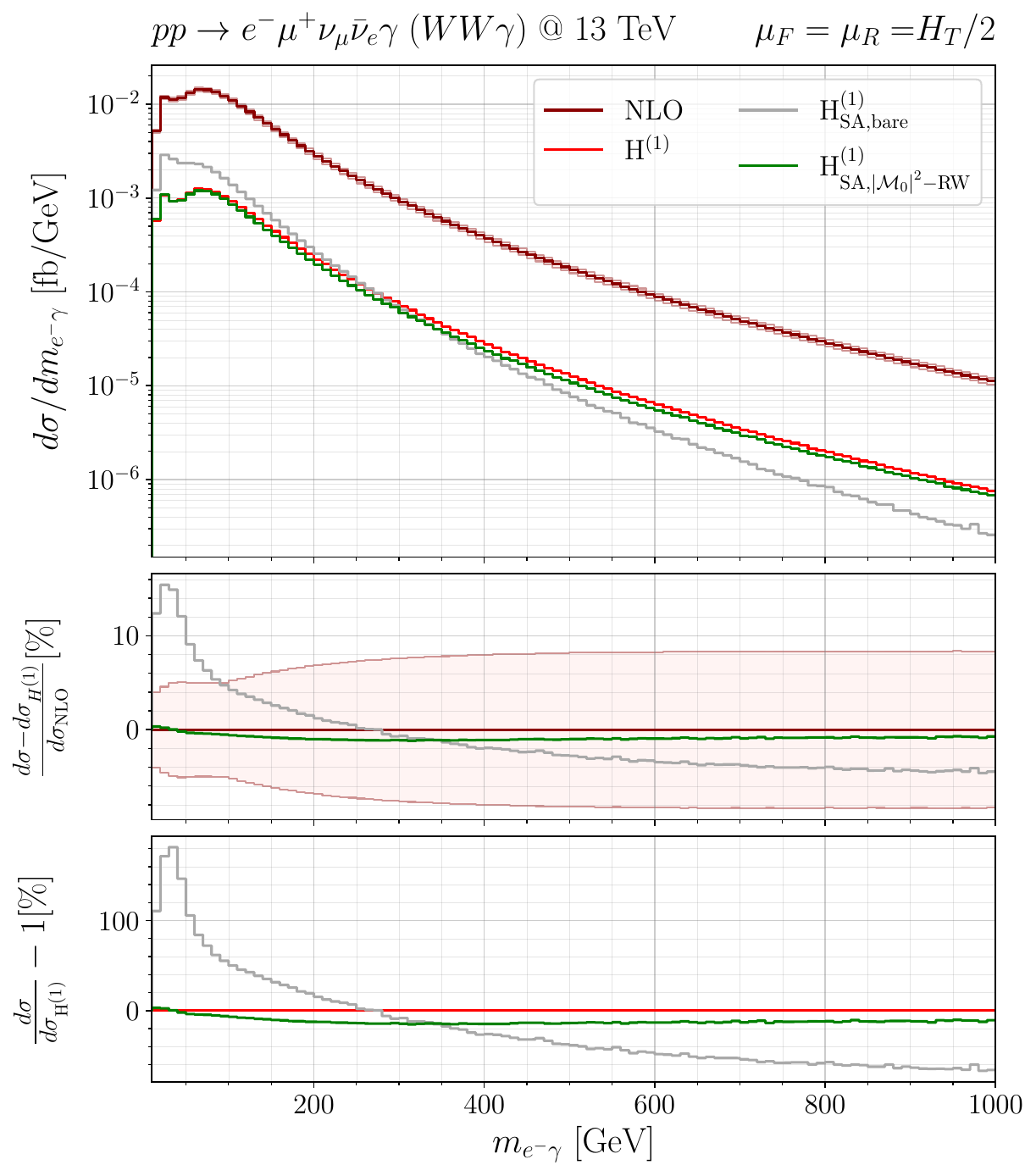}
\hfill 
\includegraphics[height=\plotheightapp]{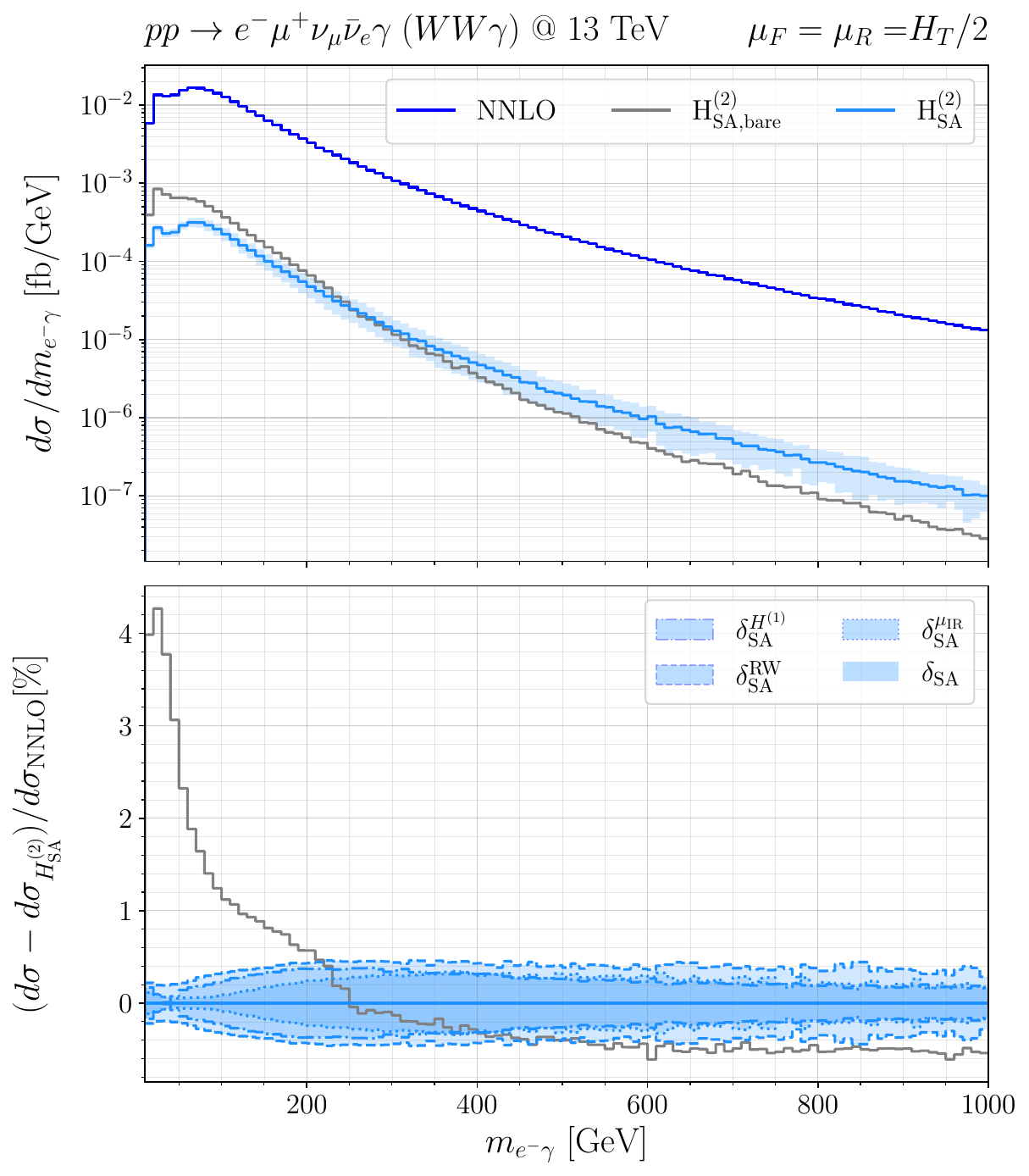}
\hfill 
\includegraphics[height=\plotheightapp]{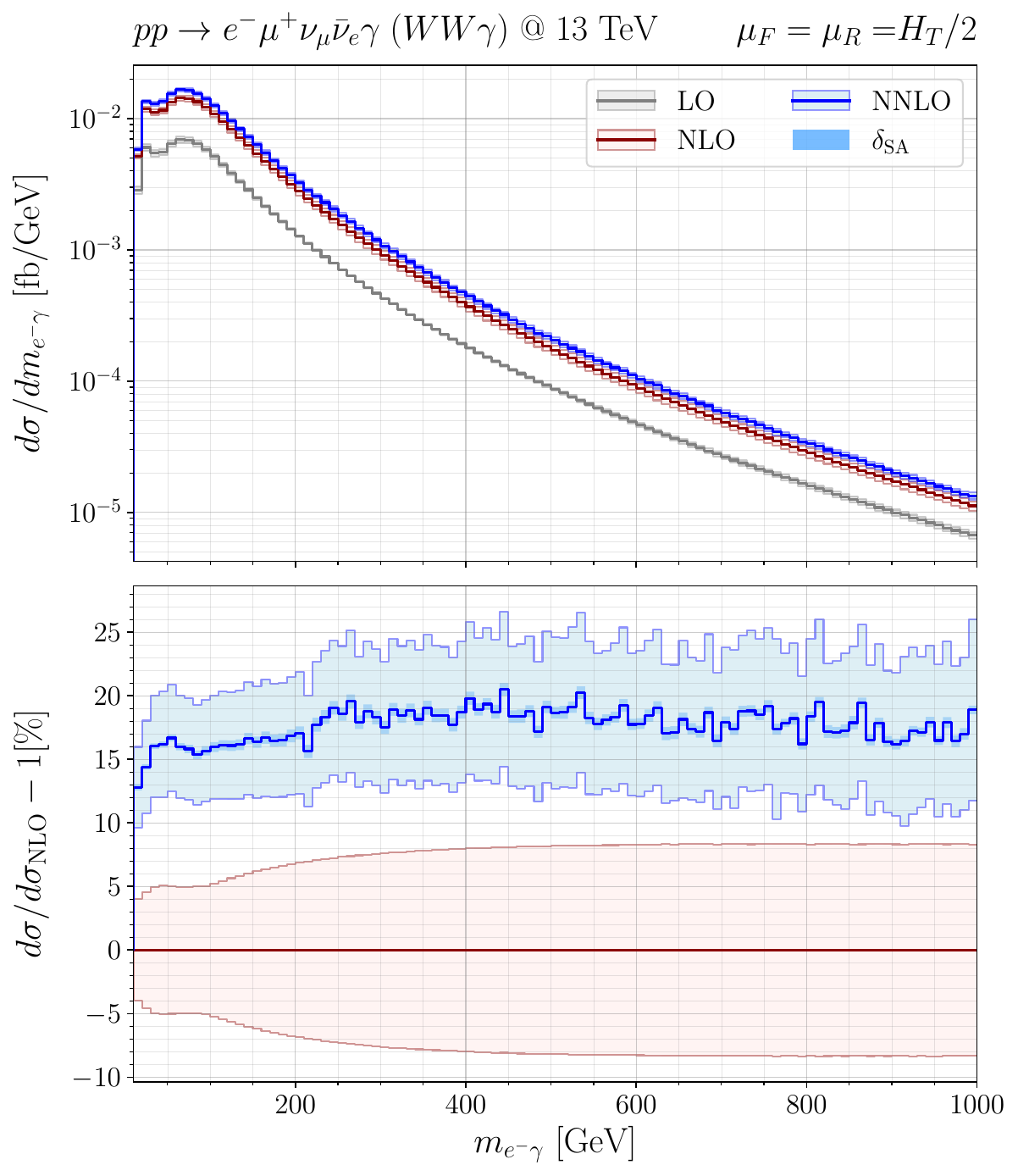}\\[2ex]

\includegraphics[height=\plotheightapp]{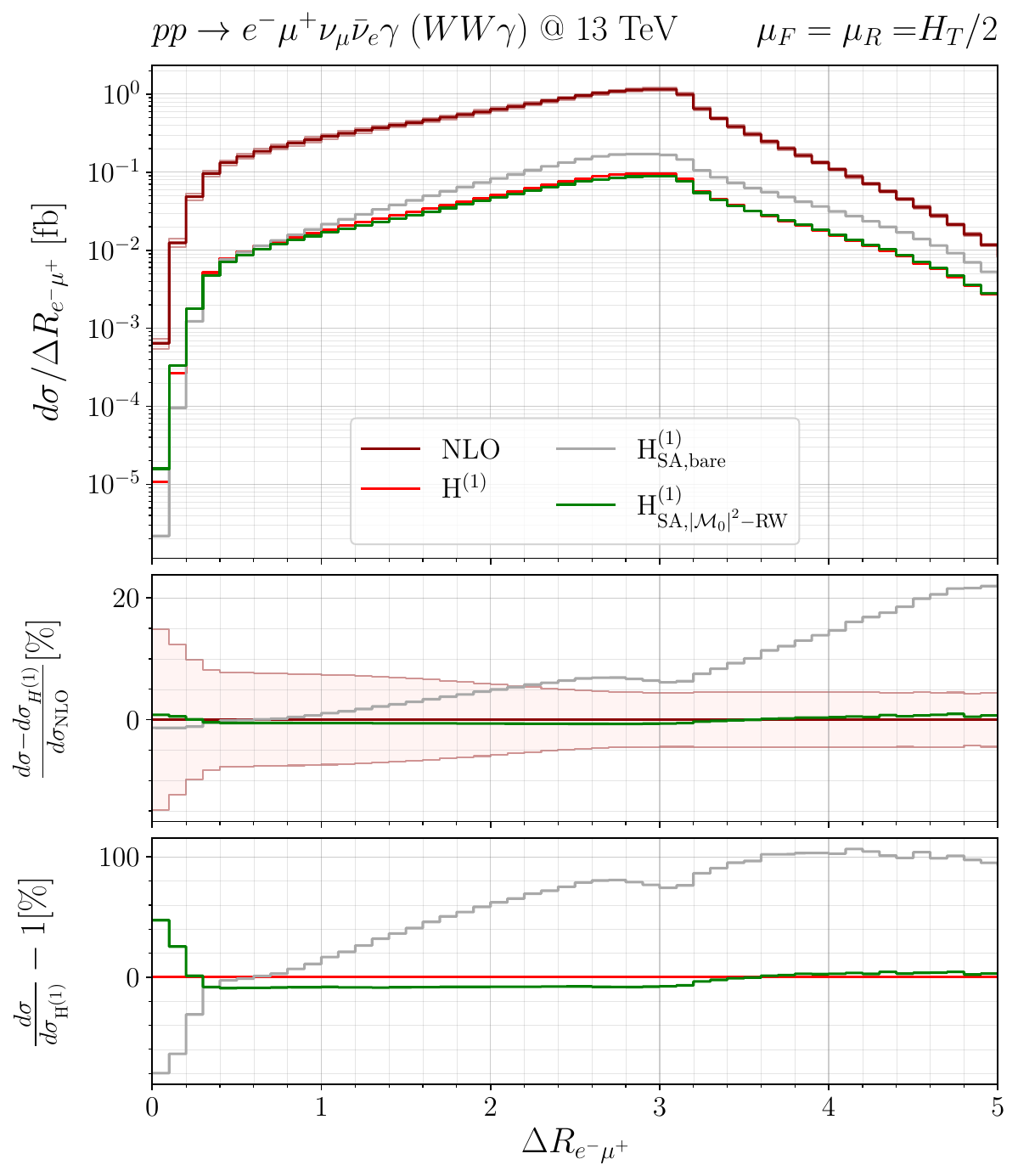}
\hfill 
\includegraphics[height=\plotheightapp]{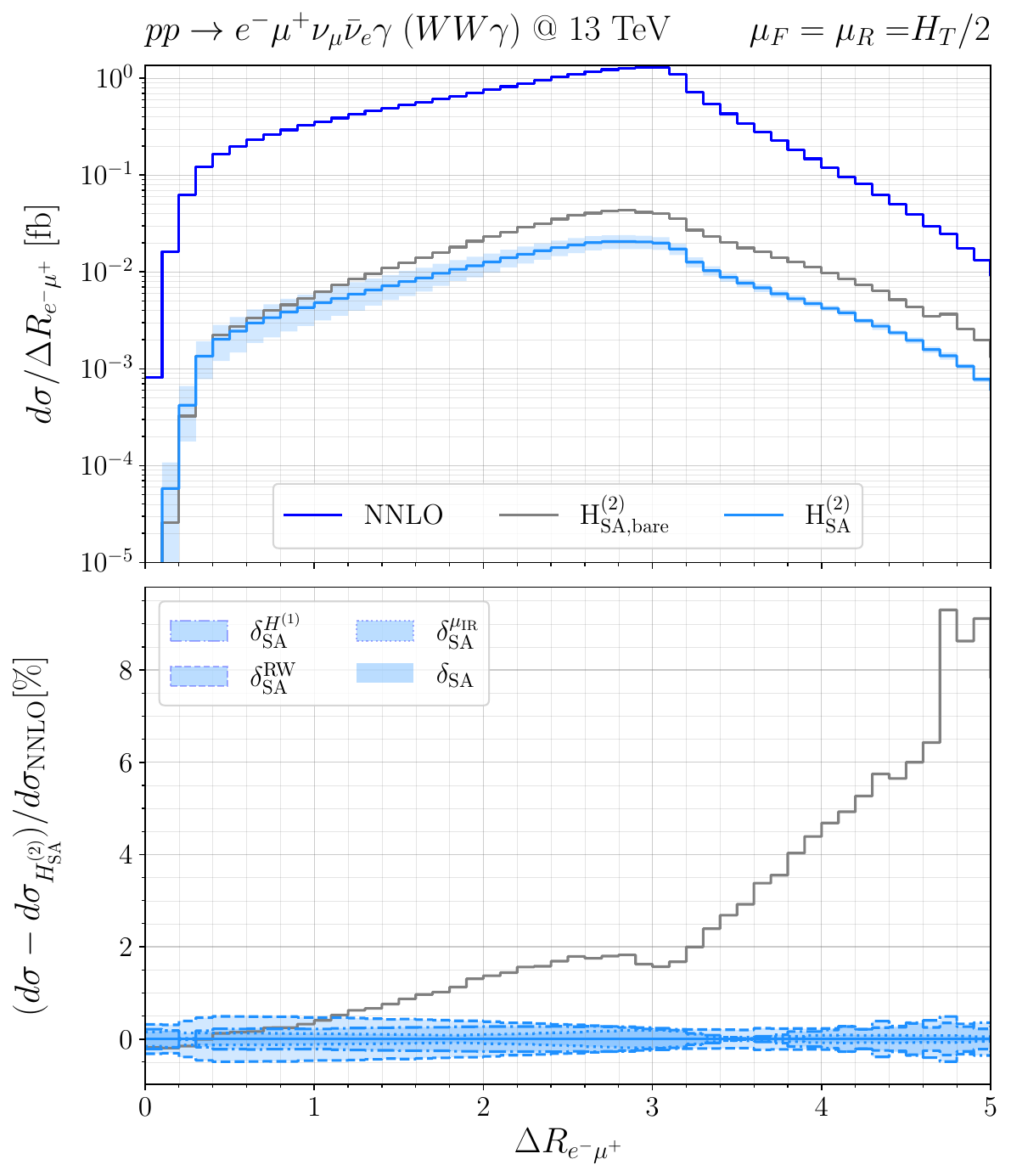}
\hfill 
\includegraphics[height=\plotheightapp]{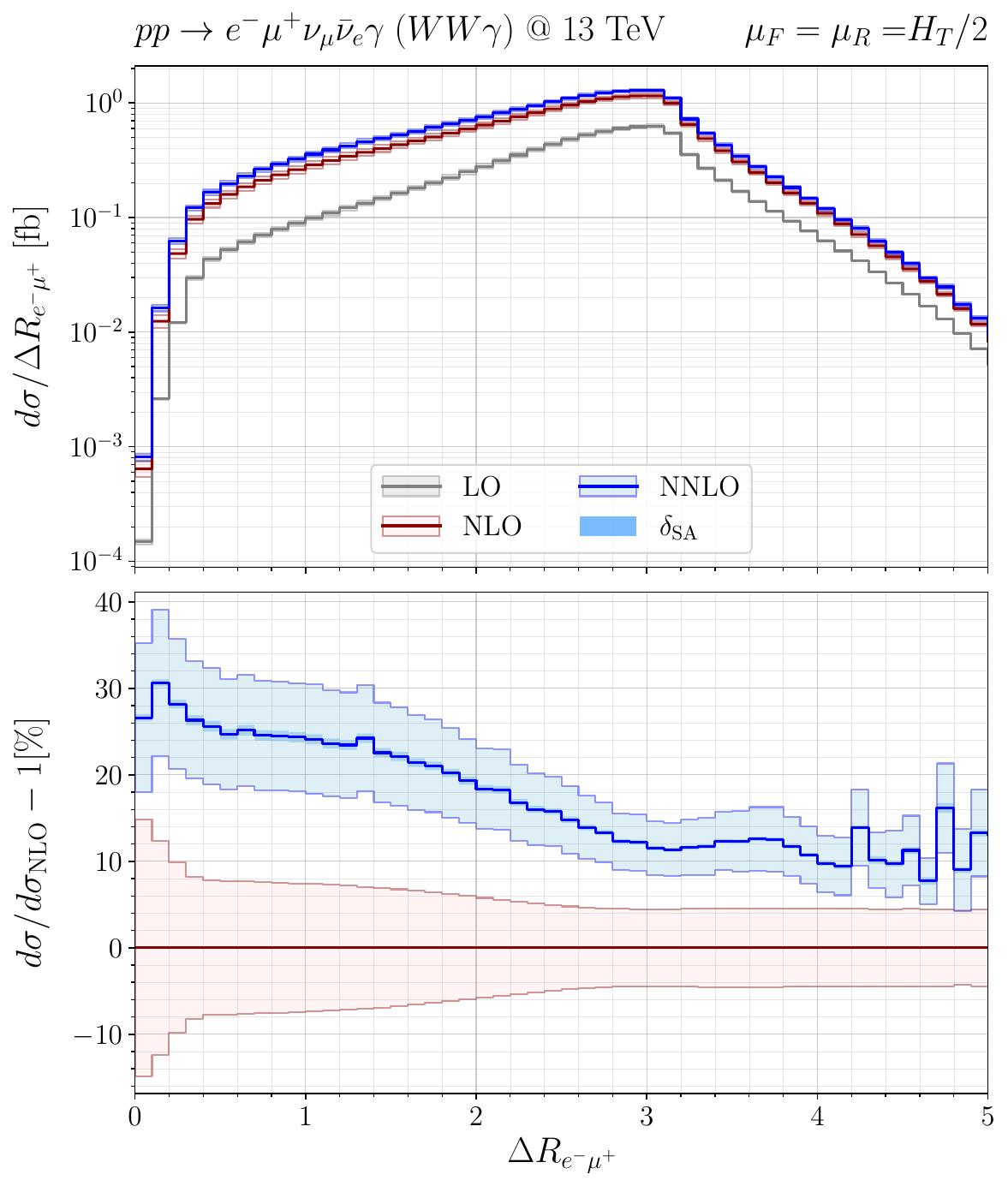}\\[2ex]

\includegraphics[height=\plotheightapp]{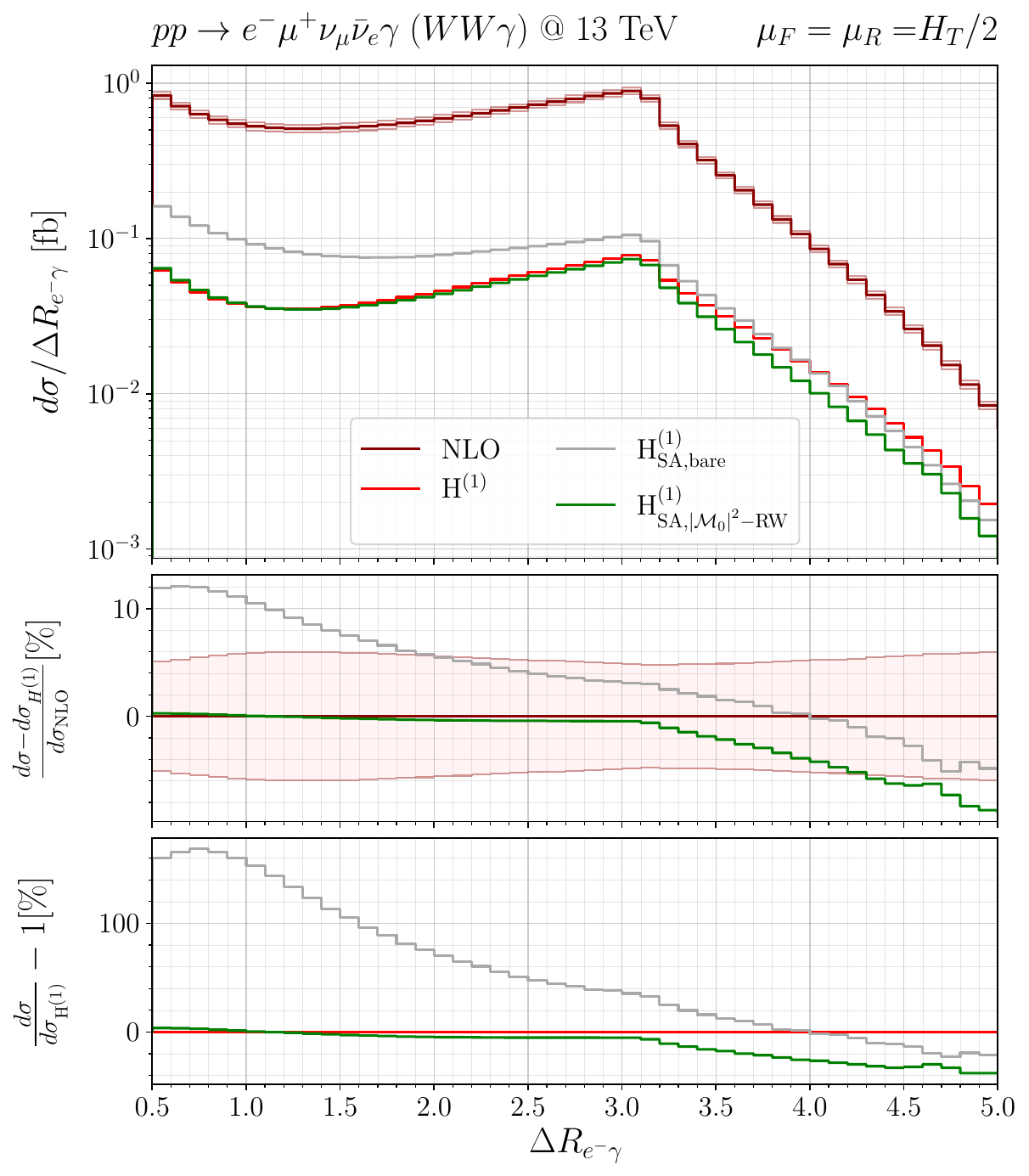}
\hfill 
\includegraphics[height=\plotheightapp]{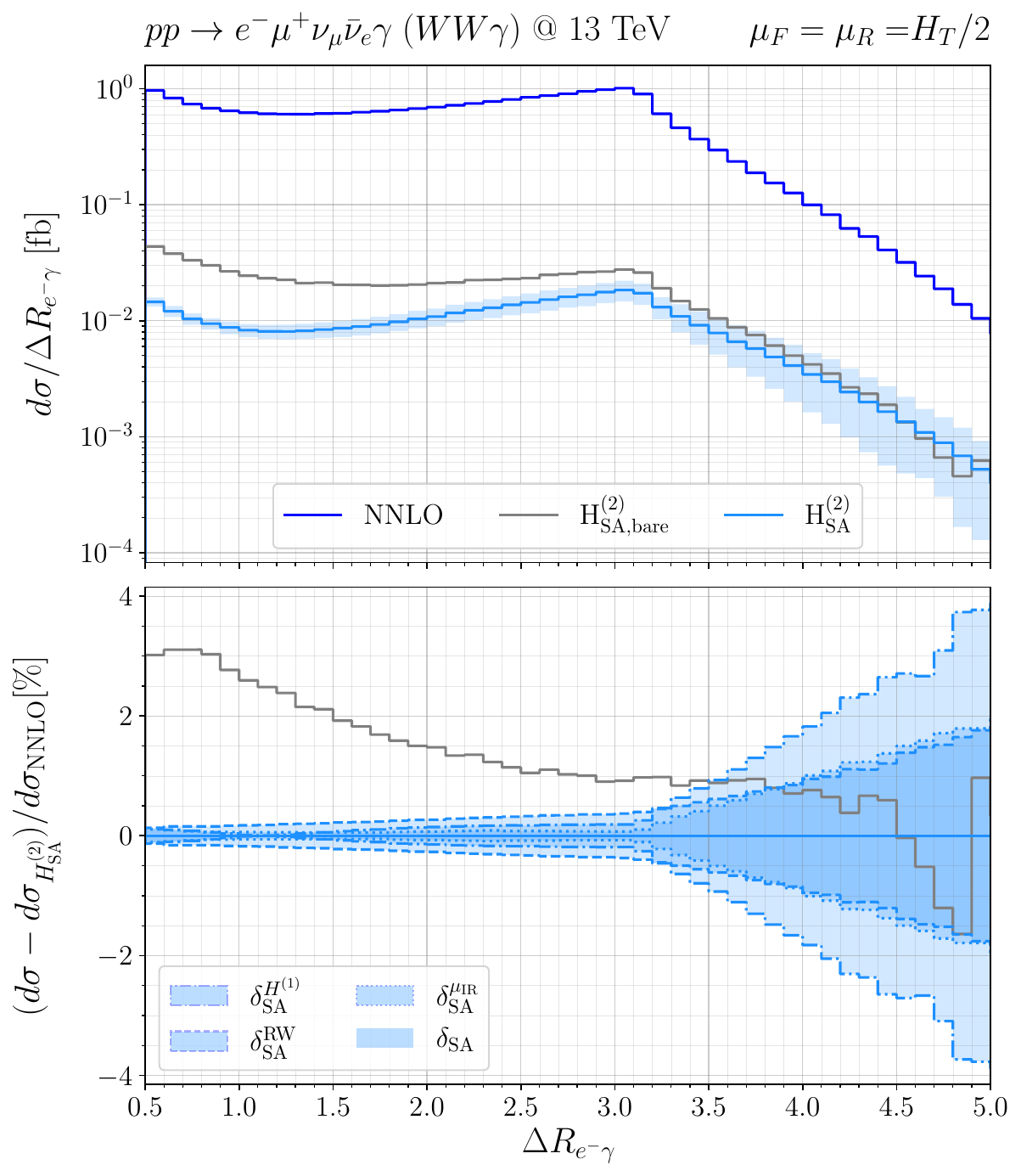}
\hfill 
\includegraphics[height=\plotheightapp]{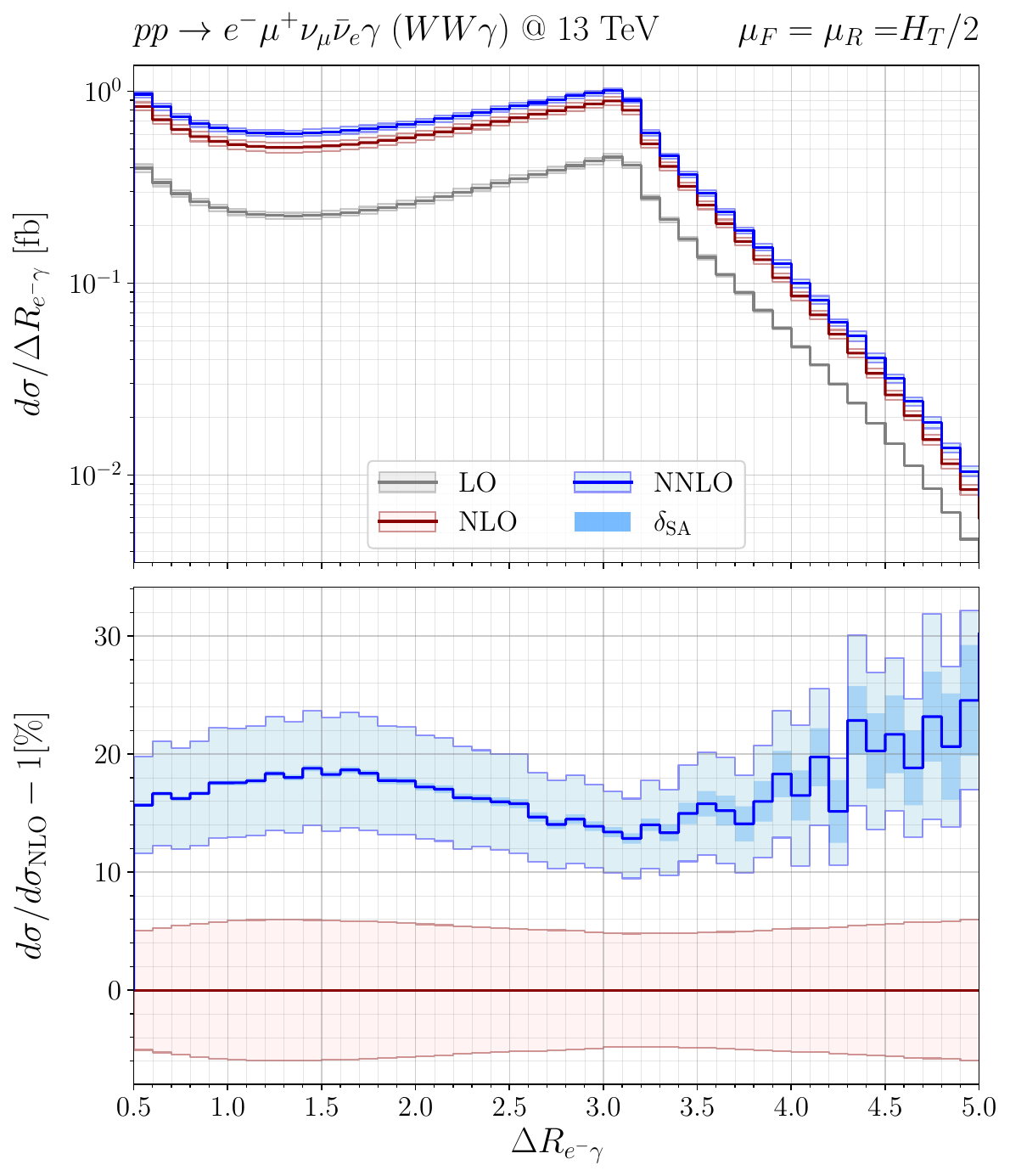}\\[2ex]

\caption{\label{WWAplots_dm_em_mup_dR_em_mup_dR_em_gamma} Distributions
  in the invariant mass of the electron--photon system, $m_{e^-\gamma}$ (first row), and
  in the distances in the $\phi$--$\eta$ plane between the two leptons, $\Delta R_{e^-\mu^+}$ (second row), as well as 
  the electron and the photon, $\Delta R_{e^-\gamma}$ (third row), for \WWgamma production. The plots follow the descriptions in
  \reffi{fig:errorWWAH1based} (left column),
  \reffi{fig:errorWWAcombined} (central column) and
  \reffi{fig:errorWWAKfactors} (right column), respectively.
  For reference, the \dsHtwoSAnoRW contribution is added in the plots of the central column.
}
\end{figure}

In detail, we provide the photon transverse-momentum distribution, \pTgamma, in \reffi{WWAplots_pT_gamma},
and the electron transverse-momentum distribution, \pTem,
the missing transverse-momentum distribution, \pTmiss,
and the transverse-mass distribution of the \WW system, $m_{T,WW}$, in \reffi{WWAplots_pT_e_pT_miss_mT_WW}.
In \reffi{WWAplots_eta_gamma_eta_em_m_lep_lep_gamma}, we show the distributions in the
absolute pseudo-rapidities of the photon, $|\eta_{\gamma}|$, and the electron, $|\eta_{e^-}|$,
as well as the invariant-mass distribution of the dilepton--photon system, $m_{e^-e^+\gamma}$.
Finally, we present the invariant-mass distribution of the electron--photon system, $m_{e^-\gamma}$,
together with the distributions in the distances in the $\phi$--$\eta$ plane
between the electron and the anti-muon, $\Delta R_{e^-\mu^+}$, and between the electron and the photon,
$\Delta R_{e^-\gamma}$, in \reffi{WWAplots_dm_em_mup_dR_em_mup_dR_em_gamma}.

\clearpage

\bibliography{biblio.bib}

\end{document}